\documentclass[12pt]{article}
\usepackage[utf8]{inputenc}
\usepackage{amsmath}
\usepackage{siunitx,letltxmacro}
\LetLtxMacro{\svqty}{\qty}
\LetLtxMacro{\qty}{\svqty}
\usepackage[font=small,labelfont=bf]{caption}
\usepackage{booktabs}
\usepackage{tabularx}
\usepackage{pgfgantt}
\usepackage{graphicx}
\usepackage{wrapfig}
\usepackage{hyperref}
\usepackage{listings}
\usepackage{color}
\usepackage{mathtools}
\usepackage{appendix}
\usepackage{mathcomp}
\usepackage[]{lipsum}
\usepackage{amssymb}
\usepackage{multirow}
\usepackage{xcolor}
\usepackage[
    style=numeric-comp, 
    sorting=none,       
    backend=biber,      
    giveninits=true,    
]{biblatex}
\DeclareFieldFormat{date}{\thefield{year}}
\DeclareFieldFormat[unpublished]{date}{\mkbibparens{\thefield{year}}}
\usepackage[export]{adjustbox}
\usepackage[letterpaper, margin=1in, bottom=0.75in, includefoot]{geometry} 
\usepackage{xcolor}
\usepackage{hepparticles}
\usepackage[T1]{fontenc}
\usepackage{setspace}
\usepackage{pdfpages}

\hypersetup{
pdftitle={DEVELOPMENT OF CRYOGENIC SCINTILLATION DETECTORS FOR THE SEARCH OF NEW PHYSICS},
pdfsubject={Topic (subtopic)},
pdfauthor={Keyu Ding},
pdflang={en}
}

\begin{document}

\color{black} 
\newpage
\begin{titlepage}
\begin{center}

\vspace*{1 cm}\textbf{DEVELOPMENT OF CRYOGENIC SCINTILLATION DETECTORS FOR THE SEARCH OF NEW PHYSICS}

\vspace{3 cm}

By

\vspace{0.5 cm}

Keyu Ding

\vspace{1 cm}

B.S., Hubei Normal University, 2019

M.S., University of South Dakota, 2022

\vspace{6.5 cm}

A Dissertation Submitted in Partial Fulfillment of 

the Requirements for the Degree of 

Doctor of Philosophy

\rule{\textwidth/2}{0.5pt}

Department of Physics
\vspace{0.8 cm}

Physics Program

In the Graduate School

The University of South Dakota

May 2024

\end{center}
\end{titlepage}

\newpage
\begin{center}

\vspace*{\fill}
Copyright by 

KEYU DING

2024

All Rights Reserved
\end{center}
\thispagestyle{empty}

\clearpage
\pagenumbering{roman}

\newpage
\addcontentsline{toc}{section}{Committee Signature page}
\includepdf[pages=-,pagecommand={}]{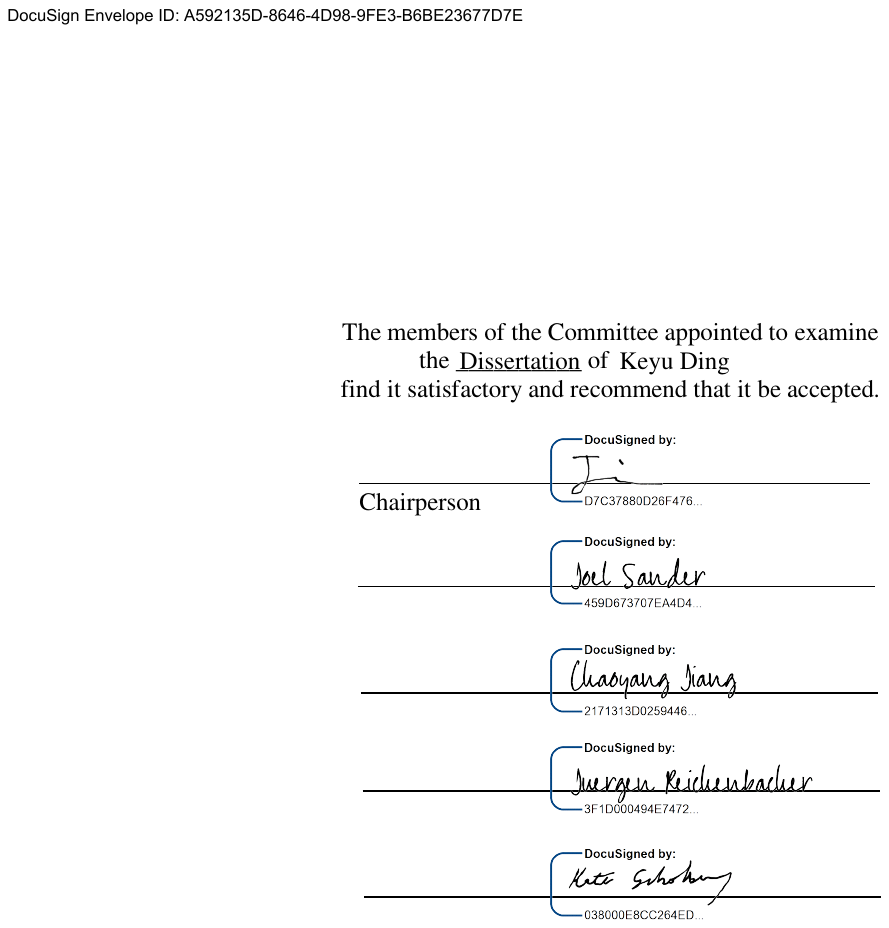}

\newpage
\addcontentsline{toc}{section}{Abstract}
\begin{center}
\section*{}
ABSTRACT
\end{center}
\noindent~~Coherent elastic neutrino-nucleus scattering (CEvNS) was first proposed in 1974. Despite having the largest cross-section among all low-energy neutrino couplings predicted in the Standard Model (SM), CEvNS detection remains challenging due to its only experimental signature being a low-energy nuclear recoil. In 2017, the COHERENT collaboration successfully observed CEvNS for the first time. A 14.6 kg low-background doped CsI at room temperature was placed 20 meters away from the 1.4 MW Spallation Neutron Source (SNS) at the Oak Ridge National Laboratory (ORNL). The SNS's pulsed proton beam provides exceptional background rejection and high-intensity neutrinos, making it ideal for CEvNS detections. 

\noindent~~CryoCsI, the proposed prototype, is a cryogenic undoped CsI scintillating detector, which has a much lower energy threshold potentially down to 0.5 keV$_{nr}$ compared to the doped CsI. This enhanced sensitivity of CryoCsI allows for the observation of more CEvNS events. Precise measurements of CEvNS can not only validate the predictions of the SM but also explore new physics. In conjunction with other COHERENT detectors, CryoCsI has the potential to achieve world-leading sensitivities in a broad range of physics topics within and beyond the SM. The sensitivities of CryoCsI to hidden-sector dark matter, non-standard neutrino interactions, and neutron radius are explored.

\noindent~~This thesis delves into the construction of CryoCsI and efforts to enhance its light yield from 20 to $50 \pm 2$ photoelectrons (PE) per keV electron-equivalent (keV$_{ee}$). It will address challenges with cryogenic SiPMs, including inferior energy resolution, optical cross-talk, and potential limitations on detecting rare events. Understanding the light yield of scintillating detectors for nuclear recoils is crucial, as explored through alpha-particle and neutron quenching factor (QF) measurements. A QF of approximately 15\% was measured using a neutron beam at the Triangle Universities Nuclear Lab (TUNL). Proposed solutions to challenges like the overshoot effect observed in PMTs will be discussed. Additionally, the thesis will explore design considerations for minimizing background noise and optimizing the CsI crystal's shape through optical simulations.

\vspace{2.5 cm}

\hspace*{\fill} \rule{\textwidth/2}{0.5pt}

\hspace{9 cm}Dissertation Advisor Approval

\newpage
\addcontentsline{toc}{section}{Acknowledgements}
\section*{Acknowledgements}
\hspace{0.5cm} As pursuing a second PhD is not in my stars so I'm seizing this one and only moment to express my profound gratitude to the countless individuals who have supported me along the way. 

First, I extend my heartfelt gratitude to my advisor, Jing Liu, whose unlimited support and guidance have been instrumental throughout my journey. Jing has not only been a mentor but also a friend, always approachable and willing to collaborate on problem-solving.

Special recognition goes to Kate Scholberg for her boundless encouragement and responsive support. Her altruistic personality, profound curiosity, extensive knowledge, and reliability serve as a constant inspiration to me. 

I am also deeply thankful to our chairman, Yongchen Sun, whose vast knowledge spans beyond physics, making our discussions diverse and enriching. My appreciation also goes to Joel Sander and Dongming Mei for their invaluable support and insights into my academic and career pursuits. I also want to express my gratitude to Chaoyang Jiang and Juergen Reichenbacher for their valuable discussions and advice on my research.

I am immensely grateful to Dmitry Chernyak for his invaluable guidance and training at the beginning of my journey. I'm also thankful to Zishen Yang for providing guidance during my work on the SiPM WLS coating. Additionally, I extend my heartfelt thanks to my fellow colleagues at USD: Qingqing Li, Conan Dettman Bock, Abbie Song Woodard, Austin Warren, Kyler Teron Kooi, Simon Maxwell Higgason, Ruslan Podviianiuk, Lukina Oleksandra, Joseph Mammo, and Tupendra Kumar Oli, for their camaraderie and collaborative spirit throughout our time together.

I am immensely grateful to the many members of the COHERENT community whom I would like to thank for creating such a supportive and collaborative environment. This experience has truly been a privilege, and I feel fortunate to have been a part of this group. Interacting with fellow members at and after conferences has been so enriching. Special thanks to Daine Markoff for her delightful presence, unwavering encouragement and informative advice, as well as her commendable advocacy for inclusivity. I extend my thanks to Diana Parno, Jason Newby, Phil Barbeau and Rex Tayloe for their guidance and support whenever needed. I am also grateful to Dan Pershey for his invaluable guidance in sensitivity studies. Charlie Prior, Adryanna Major, Emma van Nieuwenhuizen, Janina Hakenmüller, Bo Johnson, Jay Runge, Chenguang Su, Tyler Johnson, Ben Suh, Ryan Bouabid, and all the young members, I'm grateful for the wonderful times we shared, the beers we enjoyed, and the fascinating discussions we had about supernature. I hope each of you, filled with joy and abundance, continues to shine brightly in your respective fields.

Last but certainly not least, I am profoundly grateful to my parents for their unconditional love and support. Despite not having received formal education according to societal norms, they have always encouraged me to explore and expand my knowledge. Their constant love and support have been the driving force behind my journey, instilling in me the confidence and determination. My brother, you're no one's shadow. Life's measure extends far beyond a single ruler. You will carve out your own path with strength and resilience and I'm so proud of you.

\newpage
\addcontentsline{toc}{section}{\contentsname}
\tableofcontents
\newpage
\addcontentsline{toc}{section}{\listtablename}
\listoftables
\newpage
\addcontentsline{toc}{section}{\listfigurename}
\listoffigures
\newpage
\pagenumbering{arabic}
\doublespacing


\section{Coherent Elastic Neutrino-Nucleus Scattering (CEvNS)}
\label{s:cevns}
\hspace{0.5cm} Neutrinos, virtually weightless and fundamental particles in the Standard Model (SM), are abundant yet exceptionally elusive due to their infrequent interactions with other particles. CEvNS serves as a valuable tool for studying these enigmatic particles. The following sections in this chapter will provide further explanations.
\subsection{CEvNS}
\begin{figure}[ht!]\centering
\includegraphics[width=\linewidth]{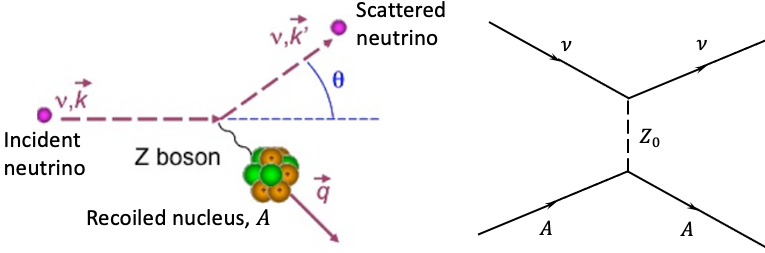}
\caption{Sketch of CEvNS interaction (left) and its corresponding Feynman diagram (right)~\cite{snowmassCO22}.}
\label{f:cevns}
\end{figure}

\hspace{0.5cm} First proposed by D. Z. Freedman in 1974~\cite{freedman74}, CEvNS is a weak neutral-current (NC) process well understood within the Standard Model of particle physics. In this process, an incoming neutrino, $\nu$, scatters off a nucleus in both a coherent and elastic way via the exchange of a $Z_0$ boson as illustrated in Fig.~\ref{f:cevns}. This interaction is characterized by a low energy and momentum transfer, making it challenging to detect. Where the neutrino arrives with energy in the tens of MeV range, its transfer of momentum to the recoiling nucleus is within a small range of sub-keV to tens-of-keV. Only in 2017, the COHERENT collaboration (see Chapter \ref{s:coherent}) first observed CEvNS using a 14.8 kg CsI[Na] detector coupled to a photomultiplier tube (PMT) at room temperature at the Spallation Neutron Source (SNS) at Oak Ridge National Laboratory~\cite{coherent17}.

\begin{figure}[ht!]\centering
\includegraphics[width=0.8\linewidth]{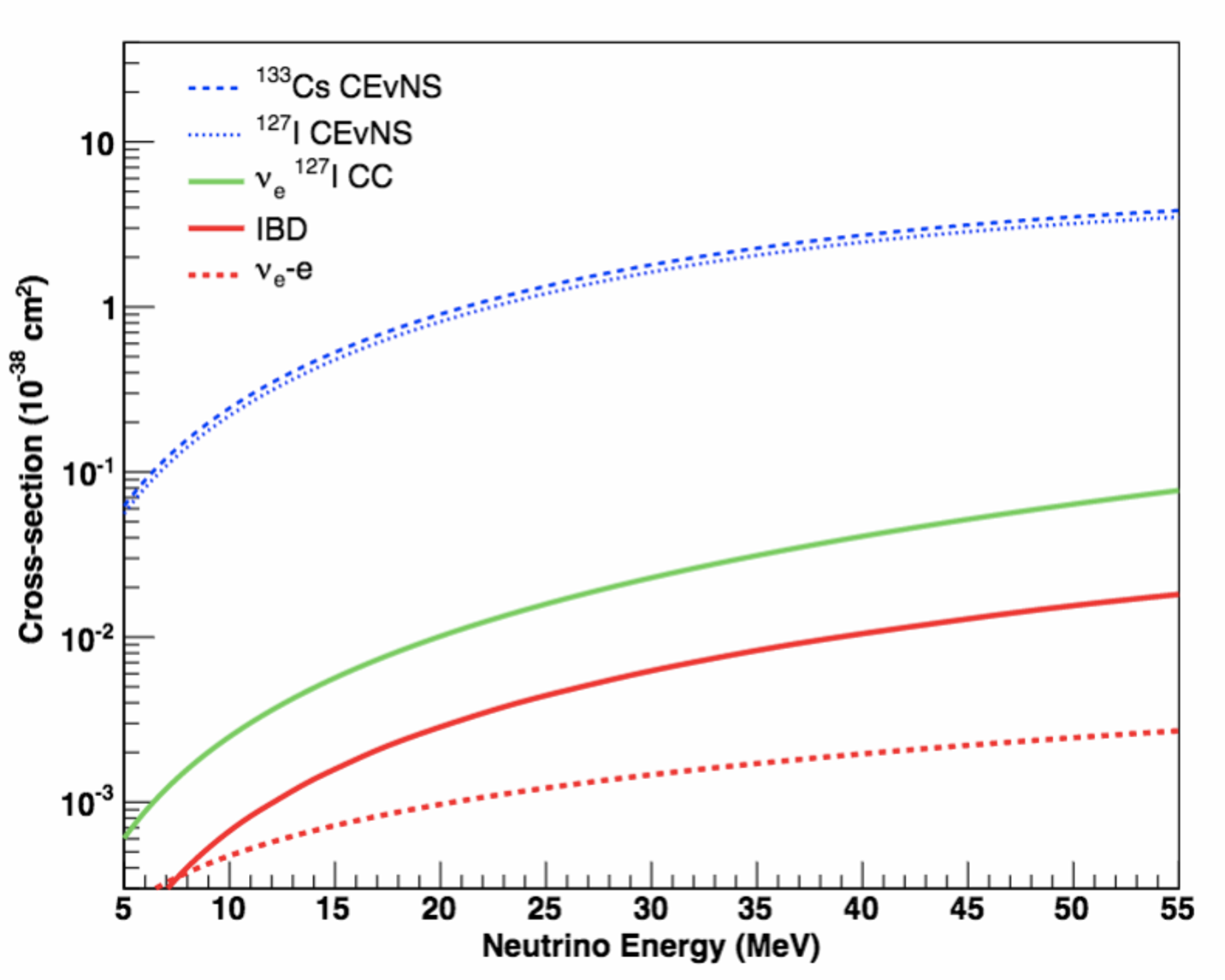}
\caption{Neutrino cross sections for neutrino energies up to 55~MeV for a CsI detector~\cite{coherent17}.}
\label{f:nuCrossSections}
\end{figure}

In CEvNS, the low-energy neutrino interacts with a nucleus as a whole. Therefore, the cross section of CEvNS is the greatest compared to when the same neutrino interacts with a nucleus/nucleon through the incoherent NC process or the Charged-Current (CC) process. This difference is even more pronounced when the neutrino interacts with an electron. CC neutrino interaction is when a neutrino/antineutrino interacts with matter via the exchange of a virtual $W$ boson. Inverse beta decay (IBD) is a CC process in which an antineutrino interacts with a proton. Predicted cross sections of relevance to a CsI detector are shown in Fig.~\ref{f:nuCrossSections}. Indeed, the cross section of CEvNS on iodine (I) is the highest. Therefore, CEvNS enables experiments with smaller mass targets or reduced budgets to observe the same number of neutrino events or to observe more neutrino events with the same target mass, compared to experiments using other neutrino interactions.

The Standard Model predicts the differential cross section of CEvNS~\cite{Barranco:2005yy} as
\begin{equation}\label{eq:CEvNS_crossSection}
    \dfrac{d\sigma}{dT}(T, E_\nu) = \dfrac{G_{F}^{2}M}{2\pi}\left[(G_{V}+G_{A})^{2}+(G_{V}-G_{A})^{2}\left(1-\dfrac{T}{E_{\nu}}\right)^{2}-(G_{V}^{2}-G_{A}^{2})\dfrac{MT}{E_{\nu}^{2}}\right],
\end{equation}
where $T$ denotes the recoil energy of the nucleus, $E_{\nu}$ represents the incident neutrino energy, $G_{F}$ stands for the Fermi constant, and $M$ corresponds to the mass of the target nucleus,
\begin{equation}\label{eq:crossSection_vectorCouplingConstant}
    G_{V}=(g^{p}_{V}Z+g_{V}^{n}N)F^{V}(Q^{2}),
\end{equation}
\begin{equation}\label{eq:crossSection_axialCouplingConstant}
    G_{A}=(g^{p}_{A}(Z_{+}-Z_{-})+g^{n}_{A}(N_{+}-N_{-}))F^{A}(Q^{2}),
\end{equation}
$g_{V}^{n,p}$ and $g_{A}^{n,p}$ represent the vector and axial-vector coupling factors, for protons and neutrons, respectively. Additionally, $Z$ and $N$ are the proton and neutron numbers, while $Z_{\pm}$ and $N_{\pm}$ correspond to the total number of nucleons with spin up or down. The terms $F^{V,A}$ refer to vector and axial nuclear form factors, and $Q$ is the transferred momentum. Given that the number of nucleons with spin up or down in a nucleus are either the same or really close, the contribution of the axial-vector term $G_A$ is negligible. 

\begin{figure}[ht!]\centering
\includegraphics[width=0.9\linewidth]{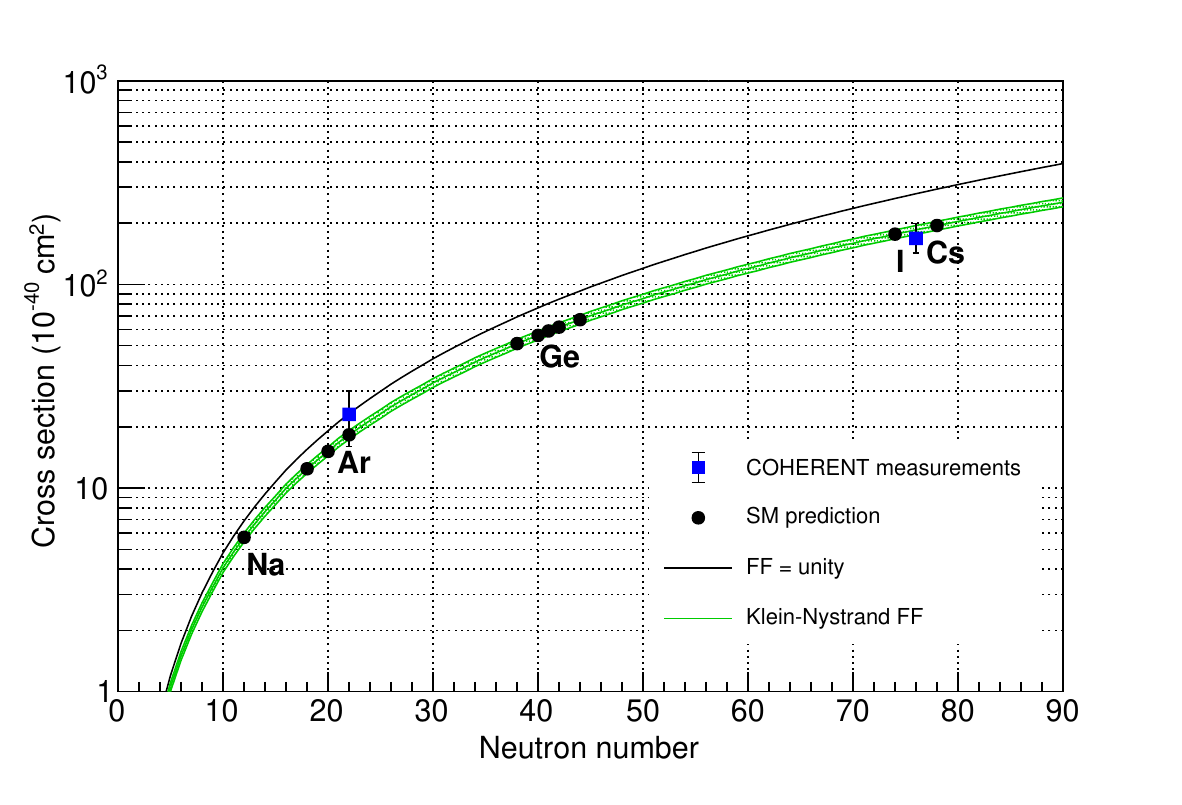}
\caption{Flux-averaged CEvNS cross sections as a function of neutron number~\cite{Barbeau:2021exu} taken from Ref.~\cite{snowmassCO22} The black line assumes unity form factor. The COHERENT collaboration has measured CEvNS using CsI~\cite{coherent17}, argon\cite{COHERENT:2020iec} and germinum (soon to be released).}
\label{f:nuxs}
\end{figure}

The vector couplings are subject to $Q$-dependent radiative corrections at the percent level, as discussed in~\cite{Tomalak:2020zfh}. The specific values~\cite{Tomalak:2020zfh} for $g_V^n$ and $g_V^p$ are approximately $-0.511$ and $0.03$ respectively. The vector contribution $G_V$ is hence mainly determined by the total number of neutrons in the target nucleus, scaling approximately as $N$, and the cross section can be simplified to be proportional to $N^2$. The cross sections as a function of the number of neutrons of COHERENT detector materials are shown in Fig.~\ref{f:nuxs}.

\subsection{Physics reach of CEvNS}
\hspace{0.5cm} The observation of CEvNS by the COHERENT collaboration has triggered significant interest, not only because it confirmed a long-predicted standard neutrino interaction that is important in the evolution of astronomical objects~\cite{janka07}, but more importantly, it demonstrates the possibility to probe a broad range of new physics through the precise detection of CEvNS, including, supernova~\cite{COHERENTsupernovae}, neutrino electromagnetic interactions~\cite{kosmas15}, sterile neutrinos~\cite{formaggio12}, non-standard neutrino interactions (NSIs)~\cite{bar05}, neutron radius~\cite{Cadeddu:2017etk}, accelerator-produced dark matter (DM)~\cite{coherentDM21} and axion-like particles~\cite{coherentAixon21}, etc. The most significant sensitivities of the proposed cryogenic undoped CsI detector, CryoCsI (see Chapter \ref{s:yummy}), will be discussed in the following. Additional sensitivities, such as sterile neutrinos, supernova neutrinos, and the $L_\mu - L_\tau$ parameter space, which could potentially elucidate the muon anomalous magnetic moment (g-2) anomaly, are discussed in more detail in Ref.~\cite{CryoCsI23}.

\subsubsection{Hidden-sector dark matter}
\begin{figure}[ht!]\centering
  \includegraphics[width=0.7\linewidth]{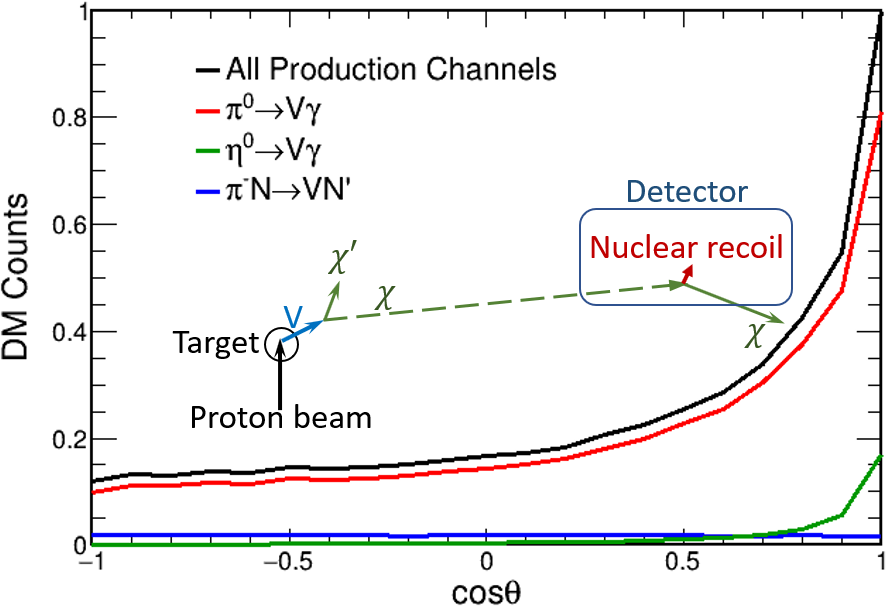}
  \caption{Dependence of the DM scattering rate as a function of the DM scattering angle relative to the COHERENT detector and the beam normalized to the on-axis prediction. The inlet sketch shows the DM production and detection mechanism at the SNS~\cite{pershey19}.}
  \label{f:ang}
\end{figure}

\hspace{0.5cm} As the sensitivity of canonical Weakly Interacting Massive Particle (WIMP) search experiments approaches the neutrino floor from a few GeV to the TeV scale, the search for low-mass DM (LDM) particles becomes increasingly interesting~\cite{maryland17}.  To avoid LDM overproduction in the early Universe, sub-GeV DM models must postulate a ``portal'' particle to mediate interactions between the relic DM candidate and SM particles~\cite{PhysRevLett.39.165, Izaguirre:2015yja}.  This weakly-coupled dark portal particle, denoted as $V$, could be produced at the SNS through the decay of mesons produced by the interactions between the proton beam and the mercury target. Subsequently, the portal particle would decay into a pair of LDM particles, $\chi^\dagger\chi$, either of which may interact with a detector, as depicted in Fig.~\ref{f:ang}.

Neutrons ranging from 1 to 10 MeV can elastically scatter with nuclei in a detector, generating a nuclear recoil signal indistinguishable from that of DM scattering events. Previous measurements in Neutrino Alley (NA) have revealed that these neutrons are all prompt, arriving in synchrony with the proton-on-target pulse. This time signature closely resembles that expected for DM interactions, except for the heaviest mass range, where a detectable delay may occur. An ongoing combined analysis of neutron data obtained from various detectors at different locations aims to provide a comprehensive understanding of the beam-related neutron (BRN) background along NA. This endeavor will prove beneficial not only for the proposed detector but also for other detectors utilized in the COHERENT experiment.
\begin{figure}[ht!]\centering
  \includegraphics[width=0.49\linewidth,trim={0 0 1.5cm 0},clip]{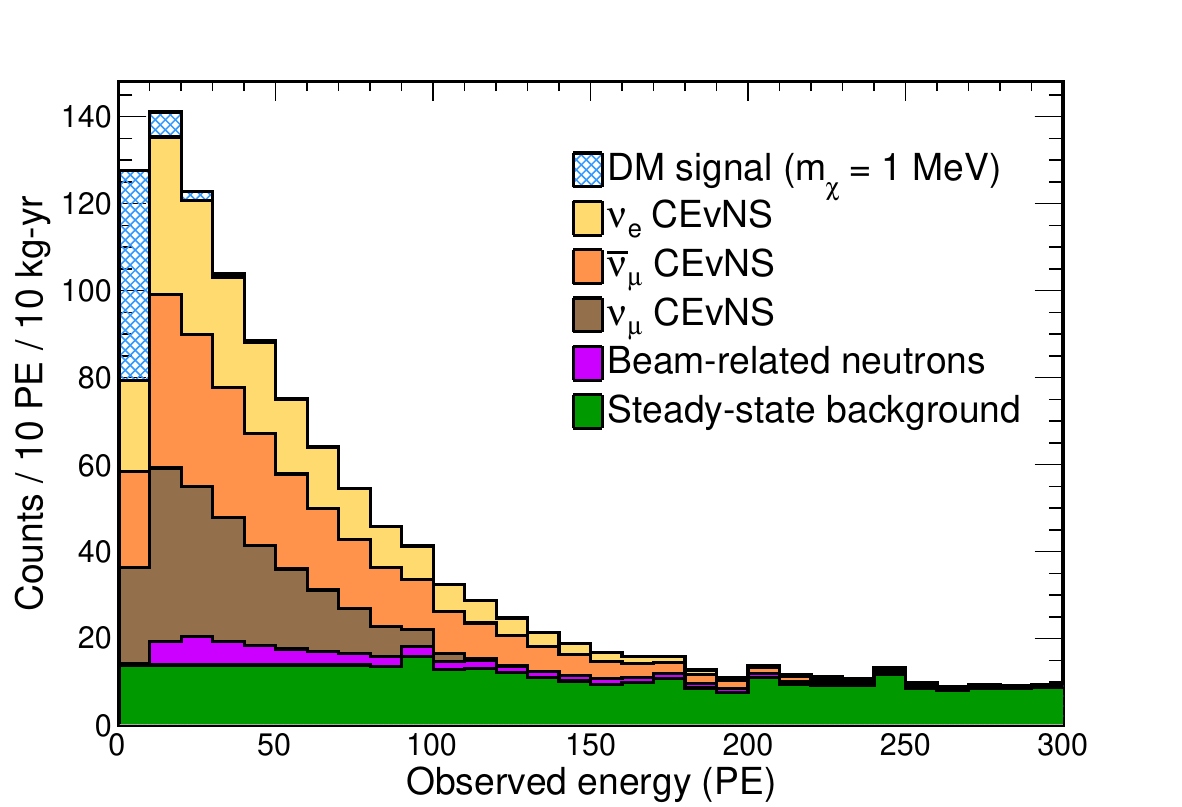}
  \includegraphics[width=0.49\linewidth,trim={0 0 1.5cm 0},clip]{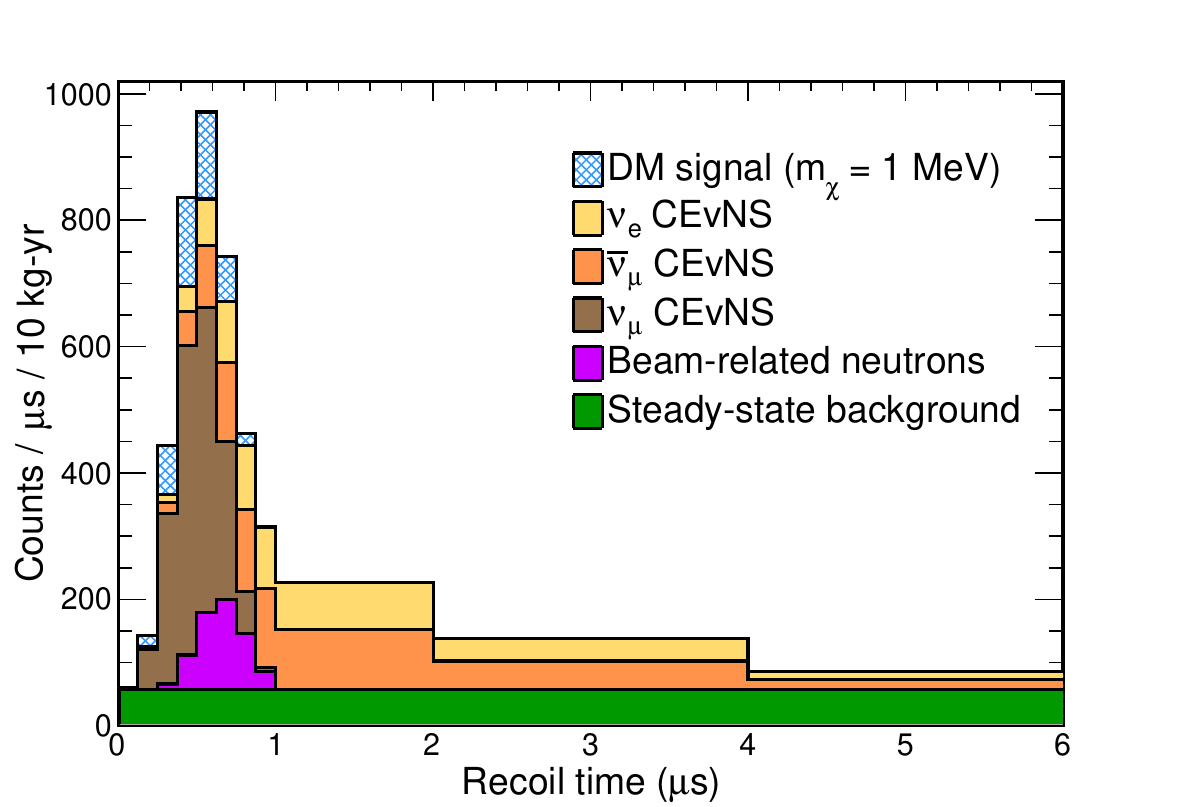}
  \caption{Expected spectra of selected events, including dark matter, CEvNS, and backgrounds, after one year of SNS running~\cite{CryoCsI23}.  A mass of 1~MeV is assumed for the dark matter.  Dark matter (blue hatched) would manifest as an additional scattering process at early times.  Later CEvNS from $\nu_e$ and $\bar{\nu}_\mu$ would improve systematic uncertainties on the detector response and neutrino flux to inform the background in the dark matter region of interest, $t_\mathrm{rec}<1\mu$s.}
  \label{f:dmSp}
\end{figure}

As CEvNS eventually imposes constraints on direct WIMP searches~\cite{Billard:2013qya}, it also serves as a background for DM searches at the SNS as depicted in Fig.~\ref{f:dmSp}. Should DM be discovered, COHERENT's multi-detector strategy offers three avenues for cross-checking the DM hypothesis. Firstly, the recoil energy spectrum is expected to exhibit differences between DM and CEvNS events (see Fig.~\ref{f:dmSp}). Secondly, the scattering cross sections rely on neutron and proton numbers, which differ for CEvNS ($\propto N^2$) and DM ($\propto Z^2$). Lastly, there will be an angular dependence of the DM flux since it originates from meson decay-in-flight, a feature that can be examined across multiple detector systems (see Fig.~\ref{f:ang}).

\begin{figure}[ht!]\centering
  \includegraphics[width=0.8\linewidth,trim={0 0 2cm 0},clip]{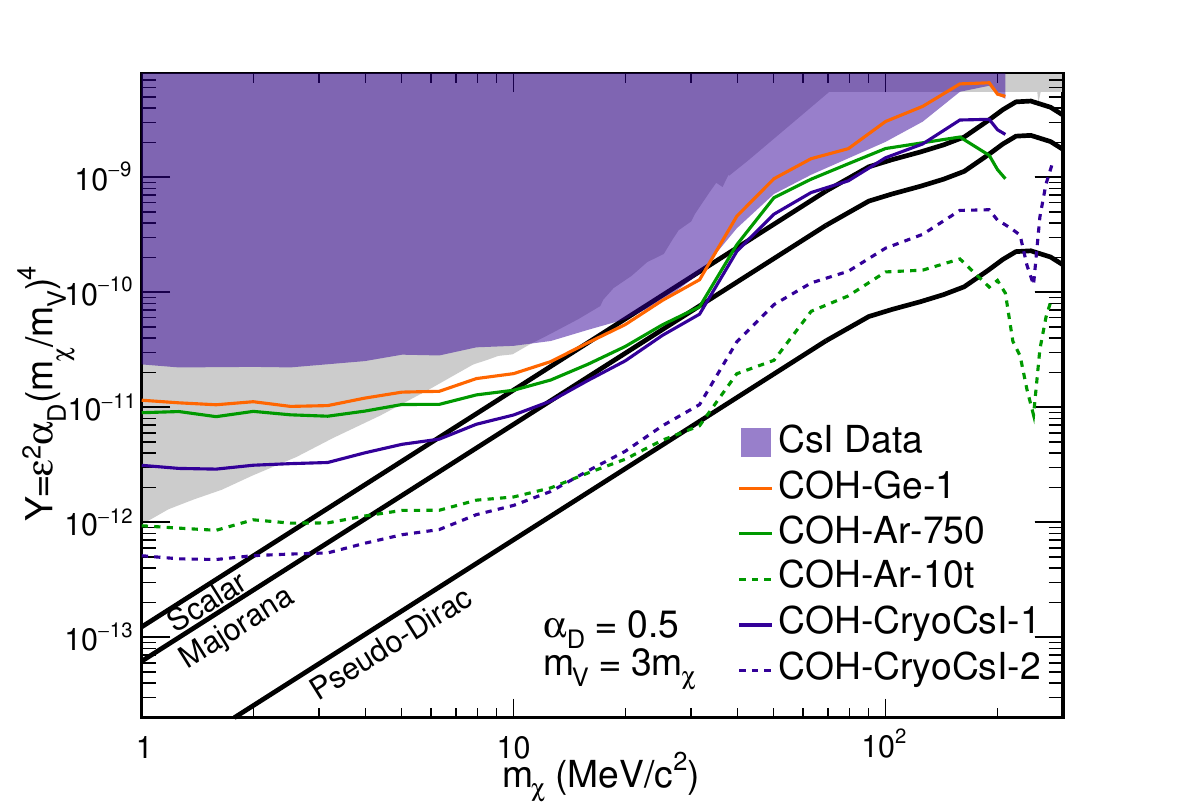}
  \caption{Predicted 90\% sensitivity to LDM production parameters in case of the vector portal theory~\cite{pershey21}. The \emph{Scalar, Majorana} and \emph{Pseudo-Dirac} lines indicate various models, where DM interactions with visible matter in the hot early Universe, explain the DM abundance today. The proposed technology is particularly efficient for probing low-mass DM due to its low threshold. COH-CryoCsI-1: 10 kg$\cdot$ 3 years; 2: 750 kg$\cdot$ 5 years.}\label{f:ldm}
\end{figure}

In this vector portal model, alongside the masses of the portal and DM particles, $m_V$ and $m_\chi$ respectively, there are two coupling constants as free parameters, denoted as $\epsilon$ and $\alpha'$. Comparing the parameters of the vector portal model to the cosmological relic density of DM can be conveniently done using the dimensionless quantity $Y=\epsilon^2\alpha'(m_\chi/m_V)^4$~\cite{Izaguirre:2015yja}; this quantity can then be compared to results from direct detection experiments. Using the model and simulation program detailed in Ref.~\cite{deNiverville:2015mwa}, the sensitivities of a 10~kg undoped crystal $\sim$20~m away from the SNS target after 3 years of data taking~\cite{csi20} and a 750~kg one after 5 years~\cite{pershey21} have been estimated and are shown in Fig.~\ref{f:ldm}. Parameters falling below these sensitivities could lead to the overproduction of dark matter during freeze-out. Notably, a 10 kg detector (COH-CryoCsI-1) exhibits immediate sensitivities to assess scalar and Majorana scenarios, while a 750 kg detector (COH-CryoCsI-2) is sensitive even in the most pessimistic case.

Compared to other DM direct detection experiments conducted at accelerators, CEvNS detectors offer significantly lower costs by reducing detector mass by orders of magnitude. The timing of the neutrino flux at the SNS also enables a unique cancellation of systematic uncertainties, a feature not available for searches in neutrino beams at higher energies like those in the SBN~\cite{Batell:2021ooj} and DUNE~\cite{DeRomeri:2023cjt}. Sub-GeV DM produced at the SNS is relativistic, resulting in much more energetic recoils (up to $100$ keV for $m_\chi=25$ MeV) compared to cosmological DM (up to $100$ meV for $m_\chi=25$ MeV)~\cite{PhysRevLett.130.051803}. Consequently, accelerator experiments are considerably more efficient at constraining light DM particles than underground direct-detection experiments utilizing similar detectors, such as COSINE~\cite{cosine18,cosine17}, SABRE~\cite{sabre15}, ANAIS~\cite{anais16}, PICO-LON~\cite{picolon16}, and COSINUS~\cite{cosinus16}. Moreover, the short duration of the proton beam spill at the SNS provides an additional advantage, reducing random background event rates by a factor of approximately $60000\times$ for DM searches at the SNS.

\subsubsection{Non-standard neutrino interactions}
\hspace{0.5cm} Neutrino-quark NSIs~\cite{wolfenstein78} naturally arise from a new force that couples feebly with SM particles. These theoretical constructs fall within the energy range explored by CEvNS experiments. The presence of NSIs would affect neutrino propagation in matter~\cite{nsi19}, critically altering neutrino oscillation dynamics that lead to experimental biases in oscillation parameters. Two neutrino-mixing solutions with differing NSI couplings are consistent with oscillation data: the large mixing angle (LMA) solution typically quoted that assumes no NSI and the LMA-Dark solution that would lead to wrong interpretations of the neutrino mass ordering~\cite{deepthi17}, CP-violating phase~\cite{liao16, flores18}, and mixing angles~\cite{miranda06} due to nontrivial NSI couplings. Data from CEvNS experiments are essential for resolving this ambiguity~\cite{Denton:2022nol}.

The NSI landscape is often expressed as an effective field theory, assuming that the mediator mass, $m_V$, significantly exceeds the momentum transfer, $\sqrt{Q^2}$. NSIs are parameterized by a tensor of vector couplings, $\varepsilon_{\alpha\beta}^{q,V}$, where $\alpha$ and $\beta$ denote incoming and outgoing neutrino flavors, respectively, and $q$ represents a quark ($u$ or $d$). The potential ambiguity between neutrino oscillations and NSIs arises from nonzero $\varepsilon_{ee}^{q,V}$ and/or $\varepsilon_{\mu\mu}^{q,V}$ couplings. Both can be directly tested with CEvNS experiments at the SNS due to its multi-flavor neutrino flux. 

\begin{figure}[ht!]
  \includegraphics[width=\linewidth]{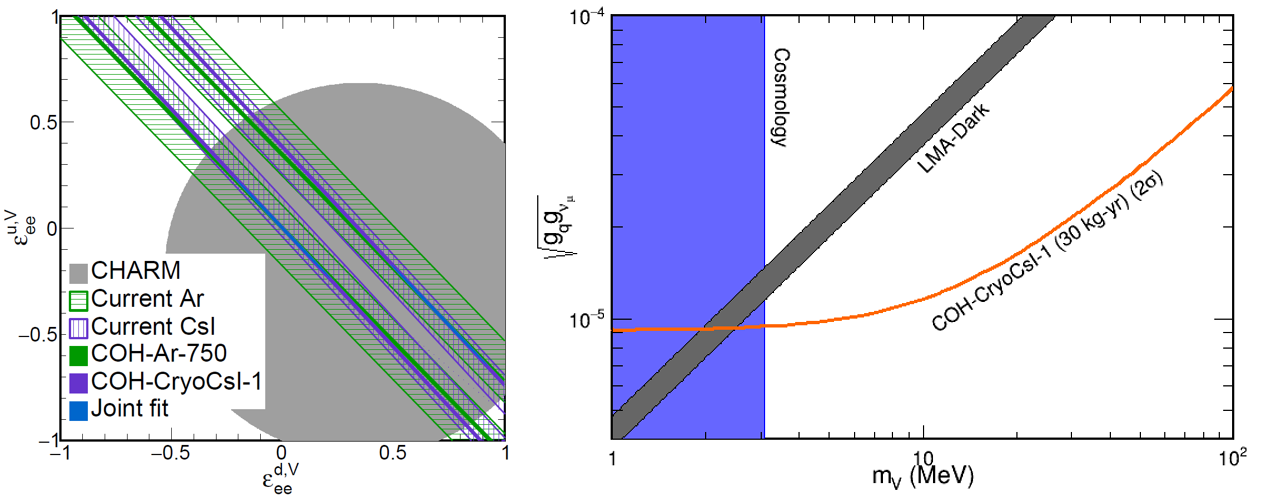}
  \caption{Left: allowed regions in the non-universal NSI parameter spaces at 90\% C.L. constrained by a 10~kg cryogenic CsI after three years of operation~\cite{ding20} compared to existing constraints~\cite{coherent17, coloma17}. Right: sensitivity to constrain the neutrino-quark coupling, $g_{\nu_\mu}g_q=\sqrt{8}G_F\varepsilon_{\mu\mu}$, as a function of mediator mass with the same undoped cryogenic CsI dataset (orange curve) compared to current constraints from cosmology (shaded blue) and NSI parameters consistent with the LMA-Dark oscillation solution.}\label{f:nsi}
\end{figure}

A sizable $\varepsilon^{q,V}_{ee}$ or $\varepsilon^{q,V}_{\mu\mu}$ would enhance or suppress the number of CEvNS events observed. If the mediator is light, this change in rate would only be observed at low recoil energies. One can therefore estimate NSI couplings by comparing the observed number of events to that predicted by the SM as a function of $Q^2$. The left plot in Fig.~\ref{f:nsi} shows the expected constraints on $\varepsilon_{ee}^{u,V}$ and $\varepsilon_{ee}^{d,V}$ compared to current constraints assuming $m_V^2\gg Q^2$~\cite{snowmassCO22}. 

In the case that the new mediator is not heavy compared to $\sqrt{Q^2}$, these couplings become functions of $Q^2$:
\begin{equation}
 \varepsilon_{\alpha\beta}^{q,V}(Q^2)=\frac{1}{m_V^2+Q^2}\varepsilon_{\alpha\beta}^{q,V}(0).
\end{equation}
Accordingly, NSIs must be constrained both in the coupling and the mediator mass. Mediators lighter than 3~MeV are disfavored cosmologically. The sensitivity to $\varepsilon_{\mu\mu}$ as a function of $m_V$ is shown in the right plot in Fig.~\ref{f:nsi}. For low-mass mediators, the proposed 10-kg detector would test all currently viable NSI parameter space consistent with LMA-Dark at $>2\sigma$ after 3 years. The proposed 10-kg detector would unambiguously clarify the degeneracy between neutrino NSI and oscillations ahead of DUNE~\cite{DUNE:2020lwj}, the next generation, US-based neutrino-oscillation experiment. 

Compared to other CEvNS experiments, such as MINER~\cite{miner}, CONNIE~\cite{connie16}, TEXONO~\cite{texono16}, CONUS~\cite{conus}, $\nu$GEN~\cite{vgen}, $\nu$-cleus~\cite{strauss17}, Ricochet~\cite{billard17}, RED~\cite{red16}, etc, sited in $\bar{\nu}_e$ reactor fluxes, the COHERENT experiment benefits from having a multiple-flavor flux, higher neutrino energies, an intense pulsed beam, and a multi-target setting. Compared to other detector systems within COHERENT~\cite{snowmassCO22}, cryogenic scintillating crystals offer the lowest energy threshold, sufficient timing resolution, and the smallest footprint~\cite{csi20}. A combined fit of data from multiple COHERENT detectors further reduces the uncertainties in the CEvNS rate and improves searches for NSI. This also improves dark-matter direct-detection experiments by solidifying predictions for the CEvNS background induced by solar and atmospheric neutrinos~\cite{Billard:2013qya}.

\subsubsection{Neutron radius}
\hspace{0.5cm} In addition to the diverse beyond the Standard Model (BSM) topics that the proposed detector can explore, CEvNS measurements offer valuable insights into nuclear physics and astrophysics by directly probing the density of neutrons within target nuclei~\cite{Cadeddu:2017etk}. Such measurements are crucial for determining the radius and equation of state of neutron stars~\cite{Salinas:2023epr}.

In the limit where $Q^2=0$, the nucleus behaves like a point charge, resulting in fully coherent CEvNS. However, for $Q^2>(10$~MeV$)^2$, the relevant kinematic region at the SNS, the coherence is only partial. The physical extent of the weak charge of the nucleus suppresses the CEvNS rate. This effect is incorporated into the CEvNS cross section through a form factor, $\lvert F(Q^2)\rvert$:
\begin{equation}
    \frac{d\sigma}{dE_r} = \frac{Q_W^2}{2\pi}\left[1-\frac{2m_NE_r}{E_\nu^2}\right]\left|F(Q^2)\right|^2
    \label{eqn:CEvNSXSec}
\end{equation}
where $E_r$ is the nuclear recoil energy, $m_N$ is the nuclear mass, $E_\nu$ is the incoming neutrino energy, and $Q_W = g_pZ + g_nN$ stands for the nuclear weak charge, where $Z$ and $N$ represent the proton and neutron numbers of the nucleus, respectively. Since $\lvert g_p\rvert=1-4\sin^2\theta_W\ll\lvert g_n\rvert=1$, CEvNS primarily depends on the nuclear neutron charge distribution. 

The neutron radius, denoted as $R_n$, which represents the root-mean-square distance to each neutron from the nuclear center. The measurement of the form factor in CEvNS interactions provides a second, independent test of $R_n$. The weak form factor has been measured through parity-violating electron scattering (PVES) experiments on $^{48}$Ca, revealing a small neutron skin~\cite{CREX:2022kgg} and $^{208}$Pb, indicating a relatively large neutron skin~\cite{PREX}. These findings exhibit a discrepancy with nuclear models at a significance level of 2~$\sigma$~\cite{PREX-2}. COHERENT experiments can probe the neutron skin thickness in light nuclei like Ar and heavier nuclei like CsI, allowing for a direct test whether the neutron skin thickness increases with nuclear mass.

\begin{figure}[ht!]
  \includegraphics[width=\linewidth]{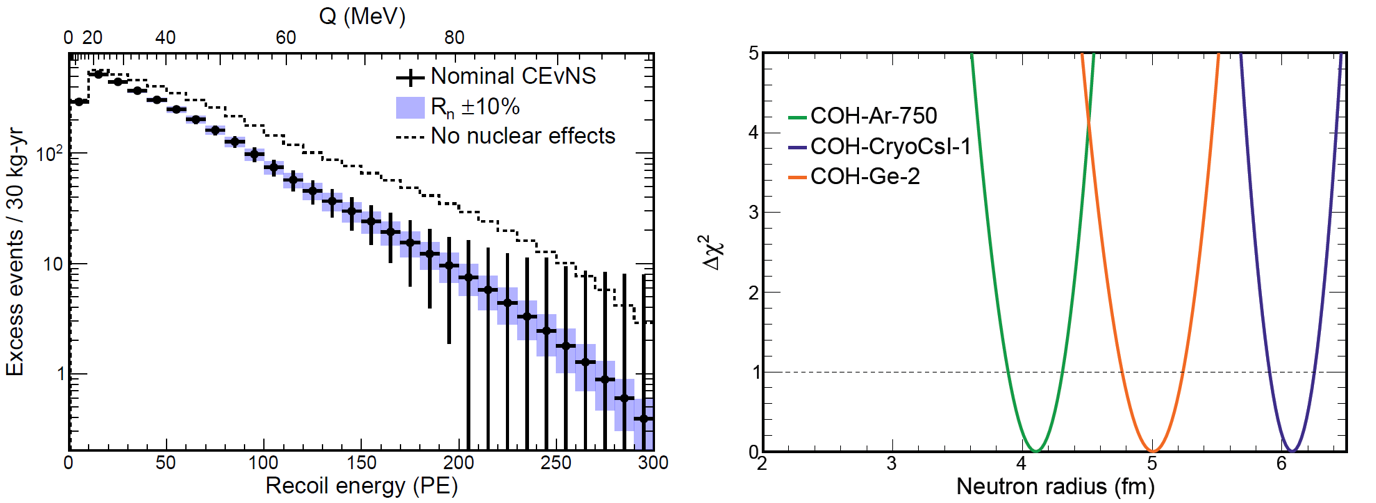}
  \caption{Taken from Ref.~\cite{snowmassCO22}. Left: The expected, background-subtracted CEvNS distribution after 3 years of running at the SNS compared to the expectation with no form factor (dashed line) and $R_n$ varied by $\pm10\%$ (solid blue) left. Right: Expected sensitivity to the neutron radii in argon, germanium, and CsI (right). A 2.9$\%$ measurement is possible in the proposed detector.}\label{f:Rn}
\end{figure}

The expected CEvNS distribution in the proposed detector after three years of operation at the SNS is illustrated in Fig.~\ref{f:Rn}. This expectation is compared to the CEvNS rate assuming no suppression due to nuclear size ($\lvert F(Q^2)\rvert=1$), and to a scenario with a modification of $R_n$ by $\pm10\%$. Through a binned log-likelihood sensitivity fit, the proposed detector exhibits a 2.9$\%$ sensitivity to $R_n$ in CsI after 3~years~\cite{snowmassCO22}, as depicted in Fig.~\ref{f:Rn}. The cryogenic undoped CsI technology stands out as the most promising among COHERENT's proposed detectors for extracting the neutron distribution. 


\include{Section2/CEvNS experiments}

\section{The COHERENT experiment}
\label{s:coherent}
\hspace{0.5cm} Pioneering CEvNS measurements, the COHERENT experiment has achieved world-first observations with detectors using CsI[Na] (6.7$\sigma$)~\cite{coherent17}, Ar ($3.1\sigma$)~\cite{COHERENT:2020iec}, and Ge (soon to be released). Detecting such rare events always necessitates high flux and careful background suppression, which the intense neutrino flux and effective background rejection capabilities of the SNS beam are particularly well-suited for. Detailed information about the SNS and the COHERENT detectors will be described in the following sections.
\subsection{The Spallation Neutron Source}
\label{s:sns}
\hspace{0.5cm} As demonstrated in Fig.~\ref{f:sns}, the SNS is the world's premier facility for neutron-scattering research, delivering pulsed neutron beams with intensities an order of magnitude brighter than any currently operating facility. At 1.4 megawatt (MW) beam power, approximately \num{1.5e14} 1-GeV protons bombard the liquid mercury target in short \SI{350}{\ns}~wide bursts at a rate of \SI{60}{\hertz}. Neutrons produced in spallation reactions with the mercury target thermalize in cryogenic moderators surrounding the target and are delivered to neutron-scattering instruments in the SNS experiment hall. As a byproduct, the SNS also provides the world's most intense pulsed source of neutrinos in an energy region of specific interest for particle and nuclear astrophysics. 

\begin{figure}[htbp]
  \includegraphics[width=\linewidth]{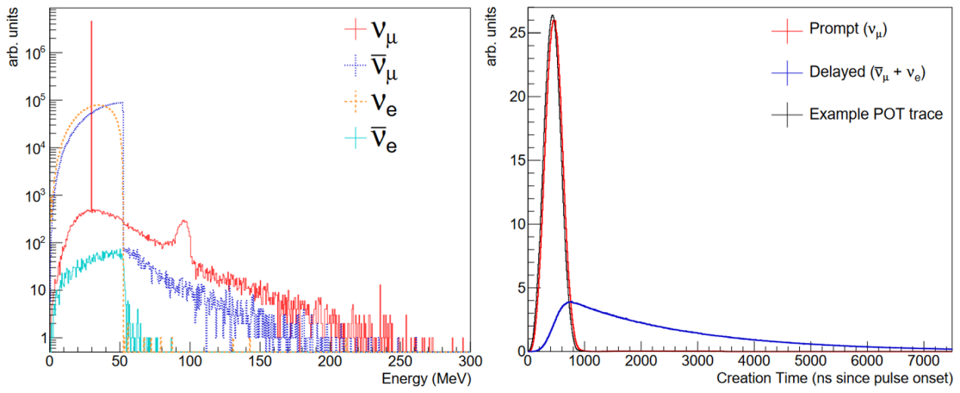}\centering
  \caption{Left: Expected $\nu$ spectra at the SNS, showing the very low level of decay-in-flight and other non-decay-at-rest flux, in arbitrary units (a.u.)~\cite{coherent18}; the integral is $4.7\times10^7$neutrinos/cm$^2$/s at 20~m~\cite{coherent21}. Right: Time structure for prompt and delayed neutrinos due to the 60-Hz pulses~\cite{coherent18}. ``Prompt'' refers to neutrinos from pion decay, and ``delayed'' refers to neutrinos from muon decay.} \label{f:vd}
\end{figure}

The SNS operates as a user facility for approximately two-thirds of the year, generating $2.81 \times 10^{14}$ neutrinos per square centimeter for each flavor at a distance of 20 meters annually. Interactions of the proton beam in the mercury target produce $\pi^+$ and $\pi^-$, along with neutrons. About 99\% of the $\pi^-$ are absorbed by nuclei in the dense mercury. In contrast, the subsequent decay-at-rest of $\pi^+$ produces neutrinos of three flavors: $\pi^+ \longrightarrow \mu^+ + \nu_\mu$ (30~MeV) at a short decay time of 26 ns, $\mu^+ \longrightarrow e^+ + \nu_e + \bar{\nu}_\mu$ at a subsequent decay time of 2.2 $\mu$s. The energy spectra of the produced neutrinos are depicted in the left plot of Fig.~\ref{f:vd}. The production of $\bar{\nu}_e$, resulting from $\pi^-$ decay in flight, is suppressed by four orders of magnitude compared to other species. In the $\nu_\mu$ spectrum, the $\sim$100 MeV bump results from muon capture ($\mu^- + p \longrightarrow n + \nu_\mu$), while the $\sim$230 MeV spike originates from kaon decay ($K^+ \longrightarrow \mu^+ + \nu_\mu$).

The precise timing structure of the SNS beam offers significant advantages for both background rejection and the precise characterization of backgrounds unrelated to the beam~\cite{Bolozdynya:2012xv}. Neutrino time distributions are presented on the right side of Fig.~\ref{f:vd}. Focusing solely on beam-related signals within a \SI{10}{\micro\s} window after a beam spill results in a factor-of-2000 reduction in steady-state background. 

\begin{figure}[htbp]
  \includegraphics[width=\linewidth]{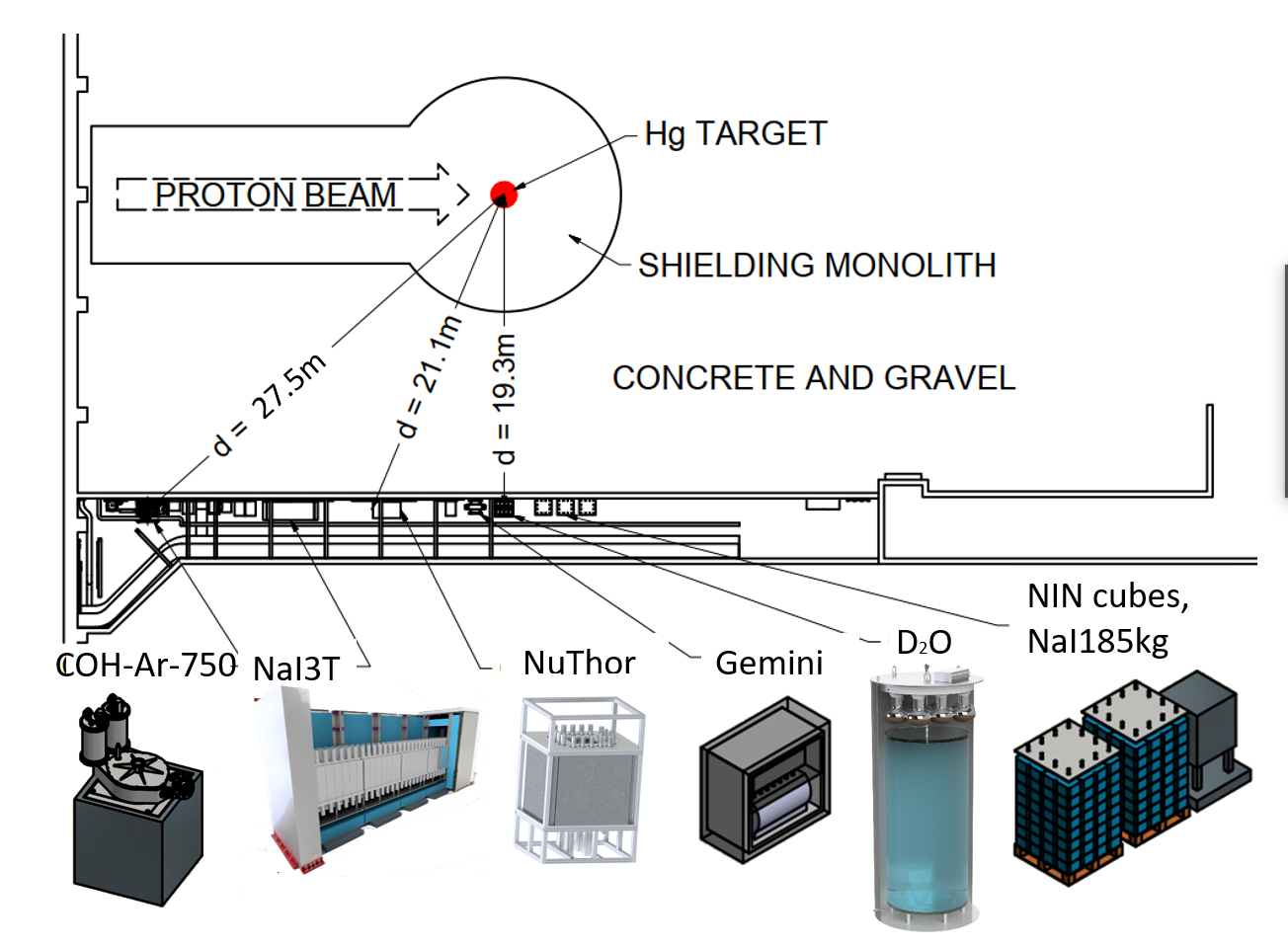}\centering
  \caption{Current sub-detector systems of COHERENT in the Neutrino Alley~\cite{NA}. } \label{f:alley}
\end{figure}

The SNS First Target Station (FTS), which had been operating at 1.4 MW of beam power, underwent an upgrade in 2023, achieving a world record beam power of 1.7 MW through an increase in the number of protons per bunch. The beam power for neutron production at the FTS is targeted to increase to 2 MW by 2027. A Second Target Station (STS) is in the planning stages at the SNS. The increased overall beam power (2.8 MW) will be divided between two targets at rates of 45 Hz to the FTS and 15 Hz to the STS. This configuration creates more advantageous conditions for suppressing steady-state backgrounds. The anticipated neutrino production per proton on target at the STS is slightly higher than that at the FTS~\cite{Barbeau:2023exu}. This expansion potentially allows COHERENT to operate larger CEvNS detectors, overcoming the size limitations of NA, as illustrated in Fig.\ref{f:alley}, and may contribute to efforts in sterile neutrino searches.

\subsection{COHERENT detectors}
\hspace{0.5cm} The COHERENT detectors are strategically positioned in NA (as shown in Fig.~\ref{f:alley}), approximately 20m from the target. This location is surrounded by contiguous shielding materials and overburden, effectively eliminating nearly all free-streaming pathways for fast neutrons, which are the primary contributors to beam-related backgrounds. The cross section as a function of the number of neutrons (different targets) is depicted in Fig.~\ref{f:nuxs}. Multiple target materials collaborate to reduce uncertainties in CEvNS measurements, including factors like the nuclear form factor, thereby laying the groundwork for exploring new physics through CEvNS detection. COHERENT physics topics and corresponding experimental requirements are depicted in Table.~\ref{table:coherent}. With significant internal funding at ORNL, the basement utility corridor is now a fully equipped and operating neutrino laboratory including dedicated electrical power and ready access to necessary utilities such as chilled water for cryogenic compressors, plant air for system actuators, facility ventilation for cryogenic boil-off, etc. 
\begin{table}[htbp]
  \centering
  \caption{COHERENT physics topics and corresponding experimental signatures and requirements from Ref.~\cite{Barbeau:2023exu}}
  \label{table:coherent}
  \begin{tabular}{@{}p{.3\textwidth}|p{.3\textwidth}|p{.35\textwidth}@{}c@{}l@{}l@{}}
   \hline
   \textbf{Topics} & \textbf{Experimental signatures} & \textbf{Detector requirements} \\ \hline
   Nonstandard neutrino interactions, new mediators & Deviation from N², deviation from SM recoil & Multiple targets, energy resolution, shape, event rate scaling, quenching factor \\ \hline
   Weak mixing angle & Event rate scaling & Multiple targets, quenching factor \\ \hline
   Neutrino magnetic moment & Low-recoil-energy excess & Low-energy threshold, energy resolution, quenching factor \\ \hline
   Inelastic CC/NC cross section for supernova and weak coupling parameters & High-energy (MeV) electrons, gammas & Large mass, dynamic range \\ \hline
   Nuclear form factors & Recoil spectrum shape & Energy resolution, multiple targets, quenching factor \\ \hline
   Accelerator-produced dark matter & Event rate scaling, recoil spectrum shape, timing, direction with respect to source & Energy resolution, quenching factor \\ \hline
   Sterile oscillations & Event rate and spectrum at multiple baselines & Similar or movable detectors at different baselines \\ \hline
  \end{tabular}
 \end{table}

\subsubsection{CEvNS detectors}
\begin{figure}[ht!]
  \includegraphics[width=0.7\linewidth]{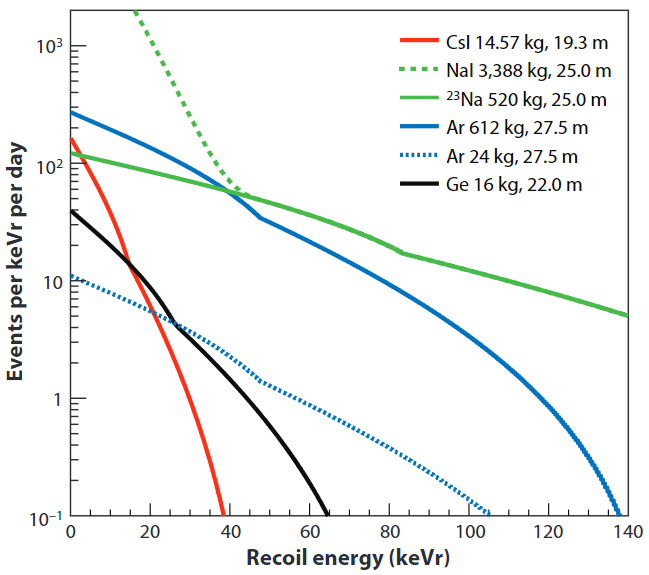}\centering
  \caption{Differential nuclear recoil spectra for COHERENT's CEvNS targets, scaled by mass and distance from the source. No quenching, thresholds, efficiencies or backgrounds are included. CryoCsI's spectrum would be scaled by the mass ratio of CryoCsI over CsI. The kinematics are manifest in that lighter targets have a lower yield overall, but more nuclei are kicked to high recoil energy~\cite{Barbeau:2023exu}.} \label{f:re}
\end{figure}

\hspace{0.5cm} The COHERENT collaboration achieved the observation of CEvNS on various detectors, starting with a 14.6-kg CsI[Na] crystal, followed by the 24-kg liquid argon detector CENNS-10. GeMini, an array of high-purity $p$-type point-contact germanium detectors, recently reported the first CEvNS measurement on germanium with an active mass of 13-kg (soon to be released). The NaIvETe, with a planned deployment of 7 modules totaling 3.4 tons, has currently deployed 3 modules for the first observation of CEvNS on sodium. Ongoing efforts involve precision measurements using detectors with larger mass and higher sensitivity. CENNS-750 is anticipated to detect 3000 CEvNS events per SNS year. Additionally, cryogenic undoped CsI (CryoCsI), benefiting from significantly increased light yield at 40-77 K compared to room-temperature CsI[Na], is part of the collaboration's exploration. Figure~\ref{f:re} shows the expected CEvNS recoil spectra for COHERENT's detectors, where CryoCsI's spectrum would be scaled by the mass ratio of CryoCsI over CsI. The thresholds of these detectors are detailed in Table~\ref{t:det}, where our prototype, CryoCsI, exhibits a much lower threshold down to 0.5 keV$_{nr}$, indicating promising sensitivities.
\begin{table}[htbp]
  \centering
  \captionof{table}{Parameters of COHERENT future detectors~\cite{coherent18}.}\label{t:det}
  \begin{tabular}{lcccc}\toprule
    Nuclear target & CsI$^a$ & Ge$^b$ & Ar$^c$ & NaI$^d$ \\\midrule
    Mass [kg] & 10 & 13 & 750 & 3400\\
    Threshold [keV$_{nr}$] & 0.5 & 2.5 & 20 & 13\\\bottomrule
  \end{tabular}\\
  $^a$ Cryogenic undoped CsI --- CryoCsI.\\
  $^b$ $p$-type point-contact high-purity Ge detector.\\
  $^c$ Single-phase liquid argon (LAr) detector.\\
  $^d$ Traditional NaI(Tl) at room temperature.\\
\end{table}

\subsubsection{CEvNS uncertainties addressing detectors}
\hspace{0.5cm} To achieve precise CEvNS measurements, various technologies are employed to tackle background and neutrino source uncertainties. The MARS neutron detector, a portable gadolinium-doped plastic scintillator, is utilized to measure BRN background~\cite{MARS}. Steady-state background is assessed by individual subsystems during periods when the beam is turned off. The neutrino cubes were deployed to measure neutrino-induced neutrons (NINs) background, via interaction of Pb($\nu_e$,Xn)~\cite{NINs}. D$_2$O subsystem, a 592-kg heavy water Cherenkov detector is deployed to benchmark the neutrino flux uncertainty (10\%) through the measurement of the $\nu_e + d \longrightarrow p + p + e^-$ reaction, for which the cross-section is well-studied (2-3\% uncertainty~\cite{vflux}).

\subsubsection{Inelastic neutrino detectors}
\hspace{0.5cm} The COHERENT experiment goes beyond CEvNS, utilizing various detectors to measure inelastic neutrino interactions. NINs cross section on lead was measured to be a factor of $0.29^{+0.17}_{-0.16}$ lower than the prediction of MARLEY event generator~\cite{NINs}. The HALO experiment may find this result intriguing as they employ this channel for the detection of galactic supernovae. A 185~kg NaI[Tl] crystal array (NaIvE) --- a prototype of NaIvETe (NaI3T), also observed the cross section being 41\% lower than the prediction from MARLEY~\cite{naive}, during the measurement of $^{127}$I$(\nu_e, e^-)$. NuThor, consisting of 52 kg of thorium-232 metal plates, is designed to measure neutrino-induced fission, an interaction never previously observed. The deployment of a second D2O module filled with light water is planned to measure the CC reaction of $^{16}$O$(\nu_e, e^-)$. This reaction holds significance for supernova neutrino detection in Hyper-Kamiokande~\cite{hyperk18}. The COH-Ar-750, a 750~kg liquid argon detector, is expected to measure approximately 340 CC $^{40}$Ar$(\nu_e, e^-)$ events and around 100 NC $^{40}$Ar$(\nu, \nu^\prime)$ events per SNS year. A $\sim$250 kg ionization detector --- LArTPC, similar to DUNE detectors, will measure the $^{40}$Ar$(\nu_e, e^-)$ cross-section, to explore neutrino-Ar interations, investigating potential systematic effects that may impact the detection of supernova neutrinos in DUNE~\cite{DUNE:2020lwj}.

\section{The key ingredients of CryoCsI}
\label{s:yummy}
\hspace{0.5cm} The high sensitivities to detect low energy events necessitates low energy threshold. Compared to other CEvNS detector systems within COHERENT, cryogenic undoped CsI scintillator stands out by offering the lowest energy threshold compared to others listed in Table~\ref{t:det}. This scintillator also boasts the smallest footprint~\cite{csi20}, a crucial consideration given the limited space in Neutrino Alley. Additionally, it complements other systems effectively, contributing to narrowing down uncertainties in CEvNS measurements. 

\begin{figure}[ht!]
  \includegraphics[width=0.7\linewidth]{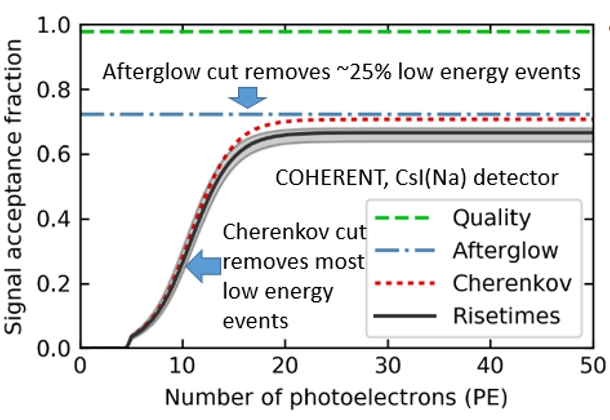}\centering
  \caption{Detection efficiency of low energy events after each event selection criterion of the COHERENT Cs(Na) detector ~\cite{coherent17}.  The most significant ones are the afterglow and Cherenkov cuts. Afterglow refers to a fraction of scintillation light that remains present for a certain time after the radiation excitation stops. Cherenkov radiation happens when electrically charged particles, such as protons or electrons, travel faster than light in a clear medium.}
  \label{f:why}
\end{figure}

\begin{figure}[ht!]
  \includegraphics[width=0.8\linewidth]{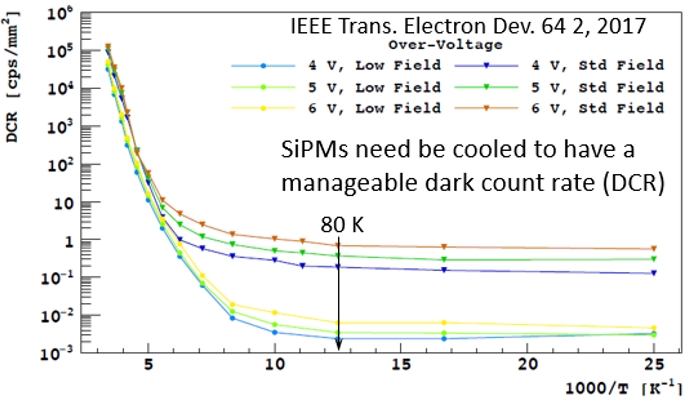}\centering
  \caption{DCR of FBK SiPMs versus operation temperature~\cite{acerbi17}.}
  \label{f:T}
\end{figure}

\begin{figure}[ht!]
  \includegraphics[width=0.7\linewidth]{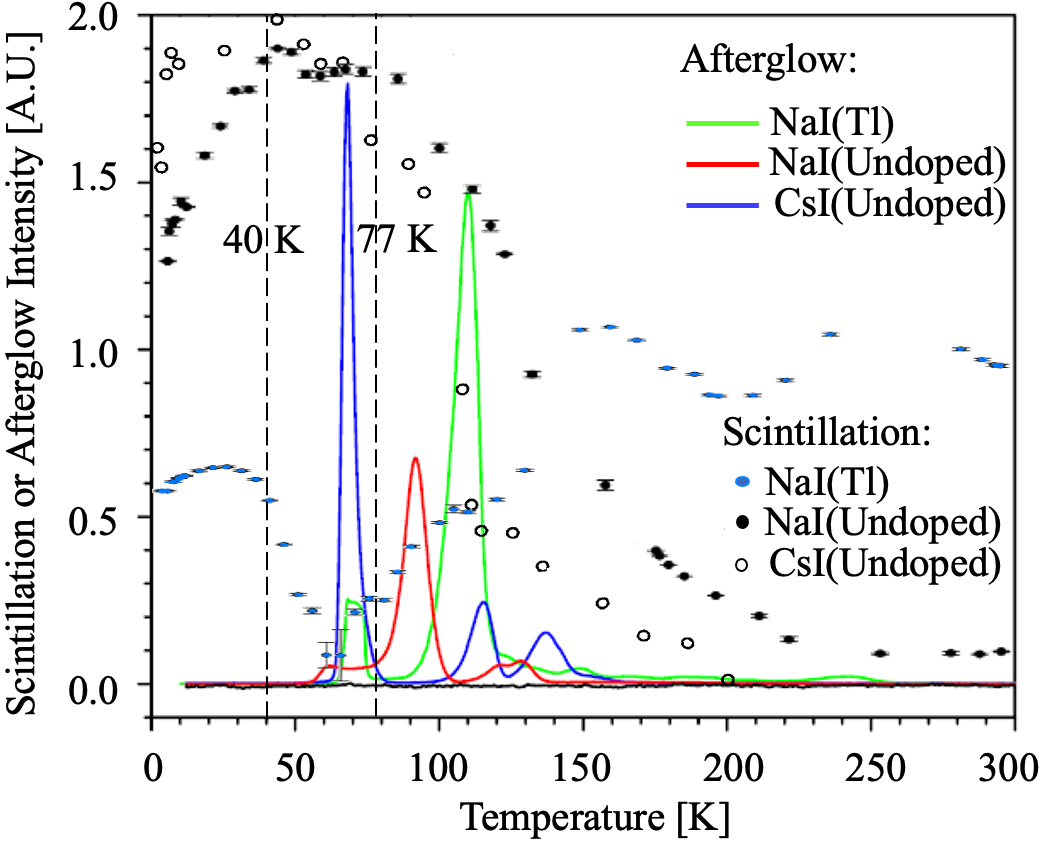}\centering
  \caption{Relative scintillation yields and afterglow rates of various crystals as a function of temperature. The scintillation yield of undoped CsI is taken from Ref.~\cite{Nishimura95}, the yields of undoped NaI and NaI(Tl) are from Ref.~\cite{Sailer12}. The afterglow rates are from Ref.~\cite{derenzo18}.}
  \label{f:ly}
\end{figure}

CryoCsI introduces several key improvements over the original CsI(Na) detector, which successfully detected CEvNS. These enhancements include:
\begin{itemize}
  \item Replacement of PMTs with SiPMs to eliminate Cherenkov light emitted from PMT windows, a factor limiting the energy threshold of the COHERENT CsI(Na) detector~\cite{coherent17} (Fig.\ref{f:why}),
  \item Cryogenic operation of SiPM arrays to reduce their dark count rate (DCR) (Fig.\ref{f:T}),
  \item Substitution of CsI(Na) at room temperature with undoped CsI at cryogenic temperature, since the latter exhibits more than twice the light yield (Fig.~\ref{f:ly}).
\end{itemize}

\subsection{Operation temperature}
\hspace{0.5cm} The temperature dependence of the light yield of undoped crystals is determined by the energy dispersion among phonons and the two types of excitons (see Fig.~\ref{f:ste})~\cite{Nishimura95}. Fig.\ref{f:ly} summarizes the relative scintillation intensities in a wide temperature range~\cite{Nishimura95, Sailer12, mikhailik15}. The figure also includes the relative scintillation intensity of thallium-doped NaI for comparison~\cite{Sailer12}. Several key points are noteworthy:
  
\begin{itemize}
  \item The light yields of undoped crystals peak around 40~K.
  \item The curves are relatively flat around their peaks, indicating that the yields at liquid nitrogen temperature are still very close to the maxima.
  \item When the temperature decreases to 4~K, the yields drop to about 2/3 of the maxima, and the missing energy must be harvested by the phonon channel.
  \item The absolute yields of undoped crystals at their maxima are about twice as high as those of doped ones at room temperature.
  \item Cooling existing doped crystals does not result in an increase in their yields.
\end{itemize}
  
In contrast to deep underground direct DM detection experiments, detectors located at the SNS are much shallower. A potential concern arises from the afterglows of crystals induced by energetic cosmic muon events. As depicted in Fig.\ref{f:ly}, both undoped CsI and NaI exhibit afterglow issues above $\sim$50K~\cite{derenzo18}. To mitigate afterglow effects, operating undoped CsI and NaI near 40K, where light yields are maximized, proves advantageous. However, maintaining this temperature is less convenient compared to the simplicity of achieving 77K using liquid nitrogen as a coolant. Interestingly, there exists a valley around 77K between the two peaks in the undoped CsI and NaI afterglow distributions. By compromising slightly on the afterglow rate, the prototype can be conveniently operated at 77K. To address both crystal afterglow and dark noise from light sensors at the single-photon level, a coincident observation of light signals in at least two light sensors can be required.

\subsection{Scintillation mechanism}
\begin{figure}[htbp]
  \includegraphics[width=0.8\linewidth]{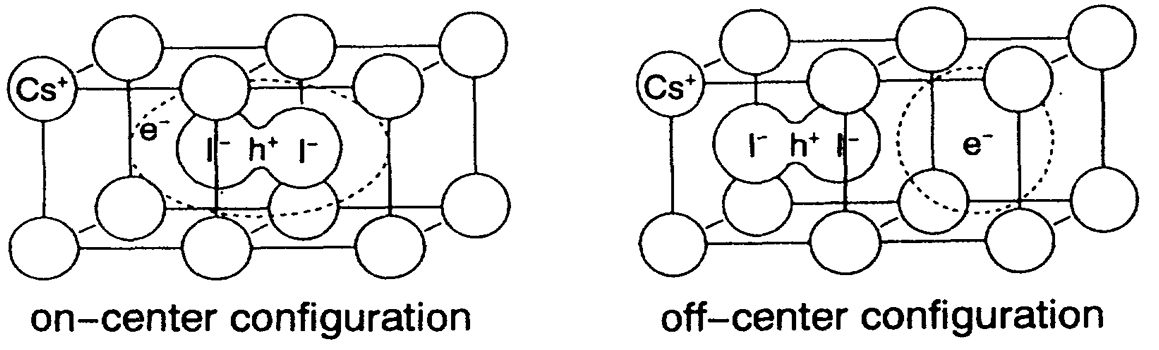}\centering
  \caption{Two types of self-trapped excitons responsible for the scintillation emission in undoped CsI, taken from \cite{Nishimura95}.}
  \label{f:ste}
\end{figure}

\hspace{0.5cm} A scintillation photon must possess less energy than the band gap width of the host crystal to avoid exciting an electron from the valence band to the conduction band, where it could be absorbed by the crystal. This requirement necessitates the presence of energy levels within the band gap. Recombination of electrons and holes in these levels generates photons with insufficient energy to re-excite electrons to the conduction band, preventing re-absorption. In Tl-doped crystals, these energy levels, known as scintillation centers, exist around the doped ions. In undoped crystals, scintillation centers are understood as self-trapped excitons rather than those trapped by doped impurities~\cite{pelant12}. Two types of excitons were identified in undoped CsI~\cite{Nishimura95}, as illustrated in Fig.~\ref{f:ste}. In both cases, a hole is trapped by two negatively charged iodine ions and can subsequently capture an excited electron, forming an exciton resembling a hydrogen atom. These excitons have energy lower than the band gap width, leading to emitted photons with insufficient energy to be re-absorbed by the host crystal.

\begin{figure}[htbp]
    \includegraphics[scale=0.9,valign=t]{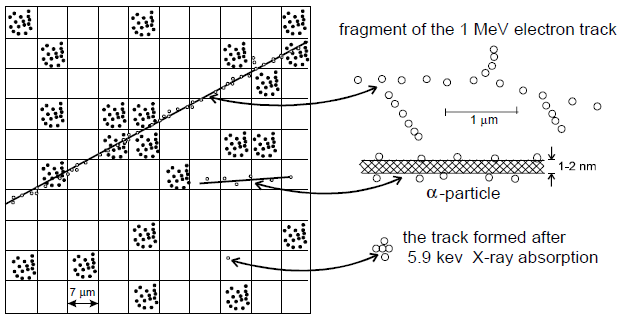}\centering
   \caption{Dopants (scintillation center) for doped crystals~\cite{trefilova02}.}
   \label{f:doped}
  \end{figure}

Scintillation centers in undoped NaI/CsI are understood to be self-trapped excitons instead of those trapped by doped impurities~\cite{pelant12}. Due to the different scintillation mechanisms (refer to Fig.\ref{f:doped} for doped crystals and Fig.\ref{f:ste} for undoped crystals), it is expected~\cite{birks1964} and observed that dopant concentration affects light yield~\cite{trefilova02, park2002csiQF} in inorganic scintillators. Microscopically, dopant is not uniformly distributed within a crystal. Charge carriers around short tracks of nuclear recoils have significantly less chance to encounter doped trapping centers for de-excite through scintillation, resulting in quenching of scintillation of nuclear recoils compared to electronic recoils.

\begin{figure}[htbp]\centering
  \includegraphics[width=0.35\linewidth]{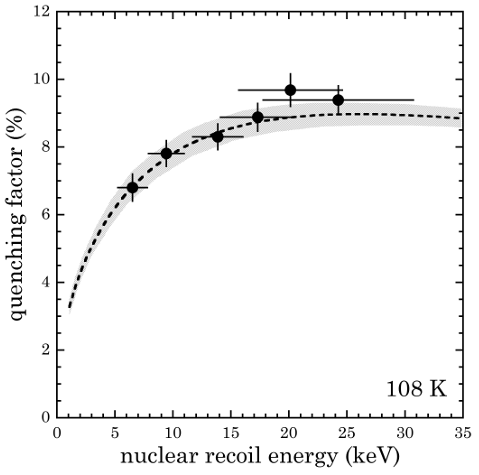}
  \includegraphics[width=0.64\linewidth]{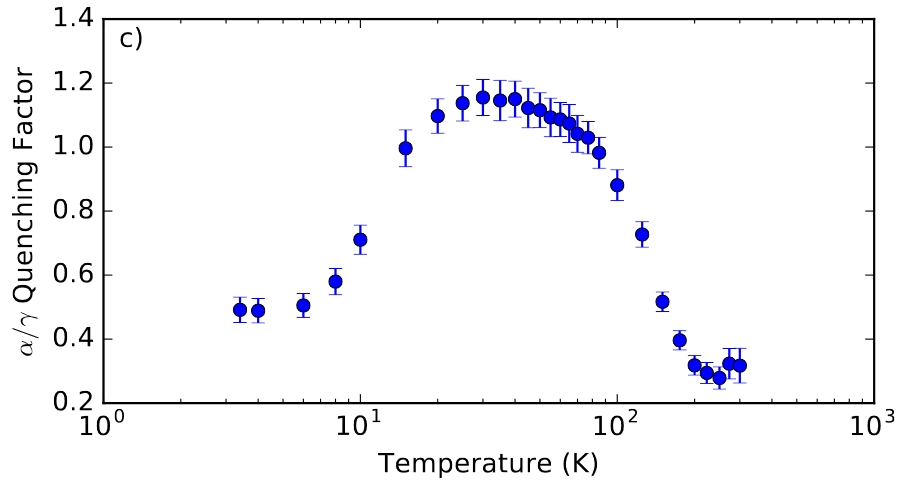}
  \caption{Left: QF of undoped CsI measured with neutrons at 108~K (Taken from Ref.~\cite{lewis21}). Right: QF (see \autoref{e:QF}) of undoped CsI measured with $\alpha$-rays in a wide temperature range (Taken from Ref.~\cite{clark18}).}
  \label{f:aq}
\end{figure}

However, this is not the case in undoped crystals, where holes can be self-trapped anywhere resulting in scintillation emission~\cite{Nishimura95}. Track lengths may not have the same effect on determining the scintillation efficiency as in doped crystals. Indeed, our measurements (see section~\ref{s:qfusd}) and several early measurements at 77~K with $\alpha$-particles reported 60\% to 100\% quenching factor (QF) in undoped CsI~\cite{Hahn53, Hahn53a, Sciver56}. A recent study~\cite{clark18} observed that the QF of $\alpha$-particles changes with the crystal temperature (see Fig.~\ref{f:aq}, right). Around 77~K, the QF is even larger than one. However, reported in another measurement at 108~K~\cite{lewis21}, QF is very similar to that of CsI(Na) at room temperature (Fig.~\ref{f:aq}, left). Possible causes of the discrepancy include different recoiled nuclei (Cs or I versus $\alpha$), deposited energies, measurement temperatures or origins of crystals. It is hence a major objective to survey the QF using neutron beams at Triangle Universities Nuclear Lab (TUNL) at around 77~K. Please refer to Chapter~\ref{s:qf} for a detailed description of the investigation.

\subsection{Scintillation wavelengths, decay times and attenuation lengths}
\hspace{0.5cm} Due to entirely distinct scintillation mechanisms, the scintillation wavelengths and decay times of undoped NaI/CsI differ significantly from those of NaI/CsI(Tl), as outlined in Table~\ref{t:RT} and Table~\ref{t:LN} for room and liquid nitrogen temperatures, respectively. Undoped NaI, being a faster scintillator than NaI(Tl), allows for a narrower coincidence time window, effectively suppressing random backgrounds. However, undoped NaI requires a stringent humidity-controlled assembly environment and is more susceptible to crushing at cryogenic temperatures compared to undoped CsI. As a result, undoped CsI was selected for our investigation. 
\begin{table}[htbp] \centering
  \caption{Scintillation wavelength $\lambda$ and decay time $\tau$ of Tl-doped and undoped NaI, CsI crystals at room temperature.}
  \label{t:RT}
  \begin{tabular}{ccccc}\hline
    Crystal & $\tau$ at $\sim297$~K [ns] & $\lambda$ at $\sim297$~K [nm] \\\hline
    NaI(Tl) & $230\sim250$~\cite{robertson61, eby54, schweitzer83} & $420\sim430$~\cite{Sciver56,Sibczynski12} \\
    CsI(Tl) & 600~\cite{Bonanomi52} & 550~\cite{Kubota88} \\
    undoped NaI & $10\sim15$~\cite{Sciver56,Sciver58,Beghian58} & 375~\cite{West70,Fontana70} \\
    undoped CsI & $6\sim36$~\cite{Kubota88,Schotanus90,Amsler02} & $305\sim310$~\cite{Kubota88,Woody90,Amsler02} \\\hline
  \end{tabular}
\end{table}
  
\begin{table}[htbp] \centering
  \caption{Scintillation wavelength $\lambda$ and decay time $\tau$ of Tl-doped and undoped NaI, CsI crystals at liquid nitrogen temperature.}
  \label{t:LN}
  \begin{tabular}{ccccc}\hline
    Crystal & $\tau$ at $\sim$77~K [ns] & $\lambda$ at $\sim$77~K [nm]\\\hline
    NaI(Tl) & 736~\cite{Sibczynski12} & $420\sim430$~\cite{Sciver56, Sibczynski12} \\
    CsI(Tl) & no data & no data \\
    undoped NaI & 30~\cite{Sciver58, Beghian58} & 303~\cite{Sciver56,Sibczynski12}\\
    undoped CsI & 1000~\cite{Nishimura95,Amsler02,csi} & 340~\cite{mikhailik15}\\\hline
  \end{tabular}
\end{table}

The attenuation/absorption length is the distance it takes for the intensity of radiation passing through a material to decrease by a factor of $1/e$ (approximately 63\%). This weakening occurs because radiation can interact with the material's atoms. In crystals, such interactions can excite vibrations within the crystal lattice, creating quasiparticles called phonons. These phonons effectively scatter the radiation, reducing its mean free path and causing it to be absorbed more readily~\cite{knoll}. Consequently, the attenuation length of a undoped CsI crystal for radiation generally increases with decreasing temperature. This is because lower temperatures lead to less intense thermal vibrations, reducing the number of phonons available to scatter the radiation.

To optimize the performance of our CsI crystal, absorption length and emission spectra at cryogenic temperatures need to be investigated~\cite{CsIProperties}. A spectrophotometer will be used to measure the absorption length of undoped CsI across various photon wavelengths at cryogenic temperatures. This will help us identify the wavelengths where radiation is strongly absorbed by the crystal (absorption band). The emission spectra of CsI using a monochrometer at cryogenic temperatures were measured by Ref.~\cite{Nishimura95}. These measurements reveal the luminescence intensity distribution of specific wavelengths of light emitted by the crystal.

\subsection{SiPM arrays}
\hspace{0.5cm} PMTs serve as excellent light sensors. However, charged particles resulting from natural radiation and cosmic rays can induce Cherenkov radiation as they traverse through a PMT quartz (or fused silica) window. With sufficient energy, a Cherenkov event can be readily differentiated from a scintillation event, as the former occurs within a shorter time window, producing a sharper current pulse compared to the latter. However, near the threshold, only a few detectable photons are generated, giving rise to a few single-photoelectron (PE) pulses that are virtually identical in shape. The efficiency of pulse shape discrimination diminishes as the energy decreases (see Fig.~\ref{f:why}). It's worth noting that employing two PMTs on the two end surfaces of a cylindrical crystal, with a requirement for coincident light detection in both, does not aid in eliminating Cherenkov events. This is because Cherenkov light created in one PMT can easily propagate to the other.

Two alternatives that avoid generating Cherenkov radiation are APDs and SiPMs since they are constructed from thin silicon wafers. APDs exhibit very high photon detection efficiencies (PDE, approximately 80\%)~\cite{ess20}. However, their gains are considerably lower than those of PMTs and SiPMs, making them insensitive to single PE events. Conversely, a SiPM, essentially an array of small APDs (microcells) operating in Geiger mode, is sensitive down to a single PE in each microcell. The size of these microcells must be small enough to prevent multiple photons from hitting the same microcell, and the spaces between microcells are not sensitive to photons. Consequently, the peak PDE of a SiPM (currently up to 56\%) is smaller than that of an APD but typically higher than the peak quantum efficiency of a PMT~\cite{jac14}.

Covering a large crystal area with a monolithic SiPM die faces challenges, primarily due to production yield limitations. As a compromise, a viable solution is to tightly tile several dies together to create an array. When considering the same active area, a SiPM array utilizes less material, occupies less space, and can be manufactured with greater radio-purity than a PMT. Table~\ref{t:sipm} outlines several SiPM arrays currently available in the market, all exhibiting sufficiently high PDE for the proposed research. Their gains are also comparable to those of typical PMTs, simplifying signal readout compared to APDs. Importantly, most of these arrays have undergone testing in liquid argon (LAr) or LN$_2$ temperature conditions (e.g., Ref.~\cite{lc08, lightfoot09, rossi16, catalanotti15, johnson18} for SensL, Ref.~\cite{otono07, akiba09, igarashi16} for Hamamatsu, and Ref.~\cite{jj16} for KETEK SiPMs). FBK SiPMs seem to be the only ones proven working down to 40~K with a good performance~\cite{aalseth17, acerbi17, giovanetti17}. The feasibility of SiPM arrays at 77~K is hence achievable, if possible, the performance of SiPM arrays from more manufacturers down to around 40~K could be verified as well.
\begin{table}[htbp]\centering
  \caption{SiPM arrays available in the market possibly suitable for the proposed research.}\label{t:sipm}
  \begin{tabular}{cccccc}\toprule
    Company & SiPM & microcell size & PDE$^\dagger$ & Largest array size & Gain$^\diamond$ \\\midrule
    SensL & J-series & 35~$\mu$m & 50\% & $50.4\times50.4$~mm$^2$ & $6.3\times10^6$\\
    SensL & C-series & 35~$\mu$m & 40\% & $57.4\times57.4$~mm$^2$ & $5.6\times10^6$\\
    Hamamatsu & S141xx &  50~$\mu$m & 50\% & $25.8\times25.8$~mm$^2$ & $4.7\times10^6$\\
    Hamamatsu & S133xx &  50~$\mu$m & 40\% & $25.0\times25.0$~mm$^2$ & $2.8\times10^6$\\
    KETEK$^\ddagger$ & AFBR-S4K33P6425B & 25~$\mu$m & 43\% & $26.8\times26.8$~mm$^2$ & $1.7\times10^6$ \\
    Broadcom & AFBR-S4N44P164M & 40~$\mu$m & 63\% & $16\times16$~mm$^2$ & $7.3\times10^6$ \\\bottomrule
  \end{tabular}
  \\\vspace{-1em}\flushleft\hspace{1.5cm}$^\dagger$ @ $420\sim 450$~nm. \hspace{1cm} $\diamond$ @ 5 volt over-voltage.
  \hspace{1cm} $\ddagger$ now Broadcom.
\end{table}
  
Potential issues with a SiPM include dark count rate~\cite{aalseth17}, afterpulses~\cite{sipm}, optical crosstalk~\cite{oto07, lc08, aki09, jan11, ding22}, and energy resolution~\cite{ding22}. Detailed discussions on mitigations and solutions can be found in Chapter~\ref{s:esipm}.

\section{The responses to electron recoils of undoped CsI measured with PMTs}
\label{s:e}
\hspace{0.5cm} Electron and nuclear recoils represent distinct pathways for energy transfer within materials. Electron recoils involve the direct excitation or ionization of individual electrons, typically observed in calibration sources where $\gamma$-rays deposit fixed energy values in the target through the photoelectric effect. In contrast, nuclear recoils result from direct collisions of incoming particles, such as neutrons, neutrinos, and dark matter, with the target's nucleus, resulting in energy deposition within the target. In the development of CryoCsI, various electron recoils and nuclear recoils measurements were conducted. Section~\ref{s:lySum} summarized all electron recoils measurements. Chapter \ref{s:e} will discuss the measurements of electron recoils using PMTs, the subsequent Chapter \ref{s:esipm} will elaborate on the measurements of electron recoils using SiPMs, and Chapter \ref{s:qf} will cover the measurements of nuclear recoils using PMTs.

\subsection{Electron recoils measurements summary}
\label{s:lySum}
\hspace{0.5cm} Our group has made consistent advancements in enhancing the light yield of undoped CsI, progressing from 20 to 50 PE per keV electron-equivalent (keV$_{ee}$) over several years. In 2016, a small undoped CsI crystal from OKEN Crystals was coupled with an R8778MODAY(AR) PMT from Hamamatsu at 77 K, achieving a yield of approximately 20 PE/keV$_{ee}$~\cite{csi}. To validate the scalability of this approach to larger crystals, two R11065 PMTs were employed to detect light from a 3-inch diameter, 5 cm tall undoped CsI crystal obtained from SICCAS (Shanghai Institute of Ceramics, Chinese Academy of Sciences), achieving a light yield of approximately 26 PE/keV$_{ee}$~\cite{csi20}. Both measurements utilized $\gamma$-ray lines ranging from 662 keV to 2.6 MeV. Additionally, the light yield in a lower energy range from an OKEN crystal was determined using an $^{241}$Am source. With improved CsI wrapping, an average yield of approximately 33.5 PE/keV$_{ee}$ was observed in this measurement range of 13.9 to 77 keV~\cite{ding20e}. More recently, a small cubic CsI crystal from Amcrys was encapsulated between two SensL SiPMs. The observed light yield in this setup was $43.0 \pm 1.1$ PE/keV$_{ee}$~\cite{ding22}. Subsequently, the SiPMs were coated with 1,1,4,4-Tetraphenyl-1,3-butadiene (TPB) to shift the 340 nm scintillation light from CsI to around 420 nm, where a typical SiPM's PDE peaks. This coating resulted in a boosted light yield of $50 \pm 2$ PE/keV$_{ee}$. The results of all measurements are summarized in Table~\ref{t:ly}. Higher light yields are anticipated with operation at 40 K and utilization of SiPMs with higher PDEs.

\begin{table}[htbp]\centering
  \caption{Achieved~\cite{csi, csi20, ding20e, ding20, ding22} and predicted light yields of undoped CsI compared to that of the COHERENT CsI[Na] detector~\cite{coherent17}. The unit of the light yield is detected PE/keV$_{ee}$.}\label{t:ly}
  \begin{tabular}{rcl} \hline
    Experiments &Type of crystals &Light yield [PE/keV$_{ee}$]\\\hline
    COHERENT 2017 & CsI[Na] & 13.5 $\pm$ 0.1~\cite{coherent17}\\
    PMT+small crystal & undoped CsI & 20.4 $\pm$ 0.8~\cite{csi}\\
    PMTs+large crystal$^\dagger$ & undoped CsI & 26.0 $\pm$ 0.4~\cite{csi20}\\
    Improved light collection$^\ddagger$ & undoped CsI & 33.5 $\pm$ 0.7~\cite{ding20e, ding20}\\
    Energy low to 5.9 keV$^\lozenge$ & undoped CsI & 33.4 $\pm$ 2.0 \\
    PMT $\rightarrow$ SiPMs$^\S$  & undoped CsI & 43.0 $\pm$ 1.1~\cite{ding22}\\
    WLS coating on SiPMs$^\S$  & undoped CsI & 50.0 $\pm$ 2.0\\
    77 $\rightarrow$ 40 K, \& SiPMs$^\star$ with 50\% PDE & undoped CsI & 60 (potential goal)\\ \hline
  \end{tabular}

$^\dagger$ See section~\ref{s:large} for detailed description\\
$^\ddagger$ See section~\ref{s:better} for detailed description\\
$^\lozenge$ See section~\ref{s:cryostat} for detailed description\\
$^\S$ See Chapter~\ref{s:esipm} for detailed description\\
$^\star$ SiPMs with high PDE
\end{table}

\subsection{Large undoped CsI coupled with PMTs}
\label{s:large}
\subsubsection{Experimental setup}
\begin{figure}[htbp] \centering
  \includegraphics[width=0.7\linewidth]{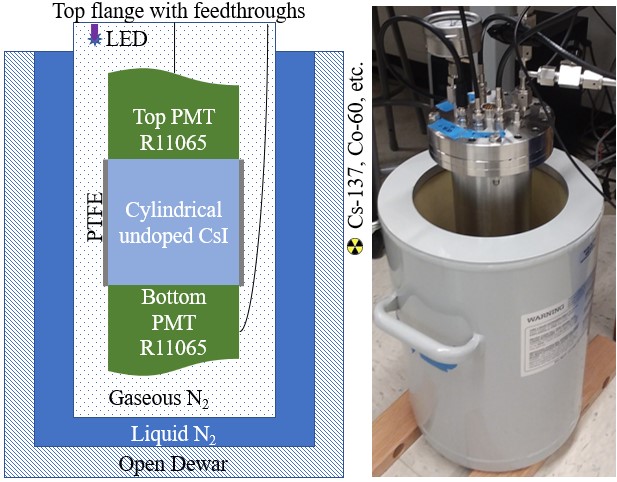}
  \caption{A sketch (left) and a picture (right) of the experimental setup.}
  \label{f:LargeSetup}
\end{figure}

\hspace{0.5cm} The right picture in Fig.~\ref{f:LargeSetup} shows an open LN2 dewar used to cool a 50~cm long stainless steel tube placed inside. The inner diameter of the tube was $\sim10$~cm. The tube was vacuum sealed on both ends by two 6-inch ConFlat (CF) flanges.  The bottom flange was blank and attached to the tube with a copper gasket in between. The top flange was attached to the tube with a fluorocarbon CF gasket in between for multiple operations. Vacuum welded to the top flange were five BNC, two SHV, one 19-pin electronic feedthroughs and two 1/4-inch VCR connectors.

\begin{figure}[htbp] \centering
  \includegraphics[width=0.8\linewidth]{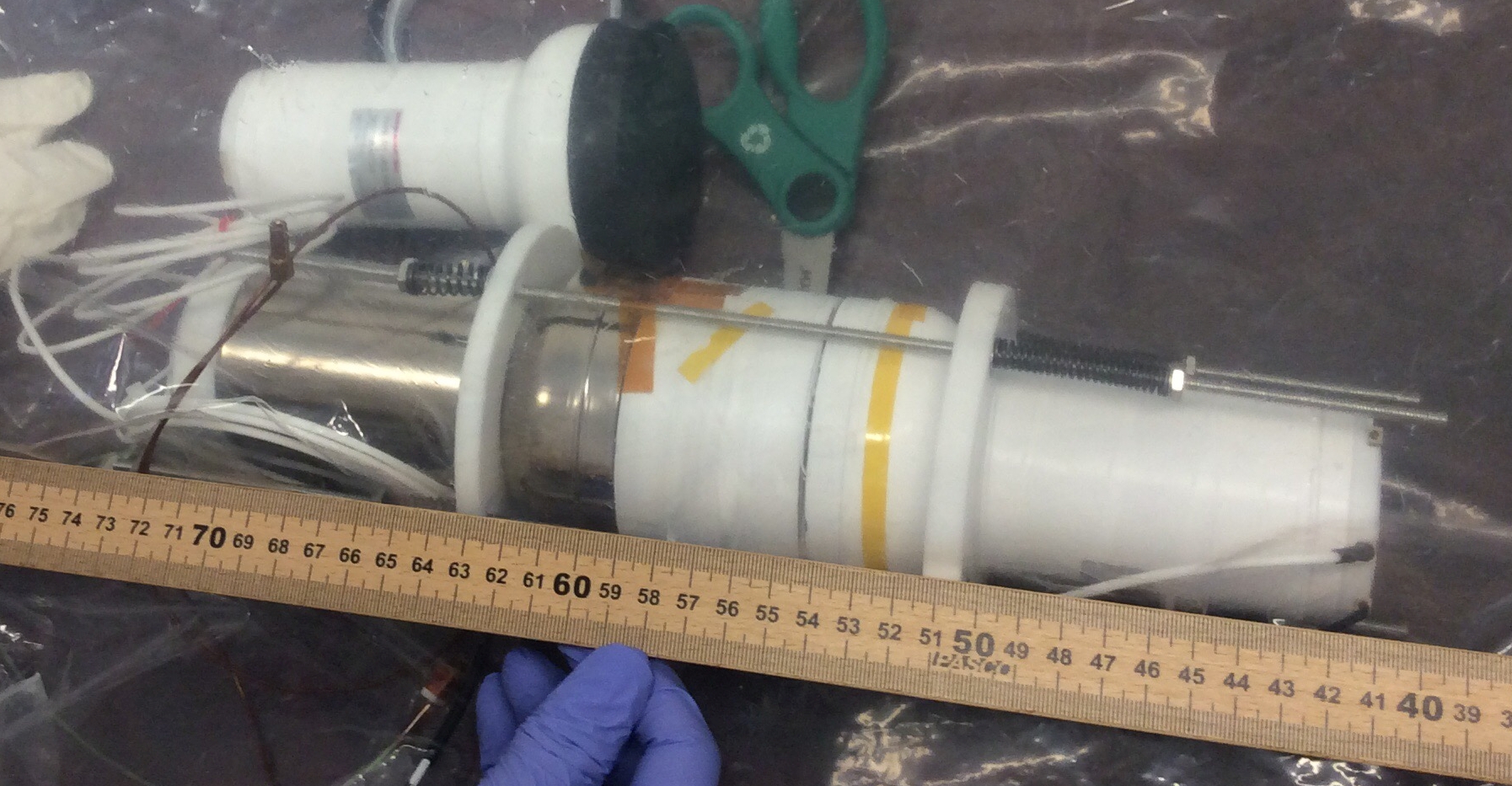}
  \caption{The detector assembly in a glove bag.}
  \label{f:det1}
\end{figure}

The left sketch in Fig.~\ref{f:LargeSetup} shows the internal structure of the experimental setup. Three different undoped cylindrical CsI crystals were used in the measurements was purchased from the Shanghai Institute of Ceramics, Chinese Academy of Sciences. It had a diameter of 3 inches, a height of 5~cm and a mass of 1.028~kg. All surfaces were mirror polished. The side surface was wrapped with multiple layers of Teflon tapes. Two 3-inch Hamamatsu R11065-ASSY PMTs were attached to the two end surfaces without optical grease. To ensure a good optical contact, the PMTs were pushed against the crystal by springs, as shown in Fig.~\ref{f:det1}. The assembly was done in a glove bag flushed with dry nitrogen gas to minimize exposure of the crystal to atmospheric moisture. The relative humidity was kept below 5\% at 22$^{\circ}$C during the assembly process.

The assembled crystal and PMTs were lowered into the stainless steel chamber from the top. After all cables were fixed beneath it, the top flange was closed.  The chamber was then pumped with a Pfeiffer Vacuum HiCube 80 Eco to $\sim1\times {10}^\text{-4}$ mbar. Afterward, it was refilled with dry nitrogen gas to 0.17 MPa above the atmospheric pressure and placed inside the open dewar. Finally, the chamber was cooled by filling the dewar with LN2. After cooling, the chamber pressure was reduced to slightly above the atmospheric pressure.

A few Heraeus C~220 platinum resistance temperature sensors were used to monitor the cooling process. They were attached to the side surface of the crystal, the PMTs, and the top flange to obtain the temperature profile of the long chamber. A Raspberry Pi 2 computer with custom software~\cite{cravis} was used to read out the sensors. The cooling process could be done within about 30 minutes. Most measurements, however, were done after about an hour of waiting to let the system reach thermal equilibrium. The temperature of the crystal during measurements was about 3~K higher than the LN2 temperature.

The PMTs were powered by a 2-channel CAEN N1470A high voltage power supply NIM module. Their signals were fed into a 4-channel CAEN DT5751 waveform digitizer, which had a 1~GHz sampling rate, a 1~V dynamic range and a 10-bit resolution. Custom-developed software was used for data recording~\cite{daq}. The recorded binary data files were converted to CERN ROOT files for analysis~\cite{nice}.

\begin{figure}[htbp]\centering
  \includegraphics[width=0.50\linewidth]{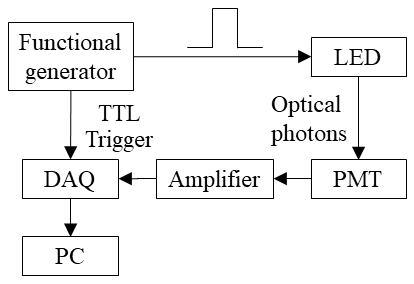}\hfill
  \includegraphics[width=0.49\linewidth]{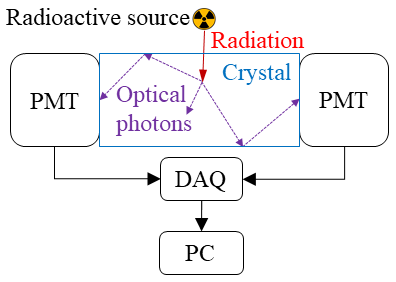}
  \caption{Trigger logics for single-photoelectron response (left) and energy calibration (right) measurements.}
  \label{f:trg}
\end{figure}

\subsubsection{Single PE responses}
\hspace{0.5cm} The single-photoelectron response of PMTs was measured using light pulses from an ultraviolet LED from Thorlabs, LED370E. Its output spectrum peaked at 375~nm with a width of 10~nm, which was within the 200-650~nm spectral response range of the PMTs. Light pulses with a $\sim$50~ns duration and a rate of 10~kHz were generated using an RIGOL DG1022 arbitrary function generator. The intensity of light pulses was tuned by varying the output voltage of the function generator so that only one or zero photons hit one of the PMTs during the LED lit window most of the time. A TTL trigger signal was emitted from the function generator simultaneously together with each output pulse. It was used to trigger the digitizer to record the PMT response. The trigger logic is shown in the left flow chart in Fig.~\ref{f:trg}.

A typical single PE pulse from an R11065 working at its recommended operational voltage, 1500~V, is well above the pedestal noise. However, the two PMTs were operated at about 1300~V to avoid saturation of electronic signals induced by 2.6~MeV $\gamma$-rays from environmental $^{208}$Tl. The consequent small single-PE pulses hence had to be amplified by a factor of ten using a Phillips Scientific Quad Bipolar Amplifier Model 771 before being fed into the digitizer in order to separate them from the pedestal noise.

\begin{figure}[htbp]\centering
  \includegraphics[width=0.7\linewidth]{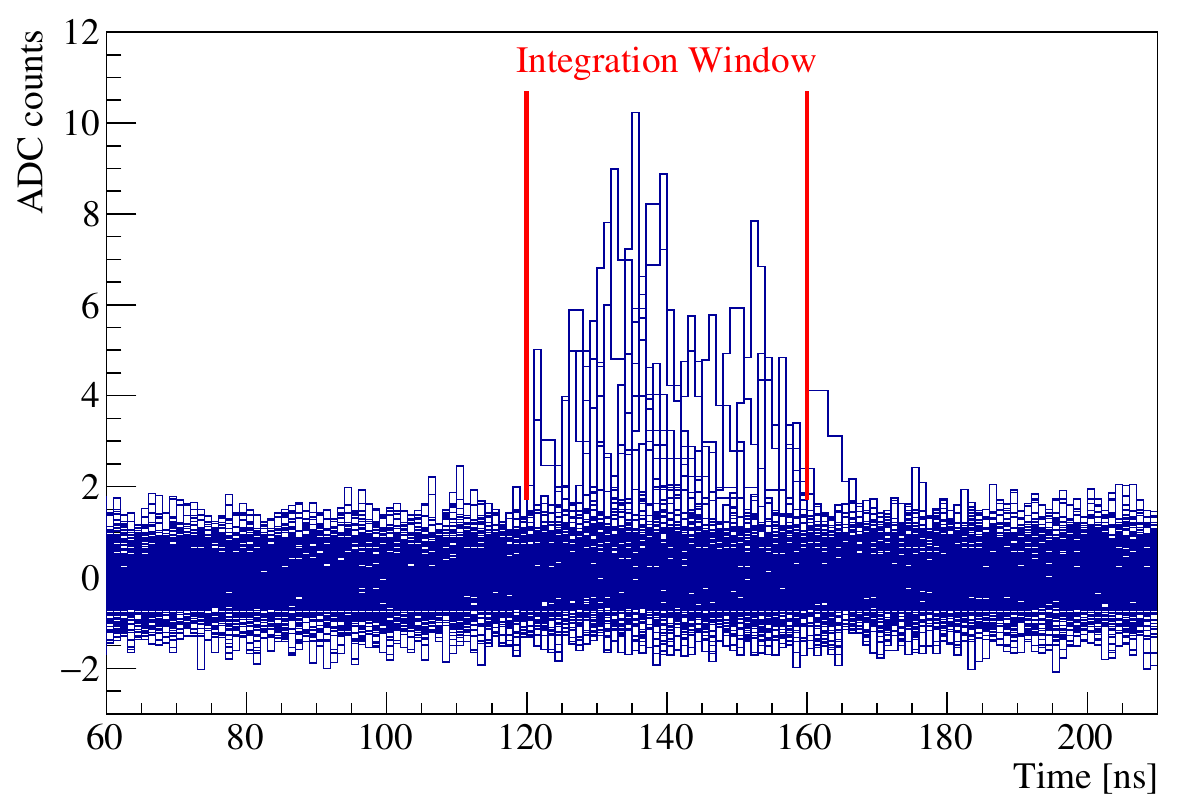}
  \caption{Two hundred consecutive waveforms from the bottom PMT overlapped with each other.}
  \label{f:spe1}
\end{figure}

\begin{figure}[htbp]\centering
  \includegraphics[width=0.8\linewidth]{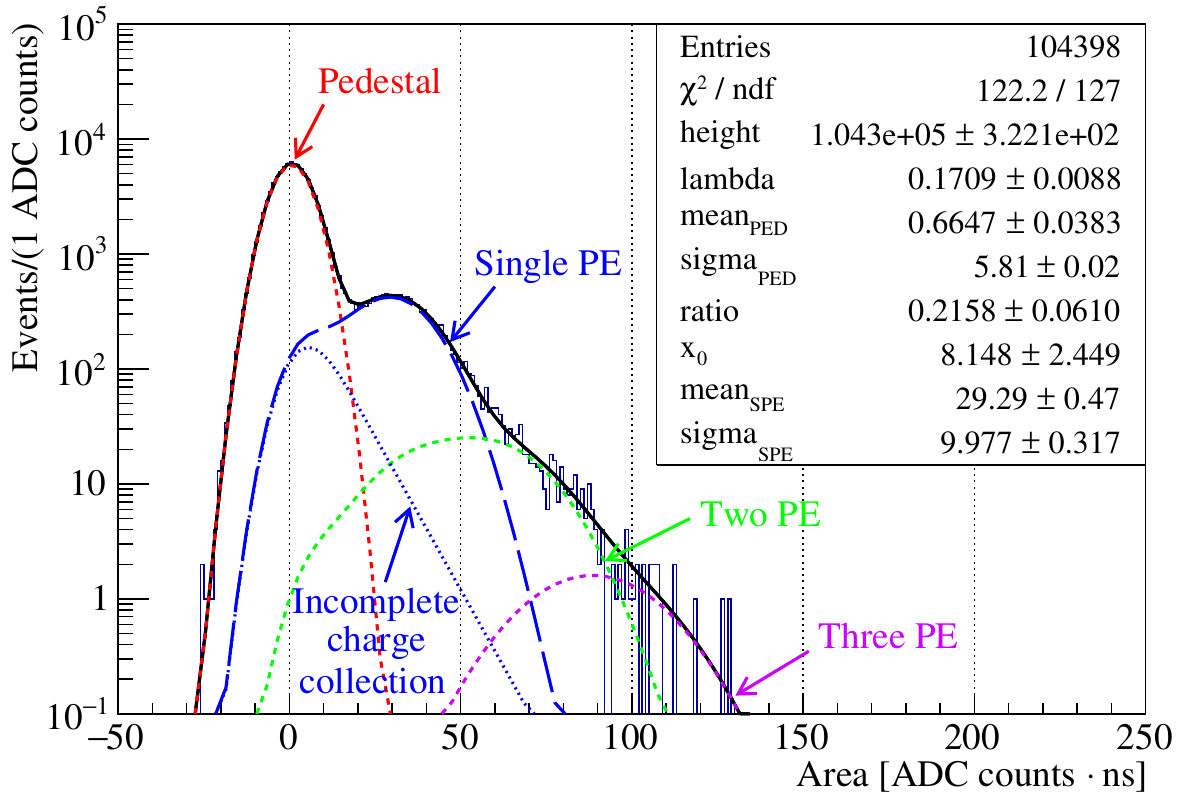}
  \caption{Single-PE response of the top PMT in logarithm scale.}
  \label{f:spe2}
\end{figure}

\begin{figure}[ht!]\centering
  \includegraphics[width=0.8\linewidth]{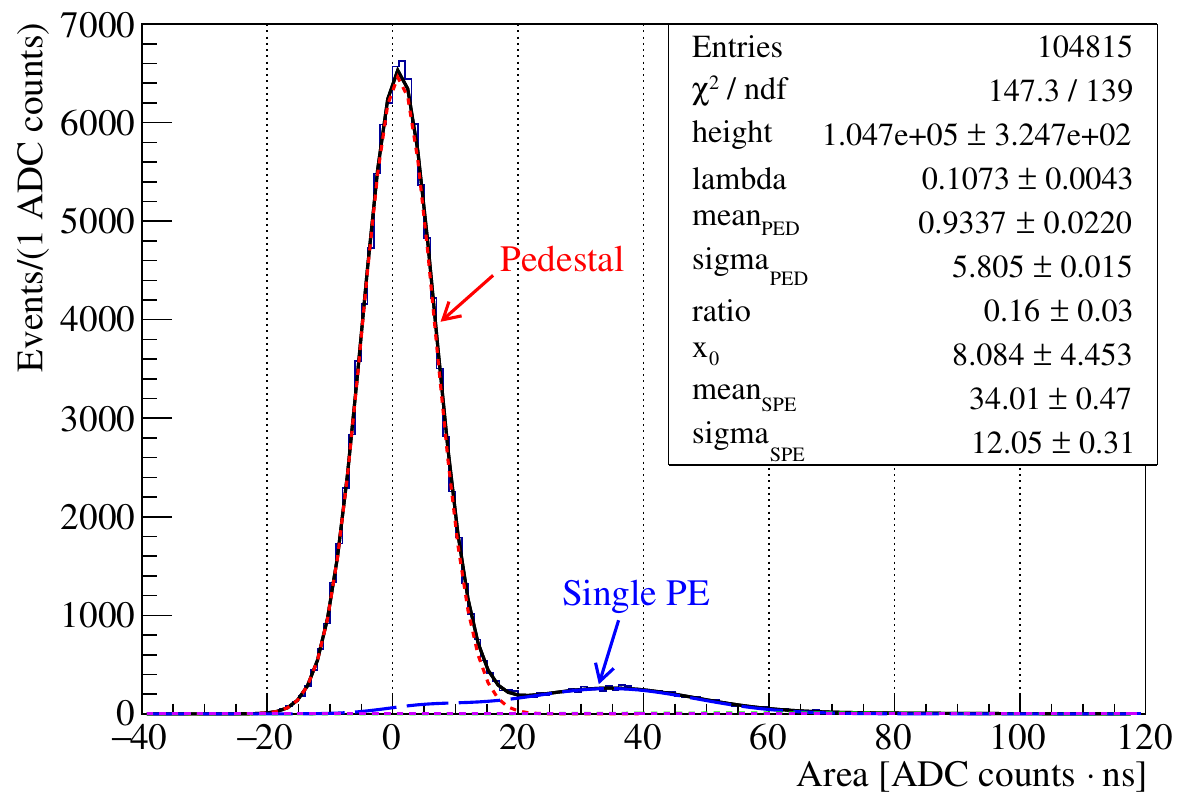}
  \caption{Single-PE response of the bottom PMT in linear scale. The two and three-PE distributions are too small to be visible.}
  \label{f:spe3}
\end{figure}

Fig.~\ref{f:spe1} shows two hundred consecutive waveforms from the bottom PMT randomly chosen from a data file taken during a single-PE response measurement. An integration window ranging from 120 to 160~ns was chosen instead of 120 to 170~ns to examine its influence on SPE response uncertainty, which is within 2\%. Approximately 20 of them contain a single-PE pulse within this integration window. An integration in this time window was performed for each waveform in the data file whether it contained a pulse or not. The resulting single-PE spectra for the top and bottom PMTs are presented in Fig.~\ref{f:spe2} and Fig.~\ref{f:spe3}, respectively.

The spectra were fitted in the same way as described in Ref.~\cite{ds1013} with a function,
\begin{equation}\label{e:Fx}
F(x)=H\sum\limits_n P(n,\lambda) f_n(x),
\end{equation}
where $H$ is a constant to match fit function to spectra counting rate, $P(n,\lambda)$ is a Poisson distribution with mean $\lambda$, which represents the average number of PE in the time window, $f_n(x)$ represents the \textit{n}-PE response, and can be expressed as
\begin{equation}\label{e:fnx}
f_n(x)=f_0(x) \ast f_1^{n\ast}(x),
\end{equation}
where $f_0(x)$ is a Gaussian function representing the pedestal noise distribution, $\ast$ denotes a mathematical convolution of two functions, and $f_1^{n\ast}(x)$ is a n-fold
convolution of the PMT single-PE response function, $f_1(x)$, with itself. The single-PE response function $f_1(x)$ was modeled as:
\begin{equation}\label{e:f1x}
f_1(x)=
  \begin{cases}
    R(\frac{1}{x_0} e^{-x/x_0})+(1 - R)G(x;\bar{x},\sigma) & x>0; \\
    0 & x\leq0,
  \end{cases}
\end{equation}
The former corresponds to the incomplete dynode multiplication of secondary electrons in a PMT, also known as incomplete charge collection, discribed by an exponential decay with a decay constant $x_0$. The latter corresponds to the full charge collection in a PMT, discribed by a Gaussian distribution $G(x;\bar{x},\sigma)$ with a mean of $\bar{x}$ and a width of $\sigma$. Where $R$ and $1-R$ defines the ratio of these two processes.

The fitting result for the top PMT is shown in Fig.~\ref{f:spe2}. The fitting function has eight free parameters as shown in the top-right statistic box in Fig.~\ref{f:spe2}, where ``height'' corresponds to $H$ in Eq.~\ref{e:Fx}, ``lambda'' corresponds to $\lambda$ in Eq.~\ref{e:Fx}, ``mean'' and ``sigma'' with a subscript ``PED'' represents the mean and the sigma of the Gaussian pedestal noise distribution, those with a subscript ``SPE'' represents $\bar{x}$ and $\sigma$ in Eq.~\ref{e:f1x}, respectively, and ``ratio'' corresponds to $R$ in Eq.~\ref{e:f1x}.  Due to technical difficulties in realizing multiple function convolutions in the fitting ROOT script, the three-PE distribution, $f_1^{3\ast}(x)$, was approximated by a Gaussian function with its mean and variance three times that of the single-PE response.

Table~\ref{t:1PE} lists means of single-PE distributions for both PMTs measured before and after the energy calibration mentioned in the next section to check the stability of the PMT gains. The average mean for the top and bottom PMT is 28.58 $\pm$ 0.51 and 33.08 $\pm$ 0.47 ADC~counts$\cdot$ns, respectively.

\begin{table}[ht!]
  \centering
  \caption{\label{t:1PE} Summary of the single-PE response for the top and bottom PMTs. Top and bottom rows for each PMT correspond to measurements before and after the energy calibration, respectively.}
  \begin{tabular}{c c c c}
    \hline
    PMT &Temperature &Temperature &Mean of single-PE \\
    &of PMT [$^\circ$C]&of crystal [$^\circ$C] &[ADC counts$\cdot$ns]\\
    \hline
     Top &
     \begin{tabular}{c} -134.3 $\pm$ 1.3 \\ -134.3 $\pm$ 1.3 \\
     \end{tabular} &
     \begin{tabular}{c} -193.3 $\pm$ 1.3  \\ -193.5 $\pm$ 1.3 \\
     \end{tabular} &
     \begin{tabular}{c} 28.53  $\pm$ 0.51 \\ 28.63  $\pm$ 0.45 \\
     \end{tabular} \\
    \hline
     Bottom &
     \begin{tabular}{c} -195.7 $\pm$ 1.3 \\ -195.5 $\pm$ 1.3 \\
     \end{tabular} &
     \begin{tabular}{c} -193.3 $\pm$ 1.3  \\ -193.5 $\pm$ 1.3 \\
     \end{tabular} &
     \begin{tabular}{c} 32.98  $\pm$ 0.47 \\ 33.18  $\pm$ 0.43 \\
     \end{tabular} \\
    \hline
  \end{tabular}
\end{table}

\subsubsection{Energy calibration}
\hspace{0.5cm} The energy calibration was performed using $\gamma$-rays from a $^{137}$Cs and a $^{60}$Co radioactive source, as well as $^{40}$K within the crystal and $^{208}$Tl from the environment. The sources were sequentially attached to the outer wall of the dewar as shown in Fig.~\ref{f:LargeSetup}. Background data were obtained before those with a source attached. The digitizer was triggered when both PMTs recorded a pulse above a certain threshold within a time window of 16~ns. The trigger logic is shown in the right flow chart in Fig.~\ref{f:trg}. The trigger rate for the background, $^{137}$Cs and $^{60}$Co data taking was 100 Hz, 410 Hz and 520 Hz, respectively, if the threshold was set to 10 ADC counts above the pedestal level.

\begin{figure}[ht!]\centering
  \includegraphics[width=0.8\linewidth]{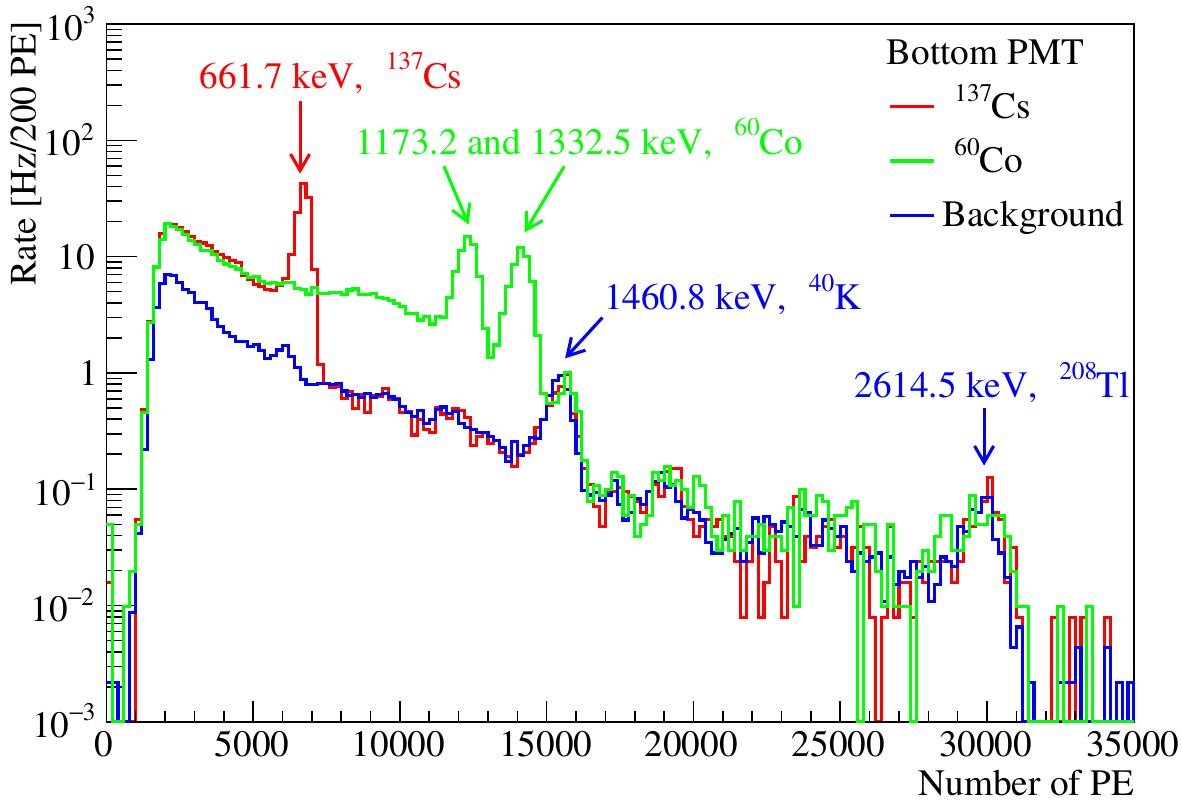}
  \caption{Energy spectra of the bottom PMT at 80~K.}
  \label{f:en1}
\end{figure}

Each recorded waveform was 8008~ns long. The rising edge of the pulse that triggered the digitizer was set to start at around 1602~ns so that there were enough samples before the pulse to extract the pedestal level of the waveform. After the pedestal level was adjusted to zero the pulse was integrated until its tail fell back to zero. The integration had a unit of ADC counts$\cdot$ns. It was converted to numbers of PE using the formula:
\begin{equation}
  \text{(number of PE)} = \text{(ADC counts} \cdot \text{ns)}/\bar{x},
  \label{e:nPE}
\end{equation}
where $\bar{x}$ is the mean of the single-PE Gaussian distribution mentioned in Eq.~\ref{e:f1x}. Its unit was also ADC count$\cdot$ns. Its value was obtained from the fittings shown in Fig.~\ref{f:spe2} and \ref{f:spe3}.  The resulting spectra normalized by their event rates recorded by the bottom PMT are shown in Fig.~\ref{f:en1}. The spectra from the top PMT are very similar.

\begin{figure}[ht!]\centering
  \includegraphics[width=0.8\linewidth]{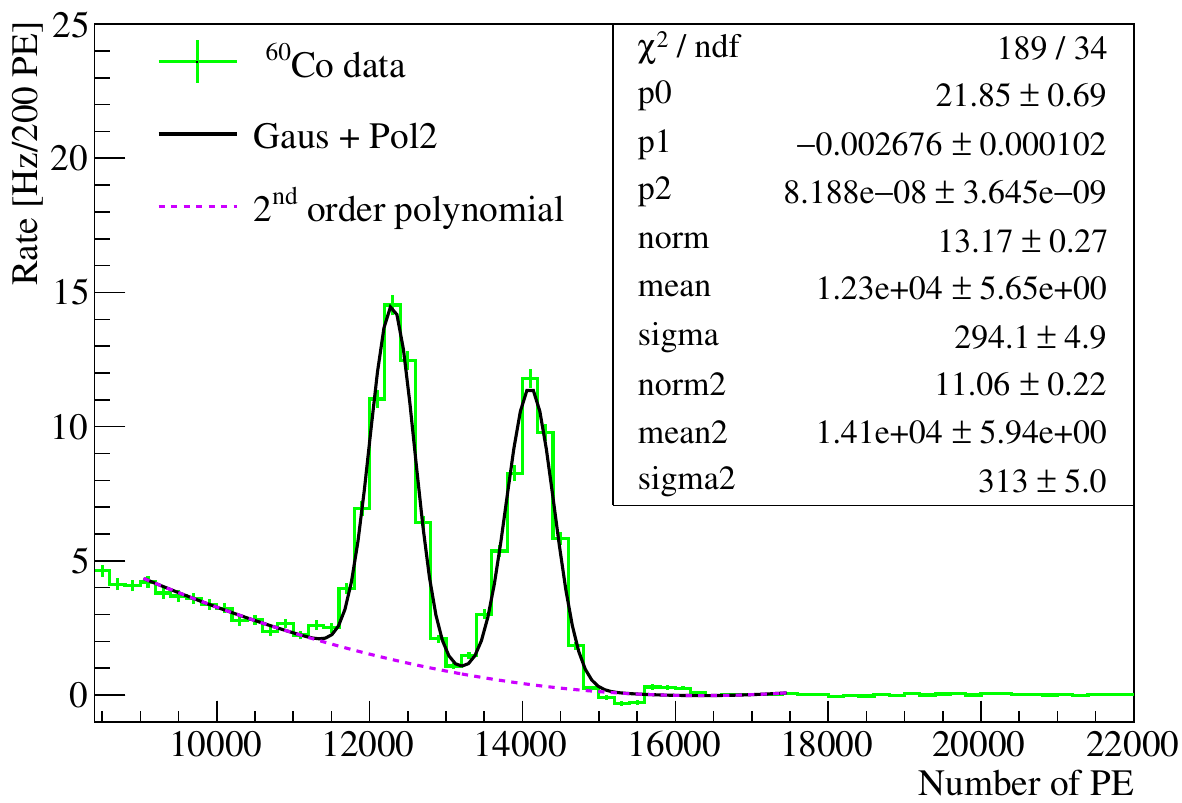}
  \caption{Energy spectrum recorded by the bottom PMT at 80~K with a $^{60}$Co source. The clearly separated peaks were fitted simultaneously with two Gaussian distributions on top of a 2$^{nd}$ order polynomial.}
  \label{f:co60}
\end{figure}

The $\gamma$-ray peaks were fitted using one or two Gaussian distributions on top of a  2$^{nd}$ order polynomial. A simultaneous fit for the 1.17 MeV and 1.33 MeV peaks from $^{60}$Co is shown in Fig.~\ref{f:co60} as an example. The peaks are clearly separated indicating an energy resolution much better than that of a regular NaI(Tl) detector running at room temperature. The means and sigmas of the fitted Gaussian functions are listed in Table~\ref{t:rPE} together with those from other $\gamma$-ray peaks.

\begin{table}[ht!]
  \caption{\label{t:nPE} Summary of $\gamma$-ray peaks in the calibration spectra.}
  \begin{minipage}{\linewidth}\centering
  \begin{tabular}{cccccc}
    \hline
   PMT & Isotope & Energy & Mean & Sigma & FWHM \\
   &  & (keVee) & (PE) & (PE) & (\%) \\
    \hline
     Top &
     \begin{tabular}{r} $^{137}$Cs \\ $^{60}$Co \\ $^{60}$Co \\ $^{40}$K \\ $^{208}$Tl \\
     \end{tabular} &
     \begin{tabular}{r} 661.7 \\ 1173.2 \\ 1332.5 \\ 1460.8 \\ 2614.5 \\
     \end{tabular} &
     \begin{tabular}{r} 9314.0 \\ 18043.1 \\ 20913.6 \\ 23137.6 \\ 42723.7 \\ 
     \end{tabular} &
     \begin{tabular}{r} 338.7 \\ 552.5 \\ 579.3 \\ 645.2 \\ 611.4 \\
     \end{tabular} &
     \begin{tabular}{r} 8.6 \\ 7.2 \\ 6.5 \\ 6.6 \\ 3.4 \\
     \end{tabular} \\
     \hline
     Bottom &
     \begin{tabular}{r} $^{137}$Cs \\ $^{60}$Co \\ $^{60}$Co \\ $^{40}$K \\ $^{208}$Tl \\
     \end{tabular} &
     \begin{tabular}{r} 661.7 \\ 1173.2 \\ 1332.5 \\ 1460.8 \\ 2614.5 \\
     \end{tabular} &
     \begin{tabular}{r} 6716.1 \\ 12297.2 \\ 14103.6 \\ 15416.5 \\ 29758.1 \\ 
     \end{tabular} &
     \begin{tabular}{r} 204.2 \\ 294.1 \\ 313.0 \\ 331.6 \\ 483.0 \\
     \end{tabular} &
     \begin{tabular}{r} 7.2 \\ 5.6 \\ 5.2 \\ 5.1 \\ 3.8 \\
     \end{tabular} \\
     \hline
  \end{tabular}
  \end{minipage}
\end{table}

\subsubsection{Light yield}
\begin{figure}[ht!]\centering
  \includegraphics[width=0.8\linewidth]{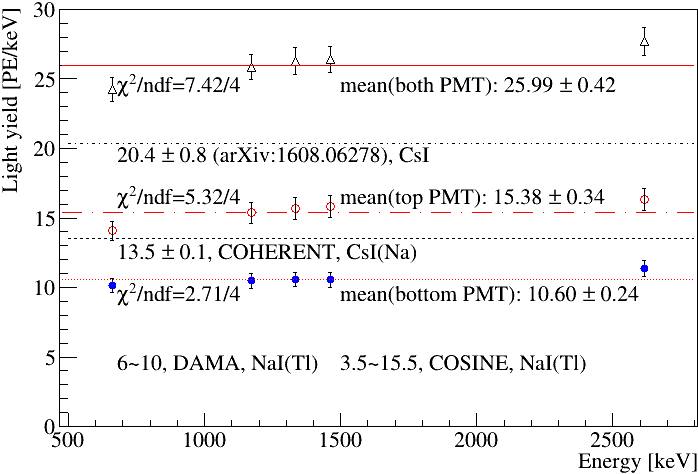}
  \caption{The obtained light yields for the top (empty circles), the bottom (filled squares) and both (empty triangles) PMTs, compared to those achieved by other experiments~\cite{coherent17, dama18, cosine19} and an earlier measurement with a smaller crystal~\cite{csi}.}
  \label{f:ly1}
\end{figure}

\hspace{0.5cm} The light yield was calculated for each PMT using the data in Table~\ref{t:nPE} and the following equation:
\begin{equation}
  \text{light yield [PE/keV$_{ee}$]} = \text{Mean [PE]}/\text{Energy [keV$_{ee}$]}.
  \label{e:LY}
\end{equation}
The obtained light yield at each energy point is shown in Fig.~\ref{f:ly1}. The light yield of the whole system was calculated as a sum of those of the top and bottom PMTs.  The uncertainties of light yields are mainly due to the uncertainties of mean values of the single-PE responses used to convert the \textit{x}-axes of the energy spectra from ADC counts$\cdot$ns to the number of PE. The data points in each category were fitted by a straight line to get an average light yield, which was 15.38 $\pm$ 0.34 PE/keV$_{ee}$ for the top PMT, 10.60 $\pm$ 0.24 PE/keV$_{ee}$ for the bottom one, and 25.99 $\pm$ 0.42 PE/keV$_{ee}$ for the system.

To understand the origin of the significant light yield difference between the two PMTs, additional measurements were performed. First, the PMT-crystal assembly were pulled from the chamber and reinserted upside down without any other change. The PMTs kept their yields unchanged. Second, the PMT with the lower yield was replaced by another R11065. No significant change was observed by the replaced PMT. Last, the crystal was flipped while the PMTs were kept in their original locations. Again, no significant light yield change was observed by the low yield PMT. Therefore, the difference in the light yields between the two PMTs was most probably due to the difference in the quantum efficiencies of individual PMTs instead of different optical interfaces or temperatures.

There seems to be a systematic decrease of the light yield as the energy decreases as shown in Fig.~\ref{f:ly1}. This may indicate some non-linearity in the energy response of the undoped CsI crystal at 77 K. If the light yield at lower energies is significantly lower. The technique under investigation may not be suitable for dark matter search. Limited by the large uncertainty of each data point, no quantitative conclusion can be drawn from this measurement. Fortunately, there exist some investigations of the non-linearity of both undoped CsI~\cite{Moszynski05} and NaI~\cite{Moszynski09} at 77~K from 5.9~keV to 1.3~MeV. The results vary with the crystals used in those studies. Some had less, others had more light yields at lower energies than that at 1.3~MeV. The difference ranges from 0 to 30\%. To verify the light yield of the crystal used in this study at lower energies, additional studies with $^{241}$Am and $^{55}$Fe were conducted. These studies yielded even higher light yields in the range of [13.9, 77] keV, which will be discussed in the following section.

\subsection{Short undoped CsI coupled with a PMT}
\label{s:better}
\subsubsection{Experimental setup}
\begin{figure}[ht!] \centering
  \includegraphics[width=0.7\linewidth]{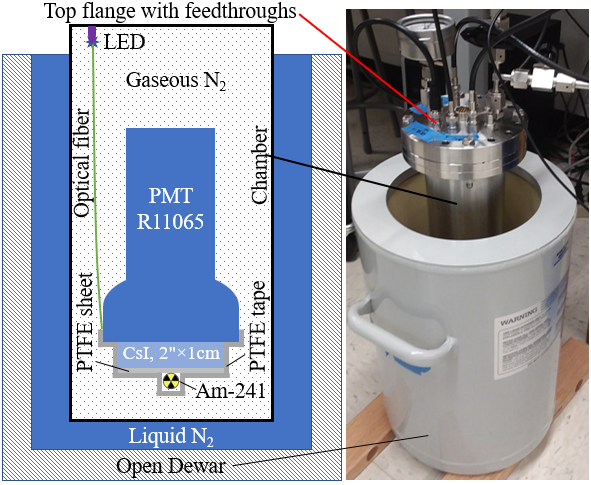}
   \includegraphics[width=0.7\linewidth]{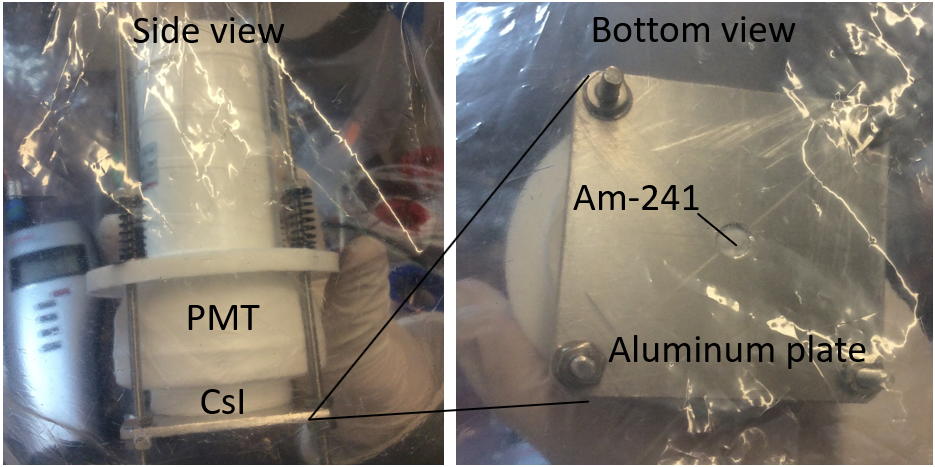}
  \caption{A sketch and pictures of the experimental setup.}
  \label{f:Bsetup}
\end{figure}

\hspace{0.5cm} Fig.~\ref{f:Bsetup} shows the internal structure of the experimental setup for the measurement of the light yield of an undoped CsI crystal. The undoped cylindrical crystal was purchased from OKEN~\cite{oken}, and had a radius of 1~inch and a height of 1~cm. All surfaces were mirror polished. It was used in an earlier measurement, where a yield of $20.4\pm0.8$~PE/keV$_{ee}$ was achieved above 662~keVee~\cite{csi}. Compared to the early measurement, the following modifications were made:

\begin{itemize}
  \item The side surface of the crystal was wrapped with multiple layers of Teflon tapes instead of a single layer to make sure that there was no light leak.
  \item The 2-inch Hamamatsu PMT R8778MODAY(AR) was replaced by a Hamamatsu 3-inch R11065-ASSY.
  \item In both setups, the PMTs were pushed against one of the crystal end surfaces by springs to ensure adequate optical contact without optical grease. However, in the previous setup, the crystal was pushed against the bottom flange of the chamber, while in this setup, the crystal was pushed against an aluminum plate with a hole in the middle, leaving space for the placement of an $^{241}$Am source.
  \item The other end surface of the crystal was pushed against a PTFE sheet in between the crystal and the aluminum plate. The $^{241}$Am source was placed on the other side of the PTFE sheet so that alpha radiation was blocked from reaching the crystal.
\end{itemize}

To minimize exposure of the crystal to atmospheric moisture, assembly was done in a glove bag flushed with dry nitrogen gas. The relative humidity was kept below 5\% at 22$^{\circ}$C during the assemble process.

The PMT-crystal assembly was lowered into a 50~cm long stainless steel chamber from its top opening. The inner diameter of the chamber was $\sim 10$~cm. The chamber was vacuum sealed on both ends by two 6-inch CF flanges. The bottom flange was blank and attached to the chamber with a copper gasket in between. The top flange was attached to the chamber with a fluorocarbon CF gasket in between for multiple operations. Vacuum welded to the top flange were five BNC, two SHV, one 19-pin electronic feedthroughs and two 1/4-inch VCR connectors.

After all cables were fixed beneath it, the top flange was closed. The chamber was then pumped with a Pfeiffer Vacuum HiCube 80 Eco to $\sim 1\times {10}^{-4}$~mbar. Afterward, it was refilled with dry nitrogen gas to 0.19 MPa above the atmospheric pressure and placed inside an open LN2 dewar. The dewar was then filled with LN2 to cool the chamber and everything inside. After cooling, the chamber pressure was reduced to slightly above the atmospheric pressure.

A few Heraeus C~220 platinum resistance temperature sensors were used to monitor the cooling process. They were attached to the side surface of the crystal, the PMT, and the top flange to obtain the temperature profile of the long chamber. A Raspberry Pi 2 computer with custom software~\cite{cravis} was used to read out the sensors. The cooling process took about 30 minutes. Most measurements, however, were taken after about an hour of waiting to let the system reach thermal equilibrium. The temperature of the crystal during measurements was about 3~K higher than the LN2 temperature.

The PMT was powered by a CAEN N1470A high voltage power supply in a NIM crate. The signals were fed into a CAEN DT5751 waveform digitizer, which had a 1~GHz sampling rate, a 1~V dynamic range and a 10 bit resolution. Custom-developed software was used for data recording~\cite{daq}. The recorded binary data files were converted to CERN ROOT files for analysis~\cite{nice}.

\subsubsection{Single PE response}
\hspace{0.5cm} The single-PE response of the PMT was measured using light pulses from an ultraviolet LED, LED370E from Thorlabs. Its output spectrum peaked at 375~nm with a width of 10~nm, which was within the 200 -- 650~nm spectral response range of the PMT. Light pulses with a $\sim$50~ns duration and a rate of 10~kHz were generated using an RIGOL DG1022 arbitrary function generator. The intensity of light pulses was tuned by varying the output voltage of the function generator so that only one or zero photon hit the PMT during the LED lit window most of the time. A TTL trigger signal was emitted from the function generator simultaneously together with each output pulse. It was used to trigger the digitizer to record the PMT response. The trigger logic flow chart is shown in the left of Fig.~\ref{f:trg}.

The PMT was biased at 1,600~V, slightly above the recommended operation voltage, 1,500~V, to increase the gain of the PMT.  Single-PE pulses were further amplified by a factor of ten using a Phillips Scientific Quad Bipolar Amplifier Model 771 before being fed into the digitizer in order to separate them well from the pedestal noise.

\begin{figure}[ht!] \centering
  \includegraphics[width=0.7\linewidth]{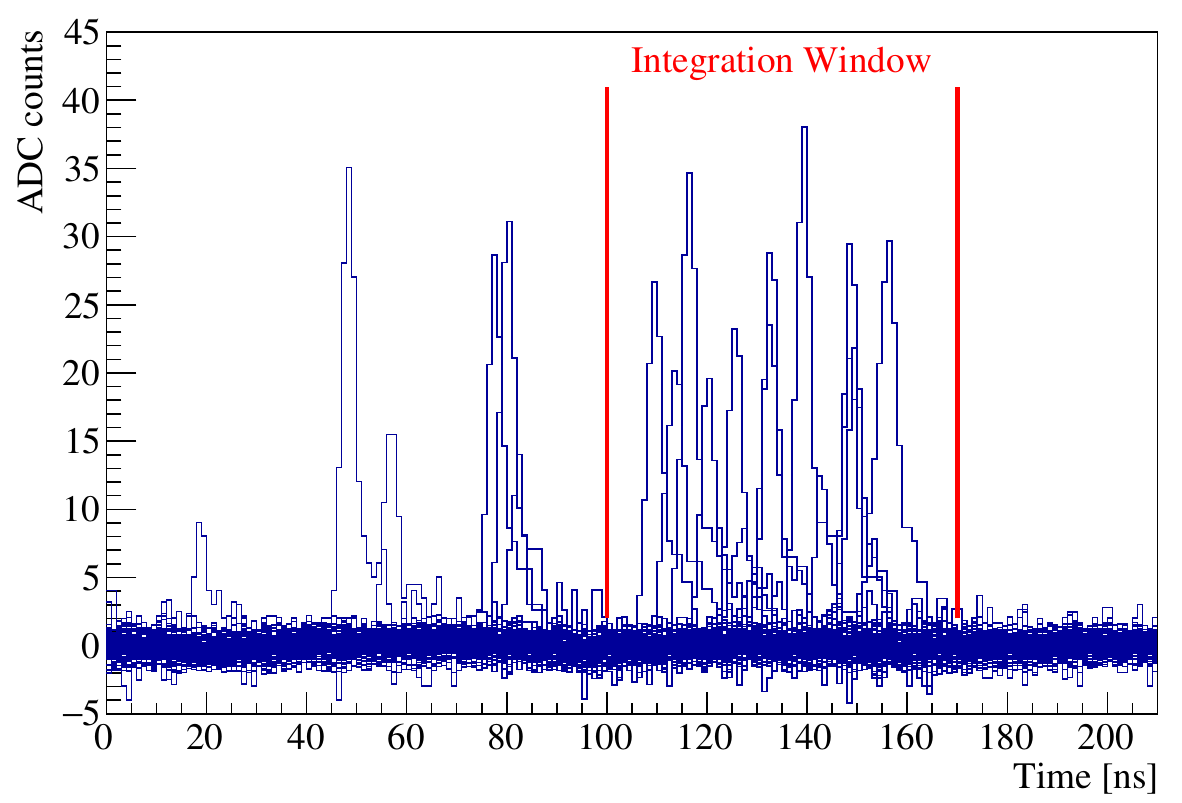}
  \caption{Two hundred consecutive waveforms from the PMT overlapped with each other measured with the crystal in place.}
  \label{f:ps}
\end{figure}

\begin{figure}[ht!] \centering
  \includegraphics[width=0.7\linewidth]{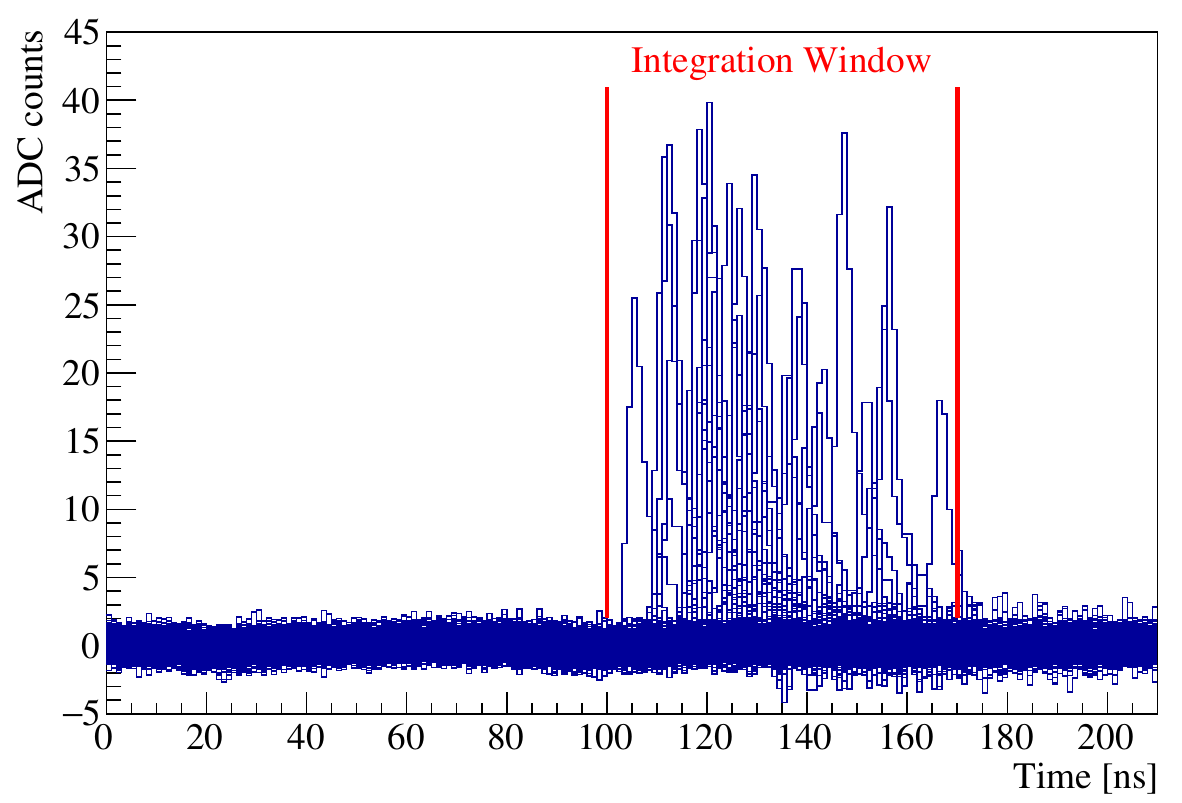}
  \caption{Two hundred consecutive waveforms from the PMT overlapped with each other measured without the presence of the crystal.}
  \label{f:psc}
\end{figure}

Fig.~\ref{f:ps} shows two hundred consecutive waveforms from the PMT randomly selected from a single-PE response measurement. The integration window marked in the figure coincided with the LED lit window. Some single-PE pulses could be seen outside of the window. They were thought to be due to scintillation of random low energy radiation in the crystal.

To verify this assumption, the same measurement was repeated without the presence of the crystal. The resulting waveforms are shown in Fig.~\ref{f:psc}, where no pulse outside of the integration window can be seen. 

\begin{figure}[ht!] \centering
  \includegraphics[width=0.8\linewidth]{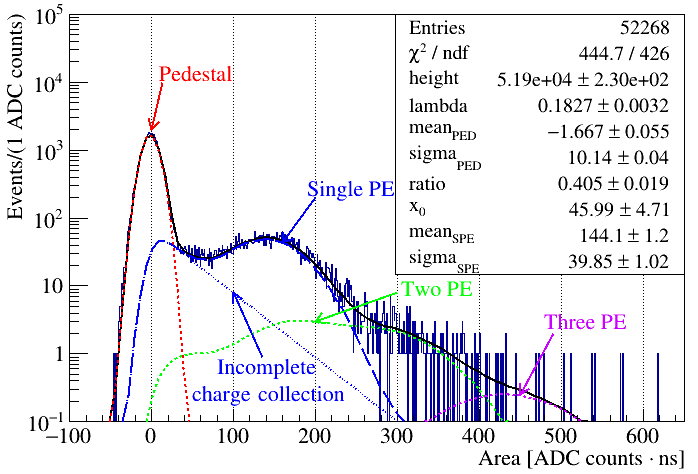}
  \caption{Single PE response of the PMT in logarithm scale.}
  \label{f:SPE}
\end{figure}

An integration in this time window was performed for each waveform in the data file whether it contained a pulse or not. The resulting single-PE spectrum is shown in Fig.~\ref{f:SPE}. The location of the single-PE peak varied within 5\% in different measurements. In the energy calibration measurement to be mentioned in following section, the single-PE spectrum with the crystal was used but with a 5\% uncertainty attached to be conservative.

The spectrum was fitted in the same way as described by equations~\ref{e:Fx}, \ref{e:fnx} and \ref{e:f1x}. The fitting function has eight free parameters as shown in the top-right statistic box in Fig.~\ref{f:SPE}, where ``height'' corresponds to $H$ in Eq.~\ref{e:Fx}, ``lambda'' corresponds to $\lambda$ in Eq.~\ref{e:Fx}, ``mean'' and ``sigma'' with a subscript ``PED'' represents the mean and the sigma of the Gaussian pedestal noise distribution, those with a subscript ``SPE'' represents $\bar{x}$ and $\sigma$ in Eq.~\ref{e:f1x}, respectively, and ``ratio'' corresponds to $R$ in Eq.~\ref{e:f1x}.  Due to technical difficulties in realizing multiple function convolutions in the fitting ROOT script, the three-PE distribution, $f_1^{3\ast}(x)$, was approximated by a Gaussian function with its mean and variance three times that of the single-PE response.

Table~\ref{t:1PE} lists the Gaussian means of single-PE distributions measured before and after the energy calibration to be mentioned in the next section to check the stability of the PMT gain. The average mean for the PMT at 1,600~V is $14.5 \pm 0.1$~ADC counts$\cdot$ns after being divided by the amplification factor, 10.

\begin{table}[ht!] \centering
  \caption{\label{t:SPE} Summary of single-PE response measurements before and after the energy calibration to be mentioned in the next section.}
  \begin{tabular}{cccc}
    \toprule
     &Temperature &Temperature &Mean$_\text{SPE}$ \\
    &of PMT [$^\circ$C]&of crystal [$^\circ$C] &[ADC counts$\cdot$ns]\\
    \midrule
     Before & -193.8 $\pm$ 1.1 & -195.7 $\pm$ 1.1 & 14.58  $\pm$ 0.12 \\
     After  & -192.8 $\pm$ 1.1 & -193.7 $\pm$ 1.1 & 14.41  $\pm$ 0.12 \\
    \bottomrule
  \end{tabular}
\end{table}

\subsubsection{Energy calibration}
\hspace{0.5cm} The energy calibration was performed using $X$ and $\gamma$-rays from an $^{241}$Am radioactive source~\cite{campbell86, toi}. The source was separated from the crystal by a PTFE sheet in between as shown in Fig.~\ref{f:Bsetup} so that $\alpha$ particles from the source could be blocked. The digitizer was triggered when the height of a pulse from the PMT was more than 5 ADC counts. As can be seen in Fig.~\ref{f:ps} and \ref{f:psc}, the typical height of a single PE pulse was around 20 ADC counts, and the baseline fluctuation was mostly within $\pm3$ ADC counts. The trigger threshold of 5 ADC counts could hence suppress most of the electronic noise spikes while let pass most of the single PE pulses. The trigger rate was $\sim 6.3$~kHz when the threshold was set to this value.

\begin{figure}[ht!]\centering
  \includegraphics[width=0.8\linewidth]{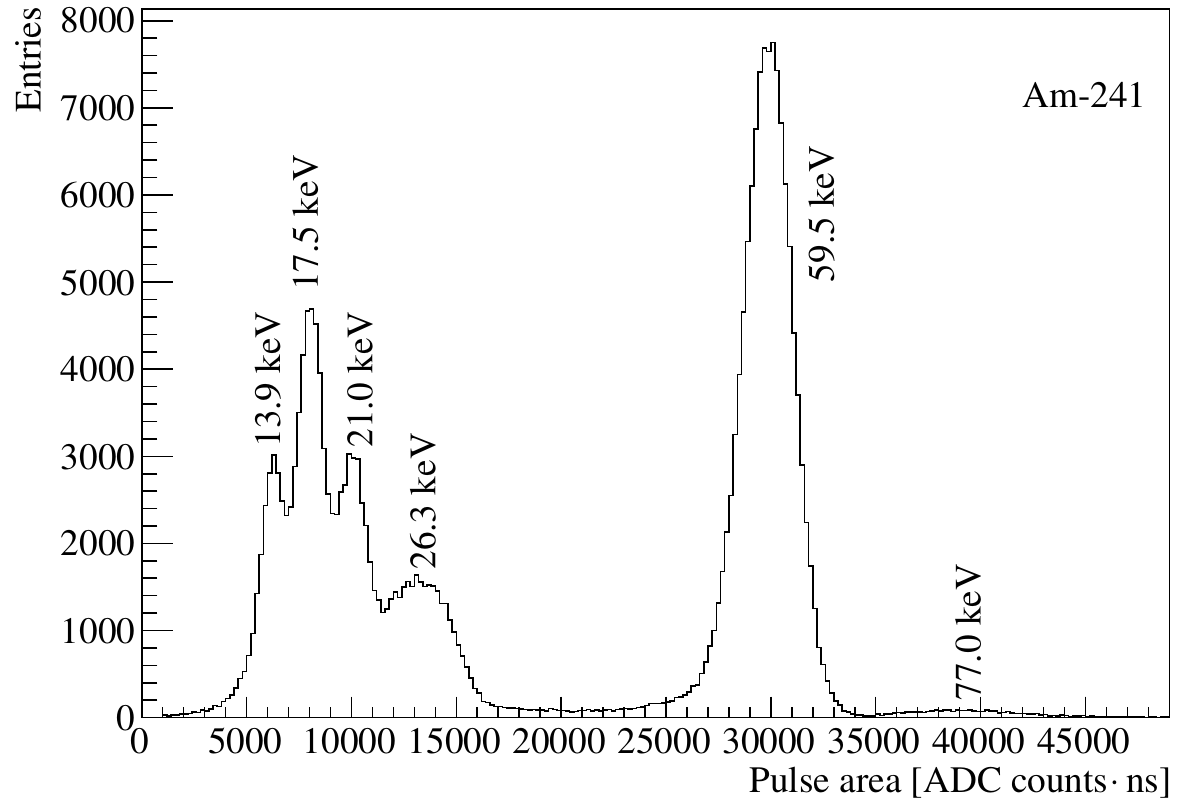}
  \caption{Energy spectrum of $^{241}$Am in the unit of ADC counts$\cdot$ns.}\label{f:SPEC}
\end{figure}

Each recorded waveform was 8008~ns long with a sampling rate of 1~GHz. About 1600~ns pre-traces were preserved before the rising edge of a pulse that triggered the digitizer so that there were enough samples before the pulse to calculate the averaged pedestal value of the waveform. After the pedestal was adjusted to zero the pulse was integrated until its tail fell back to zero. The integration had a unit of ADC counts$\cdot$ns.  The recorded energy spectrum in this unit is shown in Fig.~\ref{f:SPEC}.

The energy and origin of each peak were identified and summarized in Table~\ref{t:rPE}, based on Ref.~\cite{campbell86} and the \emph{Table of Radioactive Isotopes}~\cite{toi}.  The energy resolution achieved here is very similar to that of a typical NaI(Tl) detector at room temperature~\cite{knoll}.

\begin{table}[ht!]
  \caption{\label{t:RPE} Fitting results of $^{241}$Am peaks in the energy spectrum. The 13.9, 17.5 and 21.0~keV ones are obtained from the fitting shown in Fig.~\ref{f:fit1}. The 26.3~keV ones are obtained from the fitting shown in Fig.~\ref{f:fit2}. The 59.5~keV ones are obtained from the fitting shown in Fig.~\ref{f:fit}. The 77~keV ones are obtained from the fitting shown in Fig.~\ref{f:fit3}.}
  \begin{minipage}{\linewidth}\centering
  \begin{tabular}{cccccc}
    \toprule
    Type of & Energy & Mean & Sigma & FWHM \\
   radiation & (keVee) & (ADC$\cdot$ns) & (ADC$\cdot$ns) & (\%) \\
    \midrule
     \begin{tabular}{r} $X$-ray \\ $X$-ray \\ $X$-ray \\ $\gamma$-ray \\ $\gamma$-ray \\ Sum$^\ddagger$ \\
     \end{tabular} &
     \begin{tabular}{l} 13.9$^\dagger$ \\ 17.5$^\dagger$ \\ 21.0$^\dagger$ \\ 26.3$^\dagger$ \\ 59.5 \\ 77.0 \\
     \end{tabular} &
     \begin{tabular}{r} 6303.6 \\ 8045.6 \\ 10076.0 \\ 13202.8 \\29817.6 \\39292.9 \\
     \end{tabular} &
     \begin{tabular}{r} 639.6 \\ 571.5 \\ 815.8 \\ 1598.5 \\1206.8 \\ 2674.9 \\
     \end{tabular} &
     \begin{tabular}{r} 23.9 \\ 16.7 \\ 19.1 \\ 28.5 \\ 9.5 \\ 15.9 \\
     \end{tabular} \\
     \bottomrule
     \end{tabular}
  \end{minipage}
  $^\dagger$ Intensity averaged mean of $X$ or $\gamma$-rays near each other~\cite{campbell86, toi}.\\
  $^\ddagger$ Sum of $X$-rays and 59.5~keV $\gamma$-ray.
\end{table}

\begin{figure}[ht!]\centering
    \includegraphics[width=0.8\linewidth]{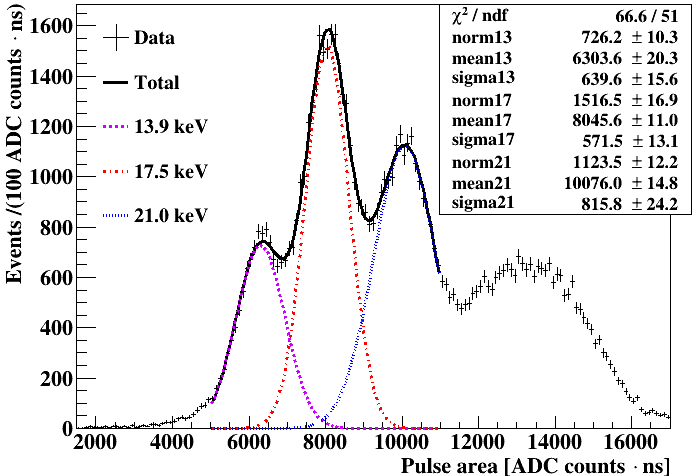}
    \caption{Simultaneous fitting of the 13.9, 17.5 and 21.0~keV $X$-ray peaks with three Gaussian functions.}
    \label{f:fit1}
\end{figure}

\begin{figure}[ht!]\centering
    \includegraphics[width=0.8\linewidth]{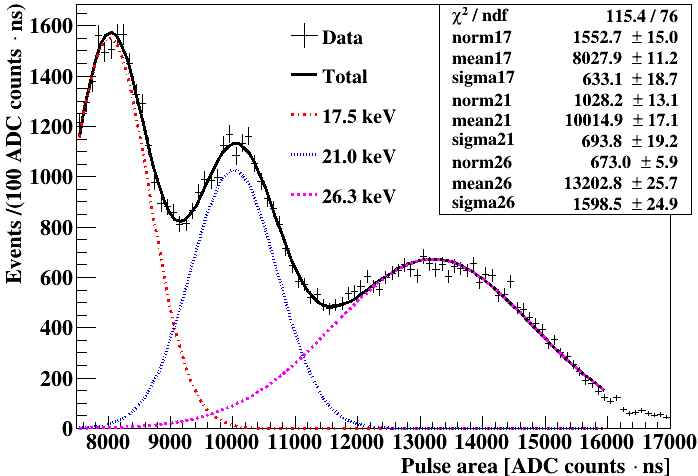}
    \caption{Simultaneous fitting of the 17.5, 21.0~keV $X$-ray and 26.3~keV $\gamma$-ray peaks with three Gaussian functions.}
    \label{f:fit2}
\end{figure}

\begin{figure}[ht!]\centering
    \includegraphics[width=0.8\linewidth]{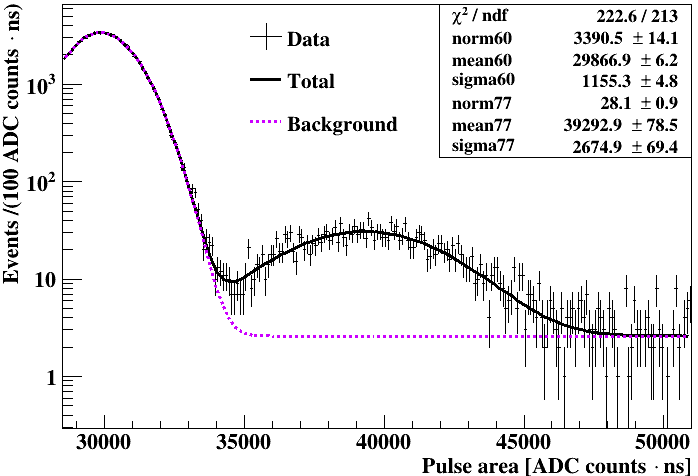}
    \caption{Gaussian fitting of the 77~keV peak on top of a flat background and the tail of the 59.5~keV peak.}
    \label{f:fit3}
\end{figure}

\begin{figure}[ht!]\centering
    \includegraphics[width=0.8\linewidth]{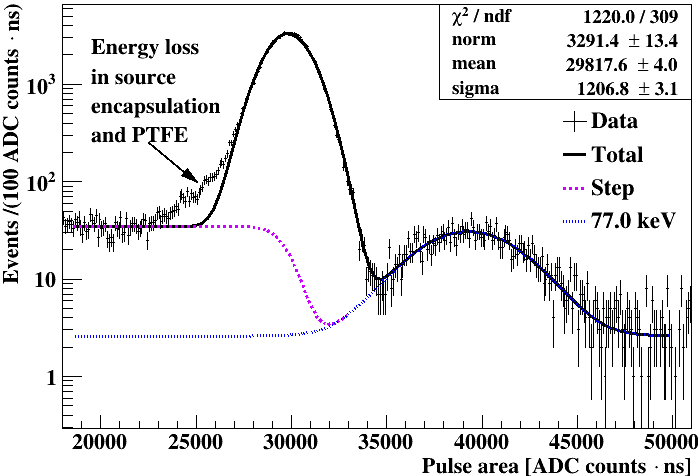}
    \caption{Gaussian fitting of the 59.5~keV $\gamma$-ray peak on top of a flap background, the tail of the 77~keV peak and a smeared step function.}
    \label{f:fit}
\end{figure}

Peaks in Fig.~\ref{f:SPEC} were fitted with combinations of simple functions as shown in Fig.~\ref{f:fit1}, \ref{f:fit2}, \ref{f:fit3} and \ref{f:fit} to extract their mean values and widths. The $X$-ray peaks at 13.9, 17.5 and 21.0~keV were fitted with three Gaussian distributions simultaneously (Fig.~\ref{f:fit1}), so were the 17.5, 21.0 and 26.3~keV peaks (Fig.~\ref{f:fit2}).  The 59.5 and 77.0~keV peaks were fitted with two Gaussian distributions on top of a horizontal line, the height of which was determined by the high energy side band of the 77.0~keV peak before the fitting (Fig.~\ref{f:fit3} and \ref{f:fit}). Part of the low energy side of the 59.5~keV peak was excluded from the fitting since it cannot be described by a pure Gaussian distribution (Fig.~\ref{f:fit3}). A Geant4-based Monte Carlo simulation~\cite{gears} revealed the origin of the tail on the low energy side of the 59.5~keV peak to be $\gamma$-rays that lost part of their energies in the source encapsulation and the PTFE plate in between the source and the crystal.

As shown in Fig.~\ref{f:fit}, a different fitting method was tried for the 59.5~keV peak to verify the mean determined by the partial Gaussian fitting (Fig.~\ref{f:fit3}). Parameters of the function are used to describe the 77.0 keV peak, and its side bands were obtained from the third fitting and fixed in this fitting. The left side of the 59.5~keV peak was partially described by a step function associated with the Gaussian function~\cite{sag} used to fit the 59.5~keV peak:

\begin{equation}\label{e:sag}
    N_0 \mathrm{erfc}\left(\frac{x-\bar{x}}{\sigma}\right) + N_1 \exp\left(\frac{(x-\bar{x})^2}{2\sigma^2}\right)
  \end{equation}  
where the height of the step, $N_0$, was determined by the left side band of the 59.5~keV peak and fixed in the fitting. The normalization factor, $N_1$, the mean, $\bar{x}$ and the width, $\sigma$, of the Gaussian function were determined by the fitting. The difference of the means determined in these two methods is less than 0.2\%, the difference of the widths is less than 5\%. The parameters obtained from the last fitting method were listed in Table.~\ref{t:rPE} and used for the light yield analysis.
  
\subsubsection{Light yield}
\begin{figure}[ht!] \centering
  \includegraphics[width=0.8\linewidth]{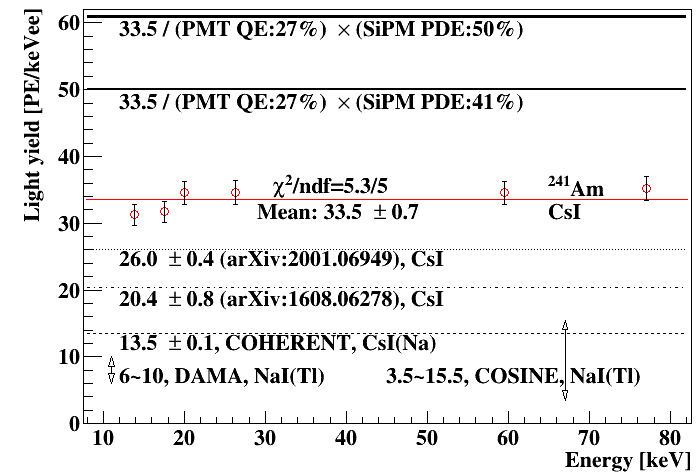}
  \caption{Currently and previously achieved~\cite{csi, csi20} light yields of undoped CsI at $\sim 77$~K together with the predicted ones with SiPM as light sensors. Those of the COHERENT CsI(Na)~\cite{coherent17}, DAMA/LIBRA NaI(Tl)~\cite{dama18} and COSINE NaI(Tl) detectors~\cite{cosine19} are plotted as well for comparison.}
  \label{f:ly2}
\end{figure}

\hspace{0.5cm} The integration area mean from Table~\ref{t:RPE} were converted to numbers of PE using the formula~\ref{e:nPE} then the light yield was calculated using the equation~\ref{e:LY}. The obtained light yield at each energy point is shown in Fig.~\ref{f:ly2}. The error bars are mainly due to the uncertainty of the mean value of the single-PE response used to convert the \textit{x}-axes of the energy spectra from ADC counts$\cdot$ns to the number of PE. The data points were fitted with a straight line to get an average light yield, which is 33.5 $\pm$ 0.7~PE/keV$_{ee}$.

In an earlier measurement with the same crystal used in this study, a yield of $20.4\pm0.8$ PE/keV$_{ee}$ was obtained in the energy range of [662, 2614]~keV$_{ee}$. One of the purposes of this study was to verify the light yield of this larger crystal at a lower energy range. An even higher yield was achieved in the range of [13.9, 77]~keV$_{ee}$. The non-linearity observed so far does not seem to be a concern for the application of undoped CsI at 77~K in neutrino and dark matter detections. Energies lower to 5.9 keV were explored and discussed in the \nameref{s:lyn}.

\section{The response to electron recoil of undoped CsI measured with SiPMs}
\label{s:esipm}
\hspace{0.5cm} This chapter describes the measurement of the light yield from a small undoped CsI crystal directly coupled with two SensL SiPMs~\cite{sipmJ} at about 77~Kelvin. The measured light yield, using X and $\gamma$-ray peaks from a radioactive $^{241}$Am source in the 18 to 60 keV range, was $43.0 \pm 1.1$ keV$_\text{ee}$. The operation of such a combination at 77~K was the first attempt in the world. With SiPMs coated with TPB, the light yield was further enhanced to $50 \pm 2$ keV$_\text{ee}$.  The chapter discusses some identified drawbacks of using cryogenic SiPMs instead of PMTs, such as inferior energy resolution and optical cross-talks between SiPMs, and explores their implications for rare-event detection, providing potential solutions.
\subsection{Experimental setup}
\label{s:exp}
\begin{figure}[ht!]
  \includegraphics[width=0.9\linewidth]{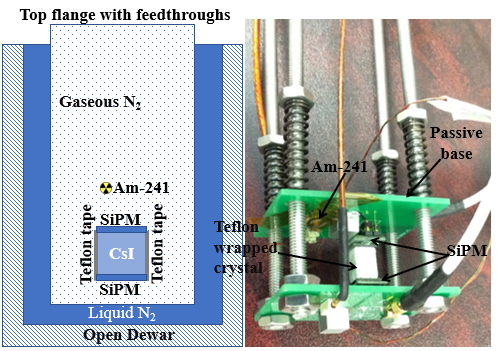}\centering
  \caption{A sketch and a picture of the experimental setup.}
  \label{f:setup}
\end{figure}

\begin{figure}[ht!]
  \includegraphics[width=0.9\linewidth]{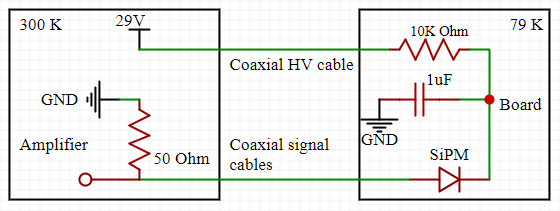}\centering
  \caption{Circuit diagram of the passive base (right) and its wiring to room-temperature devices (left).}
  \label{f:circuit}
\end{figure}

\begin{figure}[ht!]\centering
  \includegraphics[width=0.4\linewidth]{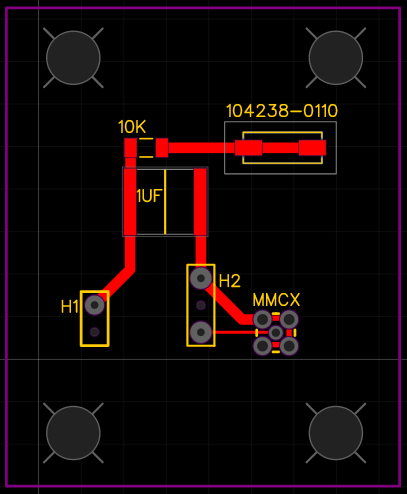}
    \includegraphics[width=0.4\linewidth]{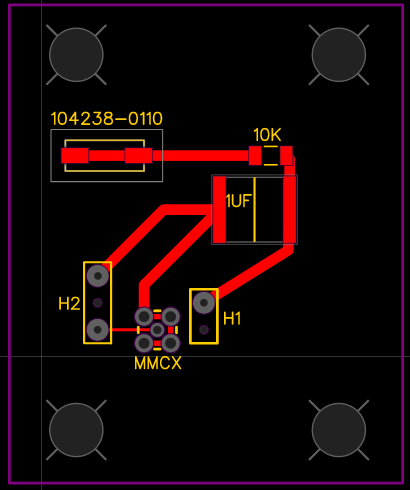}
  \caption{PCB layouts of the top (left) and the bottom (right) passive bases.}
  \label{f:bases}
\end{figure}

\hspace{0.5cm} The experimental setup for the measurement is shown in Fig.~\ref{f:setup}. The undoped cuboid crystal was purchased from AMCRYS~\cite{amcrys}, which is 6~mm in length, width and 10~mm in height. All surfaces were mirror polished. To make sure there is no light leak, side surfaces of the crystal were wrapped with multiple layers of Teflon tape. Two MicroFJ-SMTPA-60035 (sensor size: 6$\times$6~mm$^2$, pixel size: $35 \times 35 \mu$m, total number of pixels: 18980) SiPMs from SensL were used. A bias voltage of 29~V was applied to both sensors and was kept the same throughout the measurement. Breakdown voltages of these sensors were measured by SensL to be around 24.2$\sim$24.7 V at room temperature and were observed to decrease linearly with temperature at least down to -20$^\circ$C. The real over-voltage should have hence been higher than $29-24=5$ V. The two sensors were soldered on two pin adapter boards separately. Each pin adapter board was then inserted into a home-designed PCB, which only contained passive components, hence, was called a passive base. Fig.~\ref{f:circuit} shows the circuit diagram~\cite{diagram} of the base. The 1~$\mu$F capacitor was used to sustain large current during the avalanche (Geiger discharge) of a SiPM. It also forms a noise filter together with the 10~K$\Omega$ quenching resistor.  The current signal was converted to a voltage one on the 50 $\Omega$ internal impedance of the voltage amplifier used in this measurement. PCB layouts of the top and bottom passive bases are shown in Fig.~\ref{f:bases}. To ensure adequate optical contact without optical grease, the PCBs were pushed against two crystal end surfaces by springs. An $^{241}$Am source was attached on the top passive base for energy calibration.

To minimize exposure of the undoped CsI to atmospheric moisture, the assembly was done in a glove bag flushed with dry nitrogen gas. The relative humidity was kept below 5\% at 22$^{\circ}$C during the assembly process. The SiPM-crystal assembly was lowered into a stainless steel chamber from its top opening as shown in the left sketch of Fig.~\ref{f:setup}; the inner diameter of the chamber was $\sim 10$~cm, and the length is 50~cm long. The chamber was vacuum sealed on both ends by two 6-inch CF flanges. The bottom flange was blank and attached to the chamber with a copper gasket in between. The top flange was attached to the chamber with a fluorocarbon CF gasket in between for multiple operations. Vacuum welded to the top flange were five BNC, two SHV, one 19-pin electronic feedthroughs and two 1/4-inch VCR connectors.

After all cables were fixed inside the chamber, the top flange was closed. The chamber was then pumped with a Pfeiffer Vacuum HiCube 80 Eco to $\sim 1.2\times {10}^{-4}$~mbar. Afterward, it was refilled with dry nitrogen gas to $\sim 1.8$ Kgf/cm$^2$ and placed inside an open LN$_2$ dewar. The dewar was then filled with LN$_2$ to cool the chamber and everything inside. After cooling, the chamber pressure was reduced to slightly above the atmospheric pressure.

A few Heraeus C~220 platinum resistance temperature sensors were used to monitor the cooling process. They were attached to the side surface of the crystal, the top passive board, and the top flange to obtain the temperature profile of the long chamber. A Raspberry Pi 2 computer with custom software~\cite{cravis} was used to read out the sensors. The cooling process took about 20 minutes due to the small size of the crystal. Most measurements, however, were taken after about 40 minutes of waiting to let the system reach thermal equilibrium. The temperature of the crystal was -195.7 $\pm$ 0.3 $^\circ$C during measurements, which was almost the same as the LN$_2$ temperature.

The passive boards were powered by a RIGOL DP821A DC power supply~\cite{dp800}. A voltage of 29~V was applied to the SiPMs. According to their manuals, the PDE at this voltage is $\sim 50\%$ for MicroFJ-SMTPA-60035 at 420~nm. Signals were further amplified by a Phillips Scientific Quad Bipolar Amplifier Model 771, which has four channels, each has a gain of ten. Chaining two channels together, a maximum gain of 10$\times$10 can be achieved. A gain of twenty (10 $\times$ 2) was used. Pulses from the amplifier were then fed into a CAEN DT5720 waveform digitizer, which had a 250~MHz sampling rate, a dynamic range of 2~V and a 12-bit resolution. WaveDump~\cite{wavedump}, a free software provided by CAEN, was used for data recording. The recorded binary data files were converted to CERN ROOT files for analysis~\cite{towards}. 

\subsection{Single PE responses}
\begin{figure}[ht!]\centering
  \includegraphics[width=0.7\linewidth]{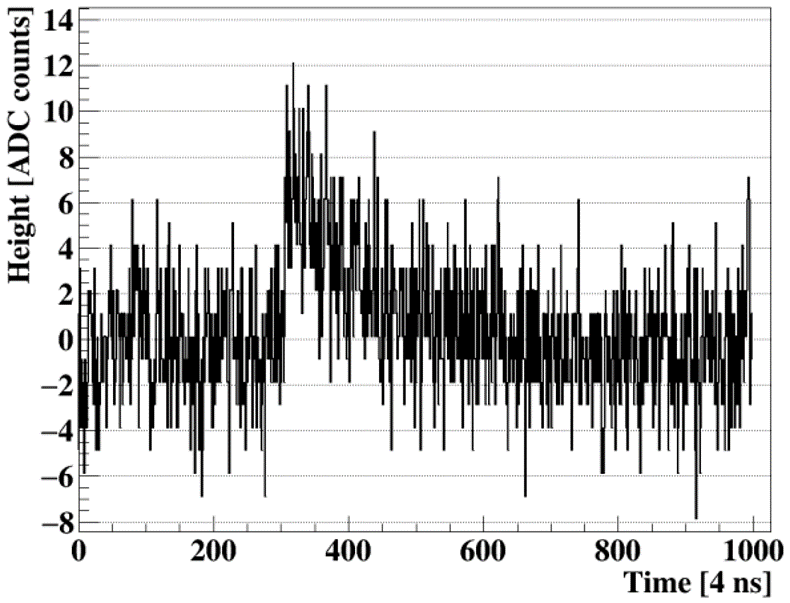}
  \caption{A random single PE waveform from the top SiPM. The ones from the bottom SiPM are very similar.}
  \label{f:single}
\end{figure}

\hspace{0.5cm} Single PE responses of individual channels were investigated using waveform data triggered by dark counts. Some pre-traces were preserved before the rising edge of a pulse that triggered the digitizer to calculate the averaged baseline value of a waveform, which was then subtracted from each sample of the waveform. Fig.~\ref{f:single} shows a typical single PE pulse. The threshold was set by eye to keep a reasonable trigger rate ($\sim 8$~kHz for the top SiPM and $\sim 5$~kHz for the bottom one) while allowing the recording of most single PE pulses.

\begin{figure}[ht!]\centering
  \includegraphics[width=0.7\linewidth]{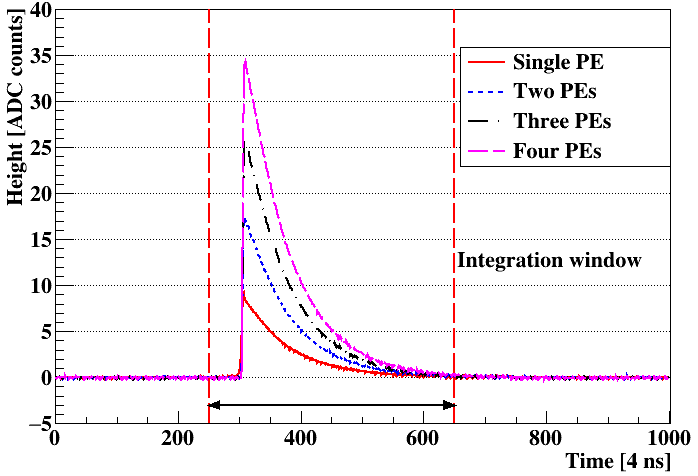}
  \caption{Average waveforms of different PEs from the top SiPM. The ones from the bottom SiPM are very similar.}
  \label{f:Ave}
\end{figure}

Overshooting or undershooting after a pulse may be hidden in a noisy baseline, especially for small PE pulses. A common way to remove the effect of electronic noise is to calculate the average waveform corresponding to the same PEs. For example, the average waveform of single PE was obtained by first adding up all single PE waveforms and then dividing the summed waveform by the total number of single PE events. The same method was used to obtain the average waveforms of higher PEs. They are all shown in Fig.~\ref{f:Ave}. No obvious overshooting or undershooting can be seen; and pulses of different PEs are well contained in the integration window. 

\begin{figure}[ht!]\centering
  \includegraphics[width=0.7\linewidth]{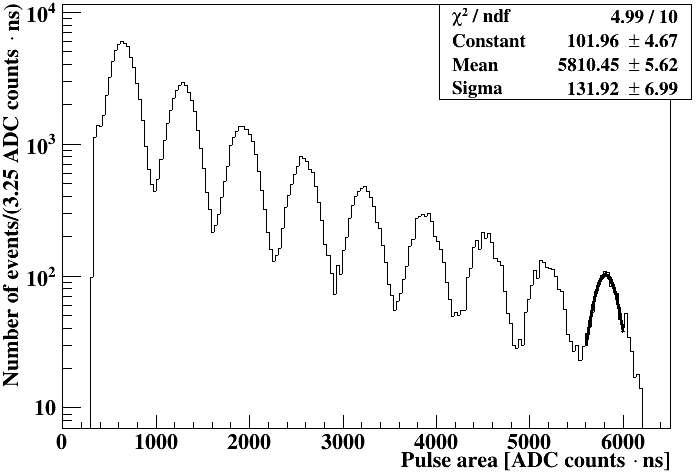}
  \caption{Single PE response of the top SiPM in logarithm scale. The ones from the bottom SiPM are very similar.}
  \label{f:1pe}
\end{figure}

Fig.~\ref{f:1pe} shows the distribution of pulse areas given by the integration, where individual PEs can be seen clearly. If the mean of the first peak is multiplied by $2, 3, 4, ...$ the results roughly match the means of the second, third, fourth, ... peaks. We hence believe that the first peak is the single PE distribution. The ninth peak was fitted using a Gaussian function to obtain its mean value and the result is shown in Fig.~\ref{f:1pe}. The same operations were done for all other peaks. The mean of single PE, mean$_\text{1PE}$, is defined as the Gaussian mean in Fig.~\ref{f:1pe} divided by the number of PEs, $n$. For example, the mean$_\text{1PE}$ for the ninth peak equals to $5810.45 / 9=645.61$ ADC counts$\cdot$ns.

\begin{figure}[ht!]\centering
  \includegraphics[width=0.8\linewidth]{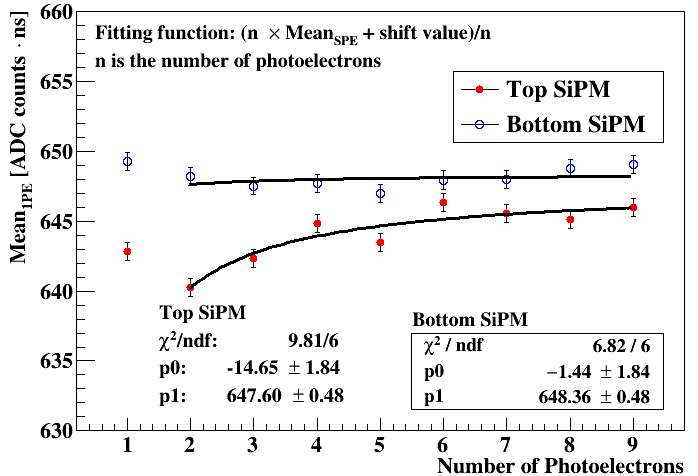}
  \caption{Mean$_\text{1PE}$ distributions obtained from top and bottom SiPMs.}
  \label{f:SPEfitting}
\end{figure}

The mean$_\text{1PE}$ as a function of the number of PEs is shown in Fig.~\ref{f:SPEfitting}. A flat line was expected while a slightly up-going curve was observed from the top SiPM. According to \cite{SiPM_fitting}, this is due to an overall shift of Fig.~\ref{f:1pe} to the left or right. To verify this idea, a function, as shown in Fig.~\ref{f:SPEfitting}, was fitted to the distribution, where, the mean$_\text{SPE}$ (p1) is the true mean of single PE before shifting, and the shift value (p0) is the amount of shift of the whole single PE response.  Note that the first points were not included in the fittings as some of the single PE pulses might not be able to pass the threshold, resulting in a slight distortion of the first peak in Fig.~\ref{f:1pe}.

According to the fitting, the means$_\text{SPE}$ (p1) of the top and bottom SiPMs are $647.60 \pm 0.48$~ADC counts$\cdot$ns and $648.36 \pm 0.48$~ADC counts$\cdot$ns respectively. The shift values (p0) are $-14.65 \pm 1.84$~ADC counts$\cdot$ns and $-1.44 \pm 1.84$~ADC counts$\cdot$ns respectively, which means that the true single PE response of the top SiPM was slightly shifted to the left, resulting in Fig.~\ref{f:SPEfitting}. However, the origin of such a small shift is still unknown as the baseline has been removed prior to the integration. The same phenomenon has been observed in Ref.~\cite{SiPM_fitting}, the cause is also not explained.

\begin{figure}[ht!]\centering
  \includegraphics[width=0.8\linewidth]{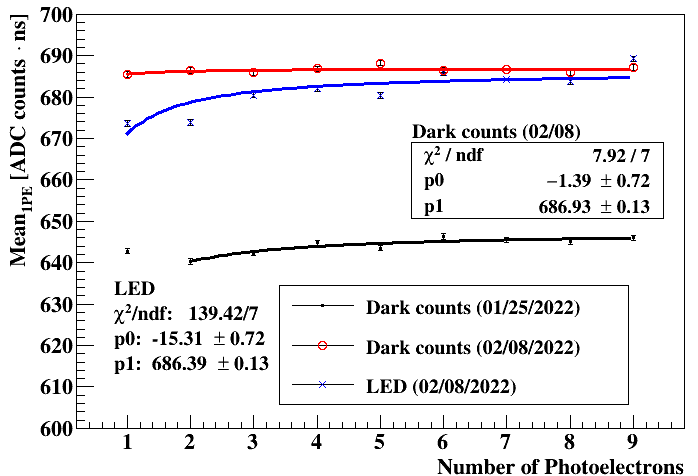}
  \includegraphics[width=0.8\linewidth]{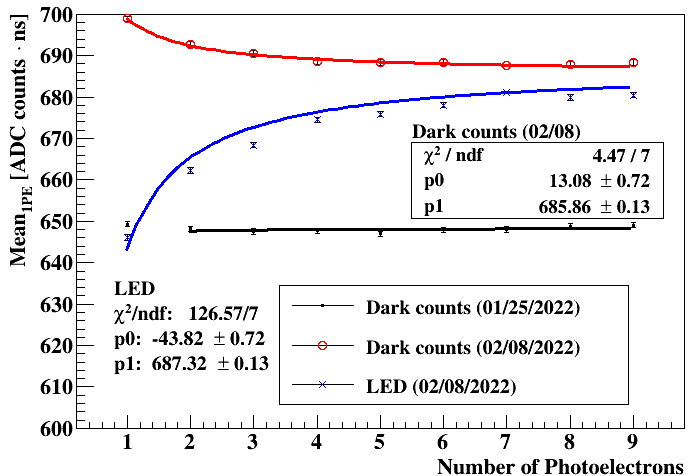}
  \caption{Mean$_\text{1PE}$ distributions of top SiPM (top) and bottom SiPM (bottom) obtained from different methods.}
  \label{f:compare}
\end{figure}

Single PE responses were also measured using an ultraviolet (370~nm) LED from Thorlabs. It was powered by a square pulse that last $\sim$50~ns and was emitted at a rate of 1~kHz from an RIGOL DG1022 arbitrary function generator. The voltage of the pulse was tuned to be around 4.55~V so that only zero or one photon hit the SiPM under study most of the time. Waveforms were recorded whenever a square pulse was generated. They were integrated in a fixed time window. The pulse area was plotted and fitted in a same way as described in the previous paragraphs. The results are compared with the dark-count-based ones and are shown in Fig.~\ref{f:compare}. Utilizing the same fitting process, the means$_\text{SPE}$ of the top and bottom SiPMs were found to be $686.39 \pm 0.13$~ADC counts$\cdot$ns and $687.32 \pm 0.13$~ADC counts$\cdot$ns. 

For fair comparison, another dark-count-based single PE measurement was done right after the LED measurement. The result is also included in Fig.~\ref{f:compare}. The means$_\text{SPE}$ from this measurement were $686.93 \pm 0.13$~ADC counts$\cdot$ns (top SiPM) and $685.86 \pm 0.13$~ADC counts$\cdot$ns (bottom SiPM). The later dark-count-based measurement gave slightly higher values than the earlier one. This indicates a possible gain shift of the SiPMs over time. 

The means$_\text{SPE}$ obtained in the earlier dark-count-based measurements (Fig.~\ref{f:SPEfitting}) were used for the light yield calculation (See Section~\ref{s:ly}), as they were done right after the energy calibration (See Section~\ref{s:ec}). The discrepancy of mean$_\text{SPE}$ in the earlier and later dark-count-based measurements was around 6.1\% for the top SiPM and 6.0\% for the bottom one. They are regarded as the uncertainties of mean$_\text{SPE}$. The results from the LED based measurements lay in between.

\subsection{Energy calibration}
\label{s:ec}
\begin{figure}[ht!]\centering
  \includegraphics[width=0.8\linewidth]{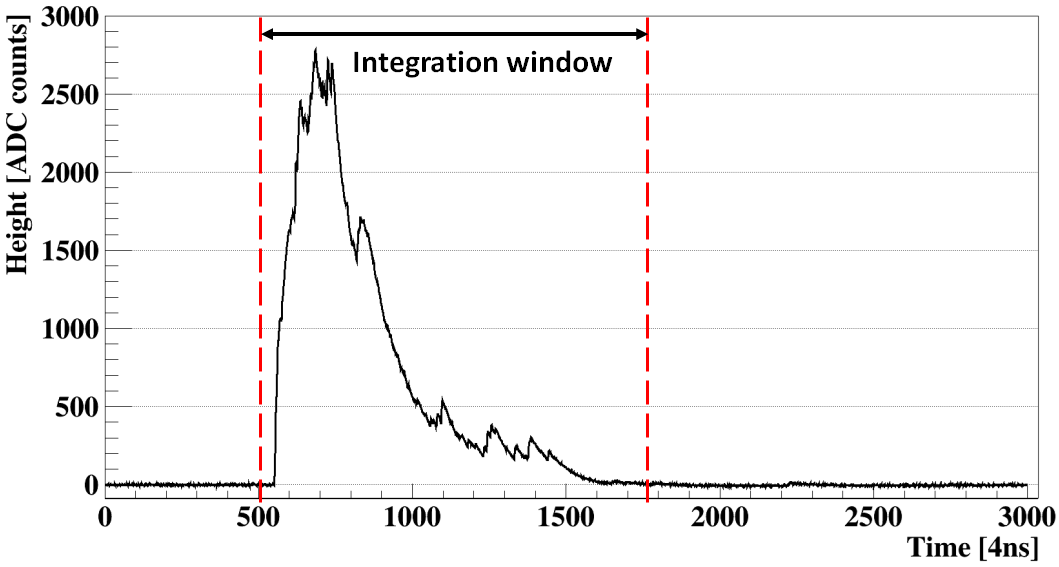}
  \caption{A randomly selected light pulse within the 59.5~keV peak from the top SiPM. The ones from the bottom SiPM are very similar.}
  \label{f:59.5}
\end{figure}

\begin{figure}[ht!]\centering
  \includegraphics[width=0.9\linewidth]{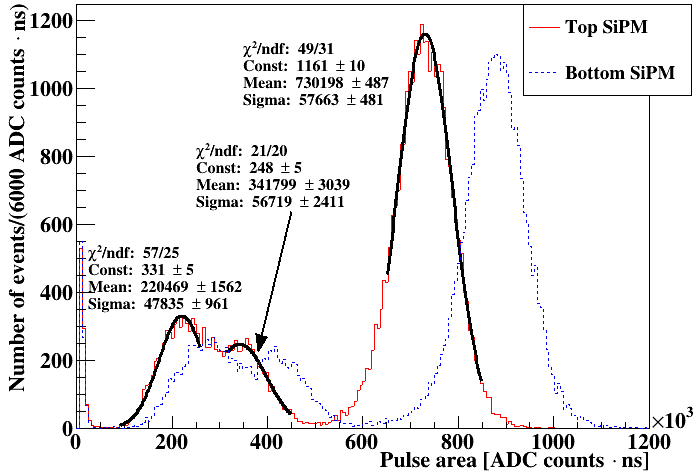}
  \caption{Energy spectrum of $^{241}$Am as the distribution of pulse areas in units of ADC counts$\cdot$ns.}
  \label{f:spec}
\end{figure}

\hspace{0.5cm} The energy calibration was performed using $X$ and $\gamma$-rays from an $^{241}$Am radioactive source~\cite{ding20e, campbell86, toi}. The source was attached to the top passive board as shown in Fig.~\ref{f:setup}. The digitizer was triggered when heights of pulses from both SiPMs were more than 80 ADC counts. Pulses induced by radiation from the source were well above the threshold. The coincident trigger rate was around $\sim 100$~Hz. The integration of a pulse starts 50 samples before the trigger position and ends 10 samples after the position where the pulse goes back to zero. The integration window of a randomly selected light pulse induced by a 59.5~keV $\gamma$-ray is shown in Fig.~\ref{f:59.5}. The integration has a unit of ADC counts$\cdot$ns. The recorded energy spectrum in this unit is shown in Fig.~\ref{f:spec}.

The origin of each peak shown in Fig.~\ref{f:spec} was identified and summarized in Table~\ref{t:rPE}, based on Ref.~\cite{ding20e}, \cite{campbell86} and the \emph{Table of Radioactive Isotopes}~\cite{toi}. The peak around $200 \times 10^3$ ADC counts$\cdot$ns is a combination of multiple $X$-rays ranging from 13.8 to 20.1 keV. The averaged mean of these $X$-rays weighted by their intensities measured in \cite{ding20e} is 17.5~keV.
 
Peaks in Fig.~\ref{f:spec} were fitted with Gaussian functions to extract their mean values and widths. Most of the right side of the 17.5~keV peak and the left side of the 26.3~keV peak were excluded from the fitting as they overlapped with multiple X-ray peaks around 21~keV in between. The 59.5~keV peak was more or less Gaussian. However, its left side was slightly higher than the right side due to the loss of energy in materials between the source and the crystal~\cite{csi20}.

The bottom SiPM received more photons than the top one. This may be due to slightly different optical coupling conditions between the crystal and the two SiPMs. The top SiPM might not be full aligned with the top surface of the crystal. This is something that can be further improved. Another possible cause would be slightly different breakdown voltages of the two SiPMs. This would result in different over-voltages given the same bias, which causes different PDEs.

\begin{table}[ht!]
  \caption{\label{t:rPE} Fitting results of $^{241}$Am peaks in energy spectra for the top (top table) and the bottom (bottom table) SiPMs.}
  \begin{minipage}{\linewidth}\centering
  \begin{tabular}{cccccc}
    \toprule
    Type of & Energy & Mean$_\text{top}$ & Sigma & FWHM \\
   radiation & [keV$_\text{ee}$] & [ADC$\cdot$ns] & [ADC$\cdot$ns] & [\%] \\
    \midrule
     \begin{tabular}{r} $X$-ray\\ $\gamma$-ray \\ $\gamma$-ray \\
     \end{tabular} &
     \begin{tabular}{l} 17.5$^\dagger$ \\ 26.3 \\ 59.5 \\
     \end{tabular} &
     \begin{tabular}{r} 220469 \\ 341799 \\ 730198 \\
     \end{tabular} &
     \begin{tabular}{r} 47835.1 \\ 56719.3 \\ 57663.1
     \end{tabular} &
     \begin{tabular}{r} 51.1 \\ 39.1 \\ 18.6 \\
     \end{tabular} \\
     \bottomrule
     \end{tabular}
  \end{minipage}
  
  \begin{minipage}{\linewidth}\centering
  \begin{tabular}{cccccc}
    \toprule
    Type of & Energy & Mean$_\text{bottom}$ & Sigma & FWHM \\
   radiation & [keV$_\text{ee}$] & [ADC$\cdot$ns] & [ADC$\cdot$ns] & [\%] \\
    \midrule
     \begin{tabular}{r} $X$-ray\\ $\gamma$-ray \\ $\gamma$-ray \\
     \end{tabular} &
     \begin{tabular}{l} 17.5$^\dagger$ \\ 26.3 \\ 59.5 \\
     \end{tabular} &
     \begin{tabular}{r} 268360 \\ 413138 \\ 879406 \\
     \end{tabular} &
     \begin{tabular}{r} 56462.2 \\ 64188.3 \\ 61193.7 \\
     \end{tabular} &
     \begin{tabular}{r} 49.5 \\ 36.6 \\ 16.4 \\
     \end{tabular} \\
     \bottomrule
     \end{tabular}
  \end{minipage}
  $^\dagger$ Intensity averaged mean of $X$-rays near each other~\cite{ding20e}.\\
\end{table}

\begin{figure}[ht!]\centering
  \includegraphics[width=0.7\linewidth]{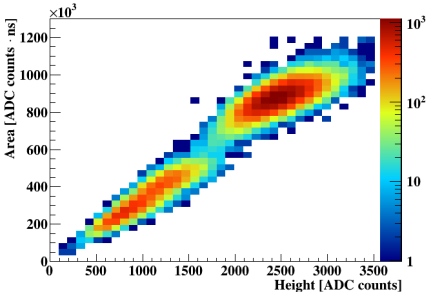}
  \caption{Pulse area versus pulse height for events collected with the $^{241}$Am source from the bottom SiPM.}
  \label{f:in}
\end{figure}

Fig.~\ref{f:in} shows the distribution of pulse areas versus pulse heights. A good linearity of the SiPM up to 59.5 keV was kept and the pulse height was controlled within the digitizer's dynamic range.

\subsection{Light yield}
\label{s:ly}
\hspace{0.5cm} The fitted means of the 17.5 keV, 26.3 keV and 59.5 keV peaks in the $^{241}$Am spectrum in the unit of ADC counts$\cdot$ns were converted to the number of PE using the formula:
\begin{equation}
  \text{(number of PE)} = \frac{\text{(Mean-shift value) [ADC counts} \cdot \text{ns]}}{\text{mean}_\text{SPE}}.
  \label{e:m1pe}
\end{equation}

The shift value is added to account for the overall shift of the energy spectrum observed in the single PE measurement. However, compared to the mean value, it is much smaller; adding it to the equation does not change the final result much.

The light yield was calculated using the data in Table~\ref{t:rPE} and the following equation:
\begin{equation}
  \text{light yield }[\text{PE/keV}_\text{ee}] = \frac{\text{(number of PE)}}{\text{Energy }[\text{keV}_\text{ee}]}.
  \label{e:ly}
\end{equation}

\begin{figure}[ht!]\centering
  \includegraphics[width=0.9\linewidth]{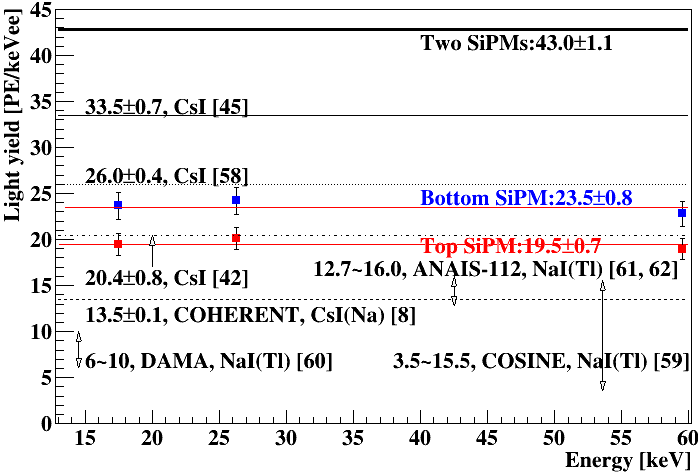}
  \caption{Light yields measured for scintillators made of iodide compounds from various experiments~\cite{cosine19, dama18, coherent17, csi, csi20, ding20e, anais19, anais19a}. All were measured with PMTs except for the one in this work. The arrows only represent ranges of light yields, their locations do not indicate energies where measurements were done.}
  \label{f:LY}
\end{figure}

A flat line was fitted to the light yields obtained from the three peaks recorded in each SiPM, as shown in Fig.~\ref{f:LY}. The uncertainties of the light yield measurements are mostly determined by the uncertainties of the mean$_\text{SPE}$. The light yield observed by the top SiPM is 19.5 $\pm$ 0.7 PE$/$keV$_\text{ee}$, and by the bottom SiPM is 23.5 $\pm$ 0.8 PE$/$keV$_\text{ee}$. The total yield is hence 43.0 $\pm$ 1.1 PE$/$keV$_\text{ee}$. This and other results from related studies are compared in Fig.~\ref{f:LY}. Since sizes of crystals in these measurements are all different, this may not be a fair comparison if the self-absorption of optical photons in these crystals were significant. However, based on our limited experience, the influence of self-absorption is not as significant as those of optical surface conditions and light collection efficiencies of light sensors, as we observed larger light yields in larger crystals coupled to PMTs with higher quantum efficiencies and better wrapping of crystals~\cite{csi,csi20,ding20e}.

\subsection{Wavelength shifter coating}
\hspace{0.5cm} In the subsequent phase of our experiment, we undertook the process of coating SiPMs with TPB to shift the 340nm scintillation light emitted by CsI to around 420nm, aligning with the wavelength where a typical SiPM's PDE reaches its maximum. The overall experimental setup and analysis procedures remained consistent, with the key modification being the application of coatings to the SiPMs. Two distinct methods of coating were explored and tested.

The first approach involved the straightforward addition of TPB to toluene (solvent) to create a solution. In the second method, TPB and polymer were introduced to toluene. To ensure a homogeneous coating, TPB was combined with a portion of polymer in toluene. The resulting solutions were carefully drop-cast onto the SiPM surfaces. Subsequently, the solution was left to stand for three hours, allowing the toluene to evaporate in a controlled environment, facilitated by a chemical fume hood. This process resulted in the formation of a transparent TPB-impregnated plastic layer. As the toluene dissipated, the polymer transformed into a thin, jelly-like layer, ensuring uniform distribution of TPB within it. 

\begin{figure}[ht!]\centering
    \includegraphics[width=\linewidth]{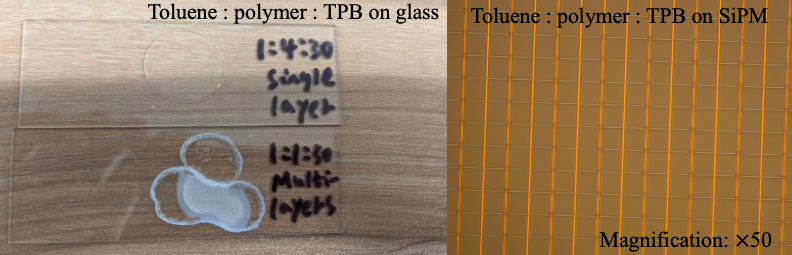}
    \caption{Left: Application of Toluene:polymer:TPB mixed solutions on glass. Right: A SensL SiPM coated with TPB in polymer under a microscope (Magnification factor: 50).}\label{f:pol}
  \end{figure}

\begin{figure}[ht!]\centering
    \includegraphics[width=\linewidth]{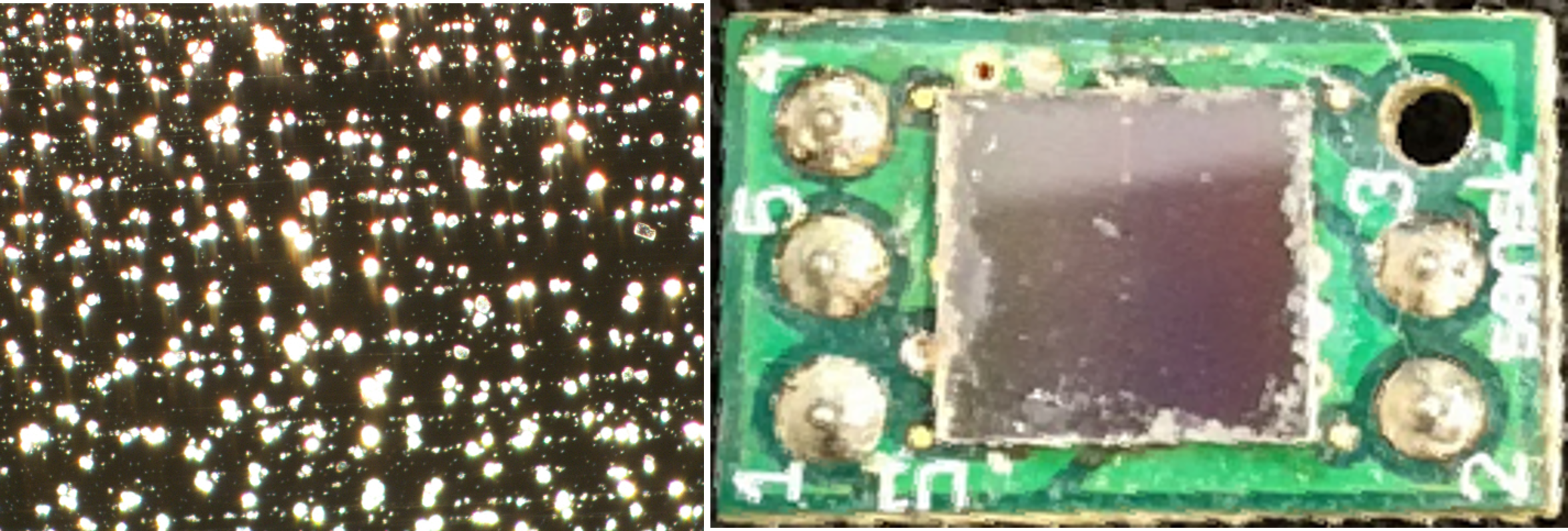}
    \caption{Left: Application of Toluene:TPB = 40:1 mixed solutions on a SensL SiPM's surface under a microscope (Magnification factor: 50). Right: The post-coated SiPM.}\label{f:tpb}
  \end{figure}

The coated SiPMs underwent meticulous examination under a microscope, and visual representations are presented in Fig. \ref{f:pol} and Fig. \ref{f:tpb}. Various mass ratios of TPB to toluene were tested, including 1:20, 1:30, 1:40, and 1:60. The 1:40 ratio yielded the highest light yield for SiPMs. However, the addition of polymer resulted in a decrease in light yield. Ultimately, through this coating process, the light yield was elevated to $50\pm2$~PE/keVee under the 1:40 ratio coating without the addition of polymer.

\subsection{SiPM challenges and solutions}
\label{s:sipm}
\subsubsection{Energy resolution}
\hspace{0.5cm} As seen in Fig.~\ref{f:LY}, the light yield obtained using SiPMs is higher than those obtained using PMTs. Assuming pure Poisson statistics, the energy resolution using SiPMs should be better than that using PMTs. However, this is not the case. The FWHM of the averaged 17.5~keV peak is $\sim$ 50\%, as shown in the last column of Table~\ref{t:rPE}. It may be explained by the merging of the 13.9, 17.8 and 21.0~keV peaks~\cite{ding20e}. The FWHM of the 26.3~keV and 59.5~keV peaks are $\sim$ 38\% and $\sim$ 18\%, respectively, while the PMT gives 28.5\% and 9.5\%, correspondingly~\cite{ding20e}. This discrepancy implies other contributions to the energy resolution in the SiPM setup, which have yet to be investigated.

The slightly worse energy resolution makes it difficult to resolve $X$-ray peak close to each other. However, it might not be a concern for low energy dark matter detection as the nuclear recoil spectrum has a shape close to an exponential decay near the threshold. The broadening of such a distribution does not necessarily reduce the number of observed events.

\subsubsection{Optical cross-talks between SiPMs}
\hspace{0.5cm} As shown in Fig.~\ref{f:spec}, there is a small increase of event rate close to the threshold ($\sim15000$ ADC counts$\cdot$ns) from both SiPMs. However, there are no $X$-ray peaks around that region from the $^{241}$Am source. Certain instrumental noise might be the cause of this small bump, for example, optical cross-talks. However, for a cross-talk event to pass the two-SiPM coincident trigger, optical photons coming out of one SiPM must hit the other. Such a phenomenon is categorized as external cross-talk in Ref.~\cite{optical_crosstalks}.
\begin{figure}[ht!]\centering
  \includegraphics[width=0.36\linewidth]{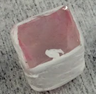}
    \includegraphics[width=0.57\linewidth]{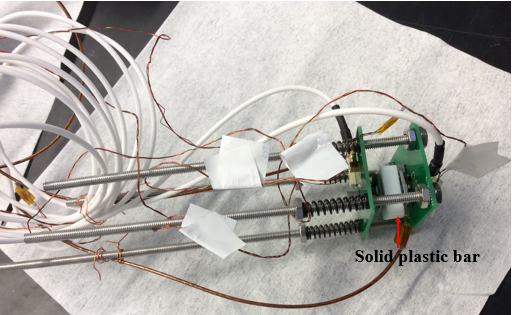}
  \caption{Left: transparent plastic cube. Right: opaque solid plastic bar in between two SiPMs.}
  \label{f:talk_set}
\end{figure}

To verify this possibility, several modifications of the experimental setup were done. First, the CsI crystal was replaced by a transparent hollow plastic cube with roughly the same dimensions, the picture of which can be seen in Fig.~\ref{f:talk_set}. Secondly, the radioactive source was removed. And finally, the plastic cube was replaced by an opaque solid plastic bar. Data were taken in coincident trigger mode with those modifications applied one by one. The coincident trigger rates are summarized in Table~\ref{t:optical}. 
\begin{table}[ht!] \centering
  \caption{\label{t:optical} Coincident trigger rates in different setups.}
  \begin{tabular}{cccc}
    \toprule
    &Experimental setups &Trigger rates [Hz]\\
    \midrule
     & CsI+$^{241}$Am & $750 \pm  50$\\
    \midrule
     & Transparent plastic cube+$^{241}$Am & $70 \pm 20$\\
    \midrule
     & Transparent plastic cube & $50 \pm 10$\\
     \midrule
     & Opaque solid plastic bar & $2 \pm 2$\\
    \bottomrule
  \end{tabular}
\end{table}

\begin{figure}[ht!]\centering
  \includegraphics[width=0.8\linewidth]{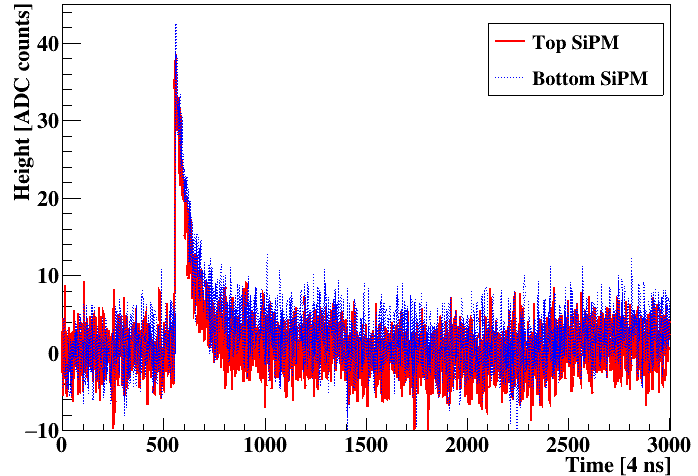}
  \caption{Waveforms from a common event taken with the transparent cube.}
  \label{f:talk}
\end{figure}

As shown clearly in Table~\ref{t:optical}, the trigger rate dropped greatly when the scintillating crystal was removed. This is easy to understand, as most of the coincidentally triggered events were due to scintillation light from the crystal. It is troublesome to see that there were still quite some coincidentally triggered events after the source and the crystal were removed. They must come from the SiPMs. The time window for coincident trigger was set to be 8~ns. If light pulses in different SiPMs are due to random dark noise, their rising edges should appear randomly within 8~ns time window. Fig.~\ref{f:talk} shows waveforms from a common event taken with the transparent cube. Light pulses from two different SiPMs went across the threshold at exact the same time, indicating that they are highly correlated. The trigger rate dropped to nearly zero when the transparent cube was substituted by the opaque solid plastic bar. All these confirm that the events close to the threshold are indeed due to external optical cross-talks between the two SiPMs. 

\begin{figure}[ht!]\centering
  \includegraphics[width=0.8\linewidth]{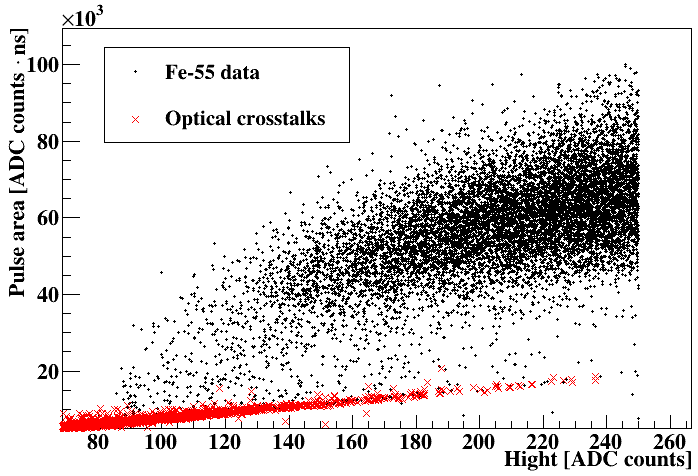}
  \caption{Area versus height of pulses in $^{55}$Fe (black dots) and optical cross-talks (red crosses) from the top SiPM.}
  \label{f:phy}
\end{figure}

One of the motivations to replace PMTs with SiPMs is to eliminate the Cherenkov radiation that would coincidentally trigger multiple PMTs. However, external optical cross-talks may coincidentally trigger multiple SiPMs as well. Some good methods to distinguish optical cross-talk events from physical ones are needed to justify the proposed replacement. A dedicated data set was taken with nothing in between two SiPMs to obtain optical cross-talk events. Another data set was taken with the CsI crystal and a $^{55}$Fe source to obtain low energy physical events. Fig.~\ref{f:phy} shows pulse area versus pulse height of waveforms in these data sets. The red crosses were from optical cross-talks, while the black dots were from $^{55}$Fe. Since pulses due to optical cross-talks are sharp and narrow (see Fig.~\ref{f:talk}), the area-to-height ratio is much smaller than that of physical events. Such a ratio can be a good parameter to remove events due to optical cross-talks.

Another method is to physically reduce the emission of optical photons from an avalanche cell, or to block them from reaching other cells~\cite{optical_crosstalks}. This is an active research area in the fabrication of SiPMs. Hopefully, this will become less a concern over time.

\subsubsection{Dark counts}
\hspace{0.5cm} One major drawback of a SiPM array compared to a PMT is its high DCR at room temperature ($\sim$ hundred kHz). Fortunately, it drops quickly with temperature, and can be as low as 0.2~Hz/mm$^2$ below 77~K~\cite{aalseth17}, while the PDE does not change much over temperature~\cite{oto07, lc08, aki09, jan11}. However, a SiPM array that has an active area similar to a 3-in PMT would still have an about 100~Hz DCR at 77~K. A simple toy MC reveals that a 10-ns coincident window between two such arrays coupled to the same crystal results in a trigger rate of about $10^{-5}$~Hz. This was also observed in this work that coincident triggers could dramatically reduce the trigger rate. If such a detector is placed near an accelerator-based neutrino source to detect neutrinos, a further time coincidence with beam pulses can be further required to make the rate negligible.

\subsubsection{Readout electronics of SiPM arrays}
\begin{figure}[ht!]
  \includegraphics[width=0.35\linewidth]{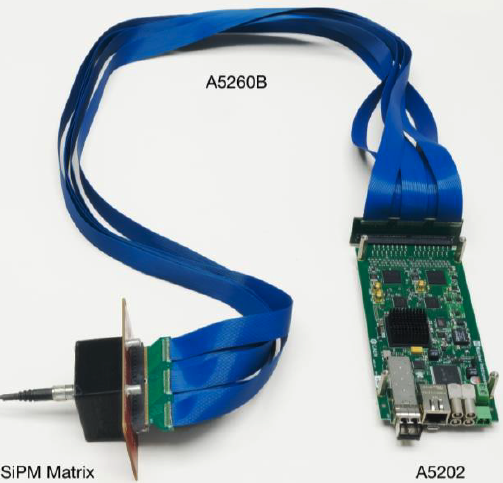}
  \includegraphics[width=0.64\linewidth]{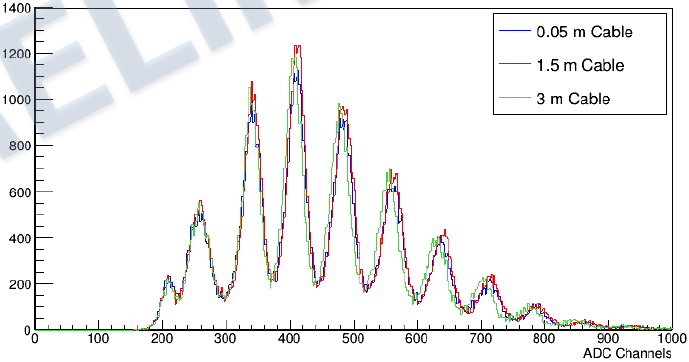}
  \caption{Left: micro coaxial cable A5260B connecting a remote SiPM array with a CAEN A5202 front-end unit~\cite{a5202}. Right: excellent single PE resolutions of the system with various cable lengths~\cite{cable}.}\label{f:readout}
\end{figure}

\begin{figure}[ht!]
  \includegraphics[width=0.95\linewidth]{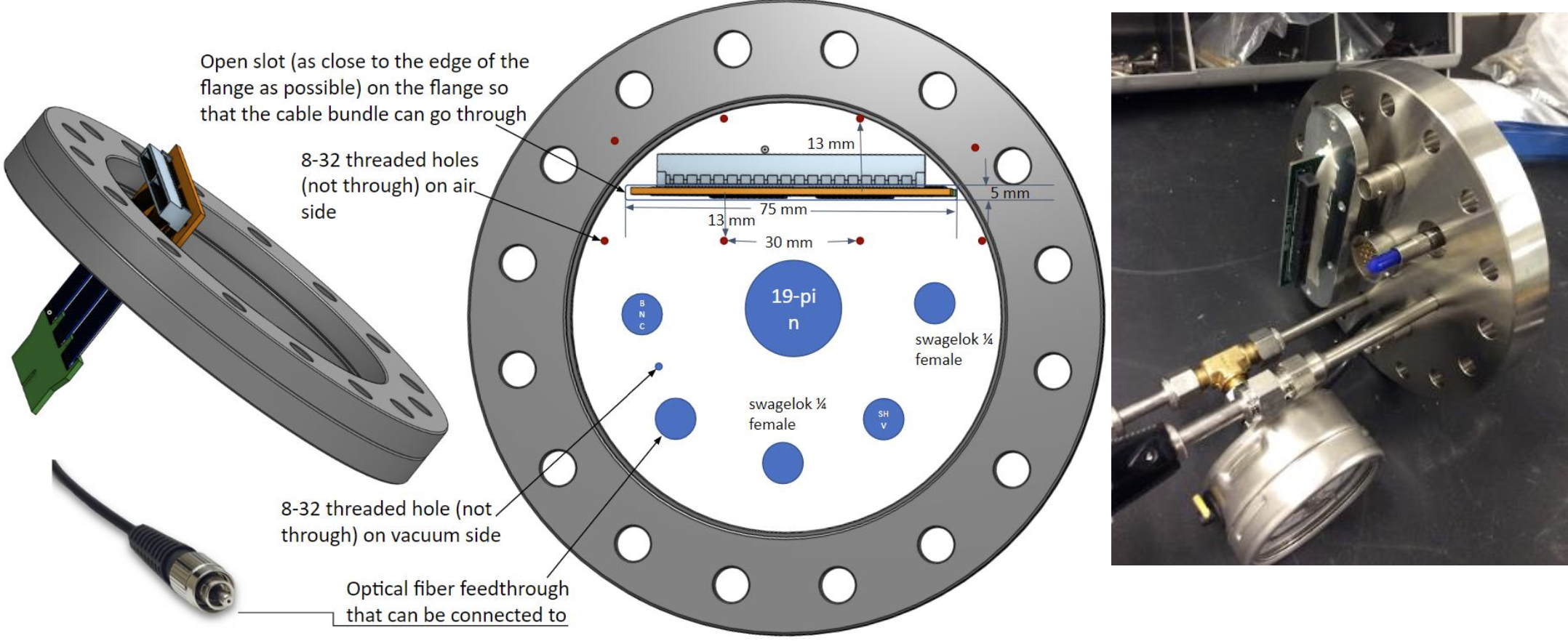}
  \caption{6$''$ CF flange to evaluate cold SiPM array + warm ASIC readout board A5202.}\label{f:flange}
\end{figure}

\hspace{0.5cm} For neutrino or dark matter detection, crystals on the scale of $\sim10$~kg are needed, the surface area of which would be in the order of $10 \times 10$ cm$^2$. To fully cover such a surface, 400 SiPMs are needed assuming each has a surface area of $5 \times 5$ mm$^2$. The electronic readout of them could be a challenge. A natural option is to group the output of a few SiPMs into one channel. However, the total number of SiPMs that can be grouped is limited by the relatively large capacitance and DCR of individual SiPMs. Another possibility is to use CMOS-based ASICS to readout many channels with a single chip. CAEN has developed a front-end system, as depicted in Fig.~\ref{f:readout}, featuring a standalone unit, A5202~\cite{a5202}. This unit houses 2 WEEROC CITIROC chips, each providing a multiplexed output of 32 SiPM channels. Additionally, it includes a flexible micro coaxial cable bundle, A5260B, which connects remote SiPM arrays to the A5202. The system exhibits excellent single PE resolution, even with a cable length of up to 3 m~\cite{cable}. The measurements conducted in this study confirmed that long cables still enable the observation of single PE pulses, as shown in the right plot of Fig.~\ref{f:readout}. Given the compact size of a scintillating crystal-based detector, this cable length is sufficient to bridge cold SiPM arrays and warm ASIC front-ends. Warm electronics are more convenient to maintain and also reduce heating inside the cryostat. A $6''$ CF flange (Fig.~\ref{f:flange}) was designed to integrate the cable bundle and assess the performance of a cold SiPM array + a warm A5202.

\section{The responses of nuclear recoils of undoped CsI}
\label{s:qf}
\hspace{0.5cm} Given that CEvNS events are inherently nuclear recoils, understanding the scintillating detector's light yield for such events is essential. QF measurements play a significant role in interpreting incoming neutrino energies deposited in the cryogenic undoped CsI crystal. Two methods are employed for QF measurements: one uses alpha-particles to simulate the process, while the other involves neutrons. Alpha-particles mimic recoiled ions, whereas neutrons kick out nuclei and create recoiled ions. These recoiled ions, containing deposited energies, further ionize other atoms in the target material. The key distinction lies in the energy scales of the recoiled ions: alpha-particles deposit energies in the MeV scale, while the ions kicked out by neutrons operate in the keV scale. This keV scale is overlapping with the recoiled energies involved in CEvNS interactions.

As depicted in Fig.~\ref{f:aq}, QF measurements for both alpha-particles and neutrons exhibit substantial differences at cryogenic temperatures. Our measurements began with the alpha-particle QF measurements at USD and then progressed to the more challenging neutron QF measurements at TUNL, requiring the design of a new cryostat to minimize neutron scattering on the materials surrounding the crystal~\cite{cryostat}. The cooling capacity was tested and confirmed to reach $\sim$77~K at USD, then transferred to TUNL to conduct the QF measurement. All measurements mentioned in this chapter utilized the same PMT, and the overshoot effect was observed. A solution was developed to correct the PMT's overshoot effect. The following sections will discuss performance of the new cryostat, QF measurements and the overshoot correction in detail.

\vspace{-0.5cm}
\subsection{Performance of a liquid nitrogen cryostat setup for the study of nuclear recoils in undoped CsI crystals}
\label{s:cryostat}
\hspace{0.5cm} The nuclear quenching measurement is normally done by putting the material under study in a neutron beam. The energy of the nuclear recoil can then be calculated from the energy of the incident neutron and the angle of the scattered neutron, assuming the neutron only scatters once within the target. High-$Z$ material should therefore be avoided around the target to minimize the occurrence of neutron multiple scatterings.

A liquid nitrogen cryostat with low-$Z$ material around a small undoped CsI crystal was developed for the nuclear quenching measurement. Its performance was compared to another cryostat previously used for light yield measurements.

A constant light yield was assumed in those sensitivity calculations. It is known, however, that the light yield of undoped CsI has non-negligible variations at different energy ranges, and it varies with crystals under investigation~\cite{LY_moszynski_energy_2005, LY_gridin_channels_2014, LY_kerisit_computer_2009}. It is hence necessary to verify that the light yields we achieved from 13 keV to 2.6 MeV still hold at lower energies with our crystals purchased from Japan and Ukraine. A $^{55}$Fe source and an $^{241}$Am source were used to achieve this utilizing  5.9 keV x-rays and 60 keV $\gamma$-rays.

\subsubsection{Cryostats}
\begin{figure}[ht!]\centering
  \includegraphics[width=\linewidth]{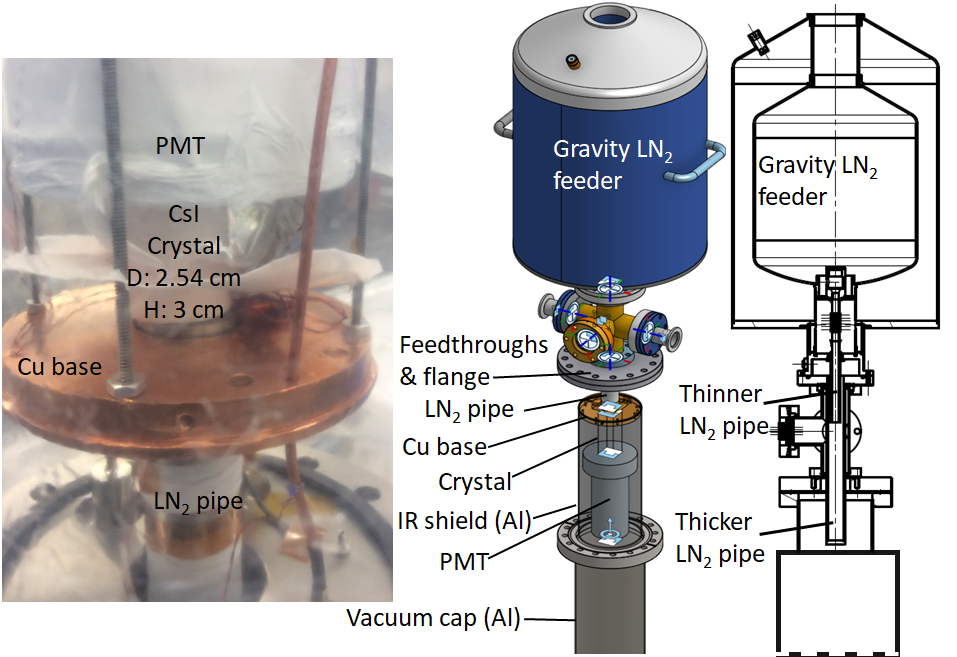}
  \caption{A sketch and photos of the cryostat setup.}
  \label{f:cryostat}
\end{figure}

\paragraph{Design based on gravity-fed liquid nitrogen dewar}\leavevmode\newline
\hspace{0.5cm} Fig.~\ref{f:cryostat} shows the experimental setup for the measurement of the light yield of an undoped CsI crystal at around 77~K. As seen in the right figure, liquid N$_2$ can drip from a dewar into a hollow pipe that is directly in contact with structures to be cooled. In the detector mounting mode, as shown in the left figure, the dewar can be taken away and this new cryostat can be flipped over and placed into a dry glove box for crystal mounting. The middle figure is the CAD drawing of the cryostat. 

The undoped cylindrical crystal was purchased from AMCRYS~\cite{amcrys}, and had a diameter of 2.54~cm and a height of 3~cm. All surfaces were mirror polished. The side surface and an end side surface of the crystal were wrapped with multiple layers of Teflon tape to make sure that there was no light leak. A Hamamatsu 3-inch R11065-ASSY PMT was pushed against the other end surface of the crystal by springs to ensure adequate optical contact without optical grease. The springs are held by the copper base. The wrapped end surface of the crystal was pushed against the cooling finger of the cryostat where liquid nitrogen was fed into the LN$_2$ pipe.

An $^{241}$Am and an $^{55}$Fe source were attached to the inner side surface of the IR shield in the cryostat. The window of the $^{241}$Am source was turned away from the crystal. This way, we could use the source shielding material to attenuate the 59.6~keV $\gamma$-rays. Aluminum foils with a total thickness of 0.048~mm were put in between the $^{55}$Fe source and the crystal to keep the intensity of the 5.6~keV peak roughly the same as the 59.6~keV peak.

To minimize exposure of the crystal to atmospheric moisture, assembly was done in a glove bag flushed with dry nitrogen gas. The relative humidity was kept below 10\% at 22$^{\circ}$C during the assembly process.

The PMT-crystal assembly was capped by an IR shield that was fixed to the copper base by three screws. The cryostat was then sealed by the vacuum cap to a 6-inch CF flange. A fluorocarbon CF gasket was put in between for multiple operations. The inner diameter of the cryostat was $\sim 10$~cm. Vacuum-welded to the flange were two BNC, two SHV, and one 19-pin electronic feedthroughs.

After all cables were fixed beneath it, the top flange was closed. The chamber was then pumped with a Pfeiffer Vacuum HiCube 80 Eco to $\sim 1\times {10}^{-5}$~mbar. The feeder was then filled with LN2 to cool the LN$_2$ pipe, then further cool everything inside. The pump was on all the time until the end of the experiment.

The PMT was powered by a CAEN N1470A high voltage power supply in a NIM crate. The signals were fed into a CAEN DT5751 waveform digitizer, which had a 1~GHz sampling rate, a 1~V dynamic range and a 10 bit resolution. WaveDump~\cite{wavedump}, a free software provided by CAEN, was used for data recording. The recorded binary data files were converted to CERN ROOT files~\cite{root} for analysis by a custom-developed software~\cite{towards}. 

\paragraph{Design based on open liquid nitrogen dewar}\leavevmode\newline
\hspace{0.5cm} We attempted to measure the temperature of the crystal attached to the liquid nitrogen pipe by wrapping temperature sensors around the side surface of the crystal with Teflon or cryogenic tapes. However, it was very hard to maintain good thermal contact between the sensor and the side surface of the crystal as both tapes were too soft. We also tried to use a stainless steel hose clamp to push a sensor tightly against the surface of the liquid nitrogen pipe. We were able to measure 77~K this way. However, a few sensors got crushed after cooling down. To avoid crushing the crystal, we did not fix a sensor on the crystal this way. Another concern about using clamps is that the additional high-$Z$ material may introduce multiple neutron scatterings in a quenching factor measurement. Instead, we compared light yields of undoped crystals operated in this cryostat and an old cryostat used in our previous measurements~\cite{csi20,ding20}.  If the light yields measured in the two are similar, we can infer that the crystals have been operated at similar temperatures.

\begin{figure}[ht!]\centering
  \includegraphics[width=\linewidth]{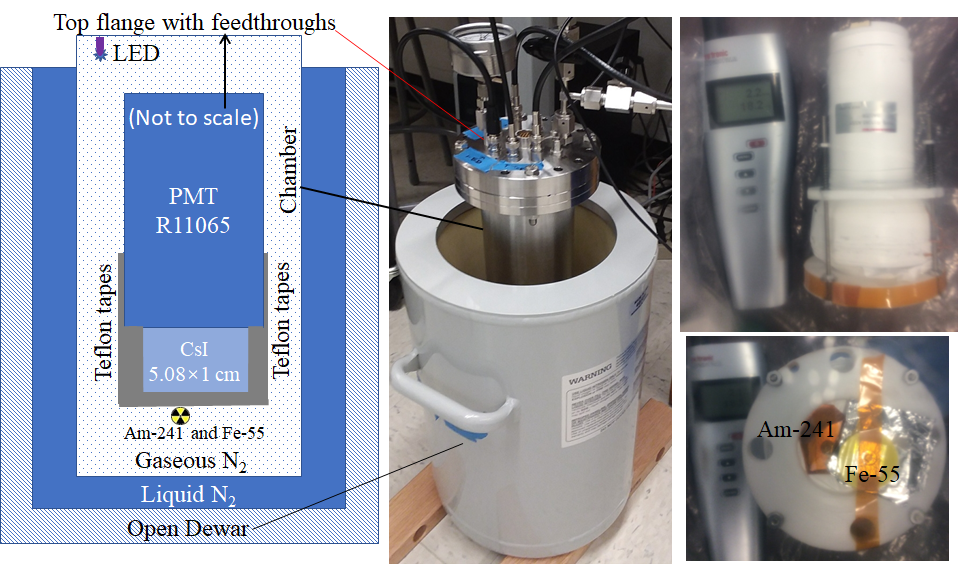}
  \caption{A sketch and photos of the cryostat setup.}
  \label{f:old}
\end{figure}

The structure of the old cryostat is shown in Fig.~\ref{f:old}. Detailed description can be found in Ref.~\cite{ding20e}. An undoped cylindrical crystal with a diameter of 5.08~cm and a height of 1~cm purchased from OKEN~\cite{oken} was used in this cryostat. There is a 1 mm copper foil in between an $^{241}$Am source and the crystal, and a 0.096~mm aluminum foil in between a $^{55}$Fe source and the crystal. The sources are directly attached to the surface of the Teflon tape wrapping. A few Heraeus C 220 platinum resistance temperature sensors were used to monitor the cooling process. They were attached to the side surface of the crystal, the PMT, and the top flange to obtain the temperature profile of the long chamber. The crystal temperature was measured to be $77.3 \pm 1.3$~K in this cryostat~\cite{ding20}.

Note that the two crystals used in the two cryostats are different in shape, which might create a difference in light collection efficiency. However, because they are all very small, paths of individual photons in them (in the order of centimeters) are much shorter than the average absorption length (in the orders of meters) in CsI. The light collection efficiency hence cannot be too different in the two measurements. In fact, a simple Geant4-based optical simulation revealed a 16\% difference in light collection efficiency, with the crystal in the new cryostat observed to have lower collection efficiency. 

The same analysis was made on two setups. Only analysis from the new setup is presented here.

\subsubsection{Single PE response}
\label{s:spe}
\hspace{0.5cm} The SPE response of the PMT was measured using light pulses from an ultraviolet LED, LED370E from Thorlabs. Its output spectrum peaked at 375~nm with a width of 10~nm, which was within the 200 -- 650~nm spectral response range of the PMT. Light pulses with a $\sim$50~ns duration and a rate of 10~kHz were generated using a RIGOL DG1022 arbitrary function generator. The intensity of light pulses was tuned by varying the output voltage of the function generator so that only one or zero photons hit the PMT during the LED lit window most of the time. The ratio of zero-photon-hit to single-photon-hit is around 10. A TTL trigger signal was provided by the function generator simultaneously with each output pulse, then was used to trigger the digitizer to record the PMT response. 

The PMT was biased at 1,600~V, slightly above the recommended operation voltage, 1,500~V, to increase the gain of the PMT. The PMT may experience saturation during exceptionally bright scintillation events; nevertheless, its linearity remains preserved within the sub-MeV range. Single-PE pulses were further amplified by a factor of ten using a Phillips Scientific Quad Bipolar Amplifier Model 771 before being fed into the digitizer in order to separate signals from the pedestal noise.
\begin{figure}[ht!]\centering
  \includegraphics[width=0.7\linewidth]{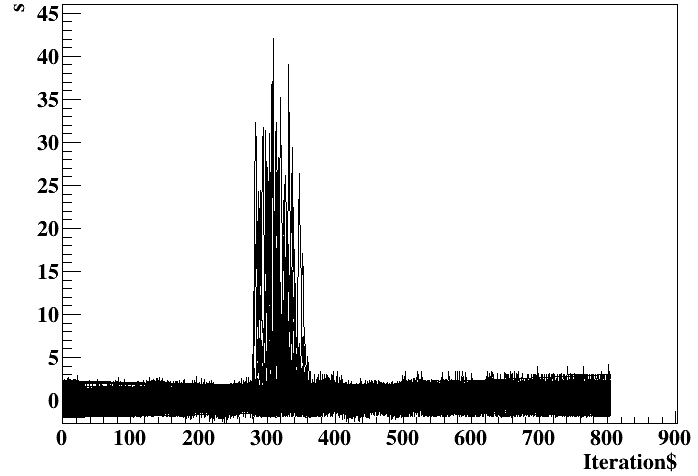}
  \caption{209 selected waveforms from the PMT overlapped with each other.}
  \label{f:OnePE}
\end{figure}

Fig.~\ref{f:OnePE} shows $\sim$210 randomly selected waveforms from the PMT response of the SPE measurement. Some quality criteria were applied to filter out noise events. Firstly, the baseline was calculated in the following steps. The values corresponding to each sample (height) were added to a summed value, then an averaged value (baseline value) was calculated using the summed value divided by the number of samples. The baseline value was then used to shift the waveforms to zero by deducting this baseline value from each corresponding height value. Secondly, the root mean square (RMS) of the baseline was also calculated. After that, averaged baseline values and RMS of the baseline were calculated in a region of 0 to 100 ns, 180 to 280 ns and 420 to 500 ns and all RMS were required to be smaller than 1 ADC count to obtain a relatively stable baseline. The averaged baseline values in the latter two regions were also set to be lower than 1 ADC count to filter out some low-frequency fluctuation events. Another restriction was that the lowest point of the waveform should be greater than $- 2$ ADC counts. Eventually, 209 SPE waveforms were selected applying those restrictions to 1300 SPE events. However, some fluctuations still exist in the pedestal which sits under the SPE waveforms. 

\begin{figure}[ht!]\centering
  \includegraphics[width=0.7\linewidth]{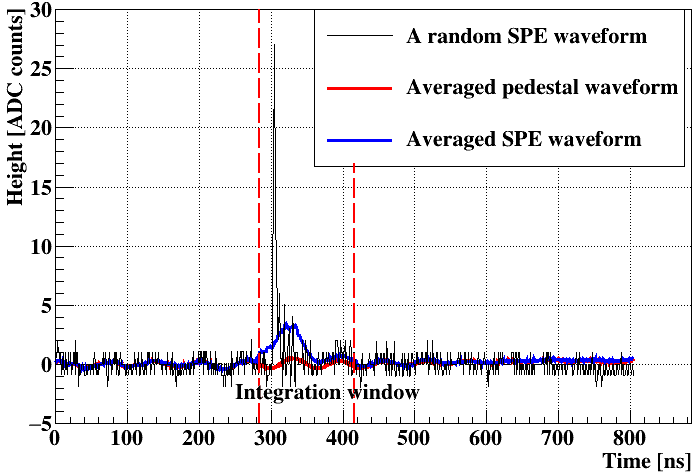}
  \caption{A random SPE waveform, averaged SPE waveform and averaged pedestal waveform.}
  \label{f:aveSPE}
\end{figure}

To examine the influence of the unstable pedestal on the SPE waveforms, the averaged SPE waveforms and the averaged pedestal waveform were studied as shown in Fig.~\ref{f:aveSPE}. The averaged SPE waveform is correlated with the averaged pedestal waveform; the baseline and averaged SPE coincide with each other except in the region where SPE initiates. Therefore, the mean area of the SPE can be obtained by subtracting the area of the averaged pedestal waveform from the area of the averaged SPE waveform. The integration window starts from and ends on both pulses crossing zero as shown in the plot. However, SPE waveforms were seen to have much higher heights as shown by the black pulse, yet the averaged SPE waveform only has a height of $\sim$5 ADC counts as shown by the red pulse. To investigate the discrepancy, we randomly selected an SPE waveform, which was much narrower than the averaged waveform. The wider width of the averaged waveform was due to the fact that the SPE pulses appeared at different locations, ranging from approximately 300 to 360 ns.

\begin{figure}[ht!]\centering
  \includegraphics[width=0.8\linewidth]{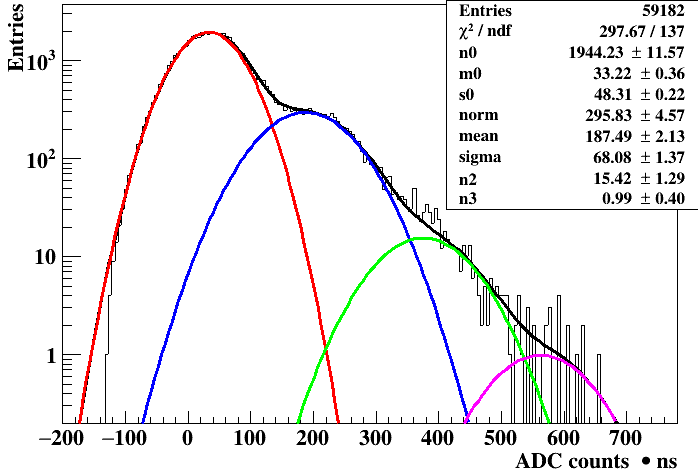}
  \caption{SPE events of the PMT in logarithm scale. The peaks are described in the text: zero-PE (red), one-PE (blue), two-PE (green), and three-PE (pink).}
  \label{f:aveSPESpec}
\end{figure}

The integration window (283-416 ns) shown in Fig.~\ref{f:aveSPE} and the same cuts applied to the selected waveforms were then used for each waveform in the data file. The resulting PE response spectra are shown in Fig.~\ref{f:aveSPESpec}. The spectra were fit in the same way as described in Ref.~\cite{ds1013,ding20e}. The red Gaussian curve is fit for the pedestal events, the blue Gaussian curve is fit for the SPE events, and two-PE (green curve) and three-PE fit (pink curve) were Gaussian functions based on the mean$_\text{SPE}$ and sigma obtained from the SPE fit. The black curve is the sum of those fit results, which matches well the single PE response. The area of single PE (mean$_\text{SPE}$) would be, as mentioned above, mean$_\text{SPE}$ = mean - m$_{0}$, where mean and m$_{0}$ are obtained from the fit result in Fig.~\ref{f:aveSPE}. The value of mean$_\text{SPE}$ is 154.27 ADC counts$\cdot$ns, and the value of m$_{0}$ is 33.22 ADC counts$\cdot$ns. These values will be further used in the light yield calculation. 

To estimate the systematic uncertainty in the determination of the mean value of the SPE distribution, multiple measurements were performed. The discrepancy is within 5\%. In the energy calibration measurements to be mentioned in the next section, the SPE spectrum with the crystal was used but with a 5\% uncertainty attached to be conservative.

\subsubsection{Energy calibration}
\hspace{0.5cm} The energy calibration was performed using an $^{55}$Fe source and an $^{241}$Am source. The digitizer was triggered when the height of a pulse from the PMT was more than 50 ADC counts ($\sim$2 PE). As can be seen in Fig.~\ref{f:OnePE}, the height of a single PE pulse was around 25 ADC counts.  The trigger threshold therefore suppresses most of the electronic noise spikes while letting pass most of the PE pulses. The trigger rate was $\sim 2.3$~kHz when the threshold was set to this value.

\begin{figure}[ht!]\centering
    \includegraphics[width=0.8\linewidth]{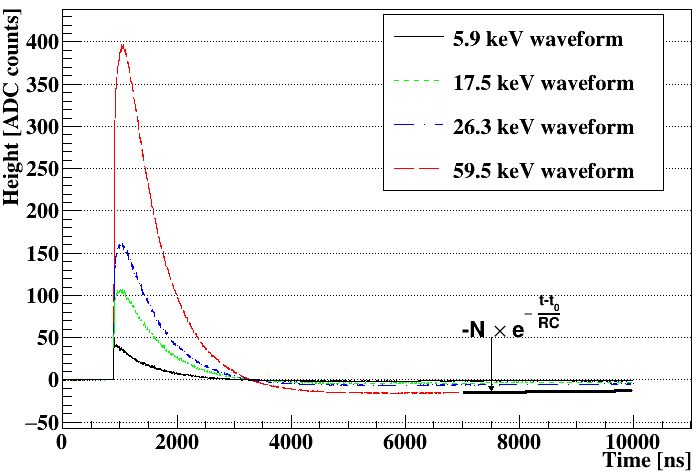}
  \caption{Averaged waveforms with negative overshoot. The tail part of the waveform was fit by an exponential function to estimate the negative overshoot effect.}
  \label{f:Am}
\end{figure}

Each recorded waveform was 10000~ns long as shown in Fig.~\ref{f:Am}. The data acquisition system was set up to record 1000 ns of pre-trigger waveform so that there were enough samples before the pulse to calculate the averaged pedestal value of the waveform, and 200 samples starting from zero were used to calculate the baseline. The pedestal was then adjusted to zero using the method described at section \ref{s:spe}.

By checking a few waveforms, negative overshoot was commonly observed. To further identify whether negative overshoot was common, averaged waveforms were examined. As shown in Fig.~\ref{f:Am}, averaged pulses were computed by summing all the waveforms first, then dividing by the number of events. The tallest one is the averaged 59.5 keV waveform, the second tallest one is the averaged 26.3 keV waveform, the second smallest one is the averaged 17.5 keV waveform, and the smallest one is the averaged 5.9 keV waveform. In the figure, the negative overshoot effect from the PMT can be clearly seen. 

\subsubsection{Overshoot correction of the PMT}
\begin{figure*}[ht!]\centering
  \includegraphics[width=\linewidth]{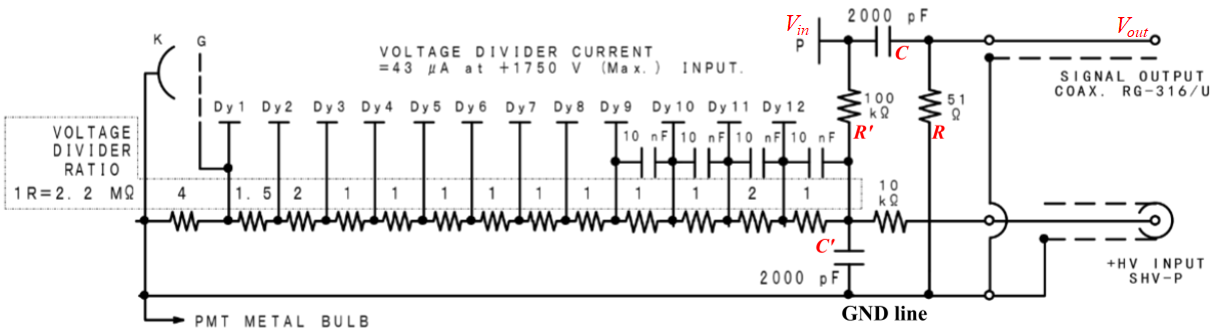}
  \caption{Circuit of the PMT from Hamamatsu.}
  \label{f:Ocircuit}
\end{figure*}

\hspace{0.5cm}Fig.~\ref{f:Ocircuit} shows the high voltage distribution circuit and the readout scheme of our PMT given by its manufacturer, Hamamatsu Photonics K.K. A 2,000 pF capacitor, $C$, is used to decouple the output line from the anode biased at high voltage. The 51 $\Omega$ load resistor, $R$, is used to match the impedance of typical oscilloscopes and digitizers. The waveform of $V_\text{out}$ can be tuned by selecting the values of $C, C', R$, and $R'$ as described in detail in Ref.~\cite{overshoot}. The values of these passive components must have been fine-tuned by Hamamatsu to reduce the negative overshoot. The previous measurements utilizing the same PMT and amplifier, we did not observe any negative overshoot. However, their values may have changed over time and through multiple thermal cycles, which resulted in the negative overshoot shown in Fig.~\ref{f:Am}. Figure~\ref{f:Am} is an example from one of the measurements in this chapter using an $^{241}$Am source to demonstrate the overshoot correction process.

As the $RC$ circuit is the origin of the negative overshoot, it is possible to correct this effect offline. A simple numerical method was developed to achieve this. Its derivation is explained step by step here. First, the current going through the load resistor $R$ can be expressed as
\begin{equation}
  I=\frac{V_{out}}{R}=\frac{dQ}{dt},
\end{equation}

where $Q$ is the charge accumulated in $C$. It can be expressed as
\begin{equation}
  Q=C(V_{in}-V_{out}).
\end{equation}

Combing the two equations, we have
\begin{equation}
  \frac{dQ}{dt}=C(\frac{dV_{in}}{dt} - \frac{dV_{out}}{dt} )=\frac{V_{out}}{R},
\end{equation}

which can be rearranged as
\begin{equation}
  V_\text{out} = RC\left(\frac{dV_\text{in}}{dt}-\frac{dV_\text{out}}{dt}\right)
\end{equation}

Numerically, this can be written as
\begin{equation} \label{e:vout}
  V_{out}[i]=RC\left(\frac{V_{in}[i]-V_{in}[i-1]}{\Delta t}-\frac{V_{out}[i]-V_{out}[i-1]}{\Delta t}\right),
\end{equation}

where $i$ is the index of individual samples in the waveform, and $\Delta t = 1$~ns is the time interval between two consecutive samples. The iterative 
\begin{equation} \label{e:vin}
  V_{in}[i]=\frac{RC+\Delta t}{RC}V_{out}[i]-V_{out}[i-1]+V_{in}[i-1].
\end{equation}
The $RC$ constant in Eq.~\ref{e:vin} was measured to be $18,044$~ns by fitting a simple exponential function to the averaged waveforms as shown in Fig.~\ref{f:Am} in the range of $[7,000, 10,000]$ ns, where the influence of scintillation decay (1000 ns~\cite{csi}) can be neglected and the influence of the $RC$ circuit persists. To estimate the uncertainty of this constant, we varied the starting point of the fitting from 7,000 to 6,000 and 8,000 ns, respectively. The $RC$ constant given by the fitting is 17574~ns in the range of ranges of $[6,000, 10,000]$~ns, and 19922~ns in $[8,000, 10,000]$~ns. The different RC constants eventually result in a less than 3\% difference in the pulse area values ($A$).

\begin{figure}[ht!]\centering
  \includegraphics[width=0.8\linewidth]{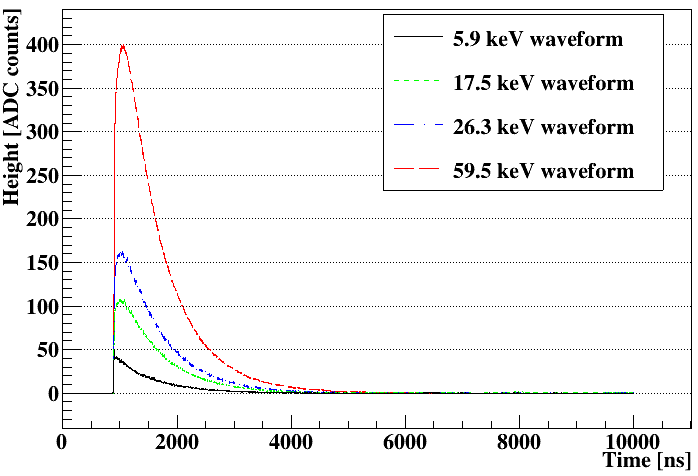}
\caption{Averaged waveforms after offline electronic correction of the PMT.}
\label{f:crt}
\end{figure}

Therefore, V$_{out}$ can be corrected sample by sample once the RC constant was identified. The averaged waveforms after correction shown in Fig.~\ref{f:crt} demonstrate the success of this correction method and the reasonableness of the RC constant. The same correction method with the same RC constant was then applied to each waveform. This method produces reasonable waveforms and doesn't make any assumptions on the input signal.

\begin{figure}[ht!]\centering
  \includegraphics[width=0.8\linewidth]{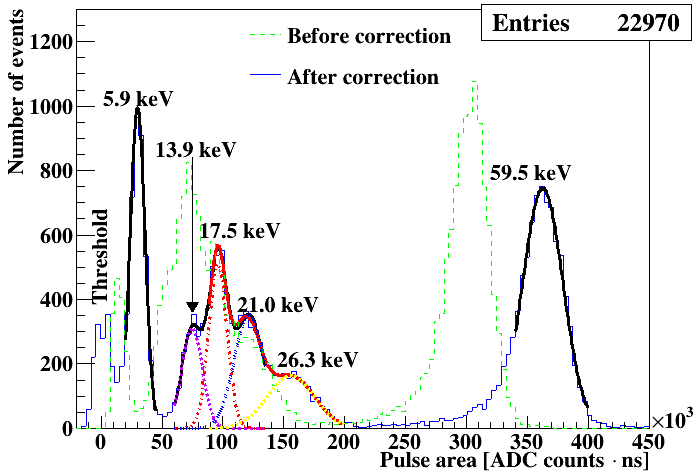}
  \caption{Energy spectra of $^{55}$Fe and $^{241}$Am in units of ADC counts$\cdot$ns.}
  \label{f:spectra}
\end{figure}

Every waveform after the correction was integrated from zero to the end. The integration had a unit of ADC counts$\cdot$ns. In Fig.~\ref{f:spectra}, the recorded energy spectra before correction are the green dashed line histogram, and the energy spectra after correction are the blue solid line histogram. A dominant 5.9 keV peak from $^{55}$Fe can be seen, 17.5 keV, 26.3 keV and 59.5 keV peaks from $^{241}$Am are shown as well. Fits were applied to the corrected spectra. The Gaussian fit was applied to the 5.9 and 59.5 keV peaks. The combined fit was applied to 13.9, 17.5 and 21.0 keV peaks, and then applied to 17.5, 21.0 and 26.3 keV peaks. Pulse area means obtained from those fits are summarized in~\autoref{t:NPE}, and whether the light response stays consistent in the energy close to the threshold is evaluated in the next section. 

\subsubsection{Light yield}
\hspace{0.5cm} The pulse area values ($A$) of the radiation pulses in the unit of ADC counts$\cdot$ns were converted to the number of PE ($N_\text{PE}$) using the formula:
\begin{equation}
 N_\text{PE} = \frac{A-m_0}{\text{mean}_\text{SPE}}.
\end{equation}

The shift value, $m_{0}$, is added to account for the overall shift of the pulses observed in the single PE measurement. However, compared to the pulse area values, the shift is small.

The light yield ($Y$) for a given energy deposited ($E_\text{dep}$) in our electron recoil measurements was then calculated using the following equation:
\begin{equation}
  Y [\text{PE/keV}_\text{ee}] = \frac{N_\text{PE}}{E_\text{dep}}.
\end{equation}

\begin{table*}[ht!]
  \caption{\label{t:NPE} Fit results along with the calculated light yield of $^{59}$Fe and $^{241}$Am peaks in the energy spectrum are shown. The top table is from the new cryostat, whereas the bottom table is from the old cryostat.}
  \begin{minipage}{\linewidth}\centering
  \begin{tabular}{l @{\extracolsep{\fill}}ccccccc}
    \toprule
    Type of & Energy & Mean (A) & Sigma & FWHM & Light yield & Uncertainty\\
   radiation & [keV] & [ADC$\cdot$ns] & [ADC$\cdot$ns] & \% & [PE/keV$_\text{ee}$] & [PE/keV$_\text{ee}$] \\
    \midrule
     \begin{tabular}{r} x-ray\\ x-ray\\ x-ray \\ x-ray \\ $\gamma$-ray \\ $\gamma$-ray \\
     \end{tabular} &
     \begin{tabular}{l} 5.9 \\ 13.9$^\dagger$ \\ 17.5$^\dagger$ \\ 21.0$^\dagger$ \\ 26.3$^\dagger$ \\ 59.5 \\
     \end{tabular} &
     \begin{tabular}{r} 29984 \\ 74709 \\ 95885 \\ 121409\\ 155757 \\ 362693 \\
     \end{tabular} &
     \begin{tabular}{r} 6138 \\ 8330 \\ 7516 \\ 12000 \\ 19322 \\ 16607 \\
     \end{tabular} &
     \begin{tabular}{r} 48.2 \\ 26.3 \\ 17.8 \\ 23.3 \\ 29.2 \\ 10.8 \\
     \end{tabular} &
     \begin{tabular}{r} 33.4 \\ 35.3 \\ 36.0 \\ 38.0 \\ 38.9 \\ 40.0 \\
     \end{tabular} &
    \begin{tabular}{r} $\pm$ 2.0 \\ $\pm$ 2.2 \\ $\pm$ 2.2 \\ $\pm$ 2.3 \\ $\pm$ 2.3 \\ $\pm$ 2.4 \\
     \end{tabular} \\
  \bottomrule
     \end{tabular}
  \end{minipage}
  
  \begin{minipage}{\linewidth}\centering
  \begin{tabular}{l @{\extracolsep{\fill}}ccccccc}
    \toprule
    Type of & Energy & Mean (A) & Sigma & FWHM & Light yield & Uncertainty\\
   radiation & [keV] & [ADC$\cdot$ns] & [ADC$\cdot$ns] & \% & [PE/keV$_\text{ee}$] & [PE/keV$_\text{ee}$] \\
    \midrule
     \begin{tabular}{r} x-ray\\ $\gamma$-ray \\ $\gamma$-ray \\
     \end{tabular} &
     \begin{tabular}{l} 5.9 \\ 26.3 \\ 59.5 \\
     \end{tabular} &
     \begin{tabular}{r} 32261 \\ 166507 \\ 354948 \\
     \end{tabular} &
     \begin{tabular}{r} 5454 \\ 14310 \\ 13264 \\
     \end{tabular} &
     \begin{tabular}{r} 39.8 \\ 20.2 \\ 8.8 \\
     \end{tabular} &
     \begin{tabular}{r} 35.9 \\ 41.6 \\ 39.2\\
     \end{tabular} &
     \begin{tabular}{r} $\pm$ 2.2 \\ $\pm$ 2.5 \\ $\pm$ 2.4\\
     \end{tabular} \\
     \bottomrule
     \end{tabular}
  \end{minipage}
  $^\dagger$ Intensity averaged mean of x-rays near each other~\cite{ding20e}. The propagation of uncertainty in the light yield is 6\% after accounting for the uncertainties associated with SPE measurements and waveform corrections. \\
\end{table*}

The calculated light yields in different setups are shown in \autoref{t:NPE}, similar light yields observed from two cryostats confirm the cooling capability of the new cryostat. 

In the new cryostat, the light yield from the $^{59}$Fe 5.9 keV radiation is 33.4 $\pm$ 2.0 PE$/$keV$_\text{ee}$, while the light yield from the $^{241}$Am 59.5 keV radiation is 40.0 $\pm$ 2.4 PE$/$keV$_\text{ee}$. The variation in the light yield at different energies is also seen in the old cryostat setup, with a light yield of 35.9 $\pm$ 2.2 PE$/$keV$_\text{ee}$ from the 5.9 keV radiation and a light yield of 39.2 $\pm$ 2.4 PE$/$keV$_\text{ee}$ from the 59.5 keV radiation. Despite the lower light collection efficiency in the crystal in the new cryostat and the intrinsic light yield would decrease if the temperature increases~\cite{Amsler02}, we measured similar light yields. All of these factors suggest that the crystal's temperature in the new cryostat cannot be significantly different from 77~K.

\subsubsection{Light yield nonlinearity discussion}
\label{s:lyn}
The observed nonlinearity of pure CsI crystal at different energies was also seen in Ref.~\cite{csi, csi20, ding20e, ding20, ding22, LY_moszynski_energy_2005, LY_gridin_channels_2014, LY_kerisit_computer_2009}. Crystals from different vendors show slightly different behavior. Light yield decreases slightly as energy goes down. One possible explanation is that low-energy radiation cannot penetrate far into the crystal, the survivability of optical photons created around the surface of the crystal depends highly on the local surface condition; while high-energy radiation on average creates optical photons deep inside the crystal, their survivability does not depend on the surface condition. 

\subsection{Quenching factor measurements at USD}
\label{s:qfusd}
\begin{figure}[ht!]
  \includegraphics[width=\linewidth]{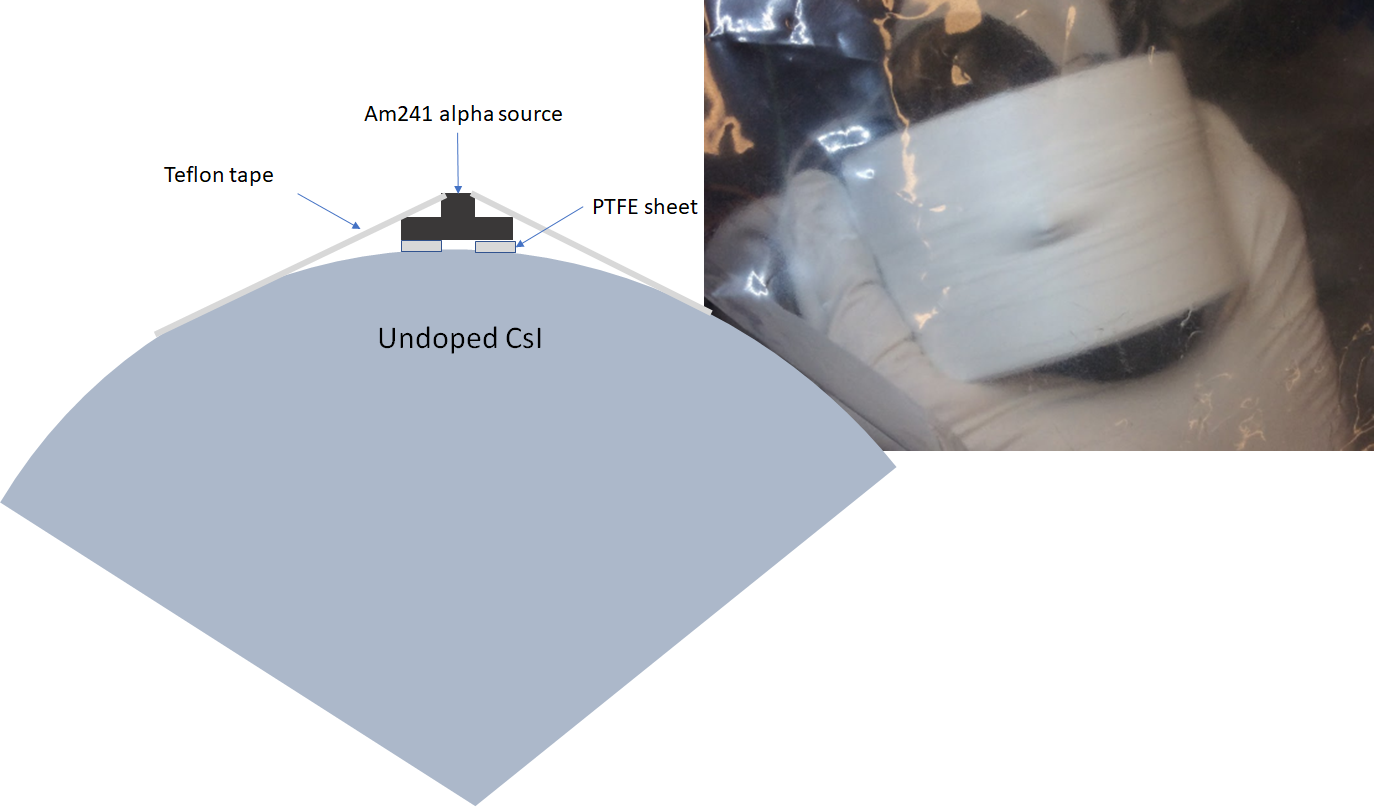}
  \caption{The nuclear quenching factor measurements from alpha-particles.}\label{f:alpha}
\end{figure}

\hspace{0.5cm} The QF measurements for alpha-particles in undoped CsI crystals were conducted using an $^{241}$Am source, as depicted in Fig.~\ref{f:alpha}. A piece of PTFE sheet was carefully drilled to create a small hole, enabling some alpha particles to penetrate into the undoped CsI crystals. Subsequently, the crystals and the source were enveloped in three to four layers of Teflon tapes. 

\begin{figure}[ht!] \centering
  \includegraphics[width=\linewidth]{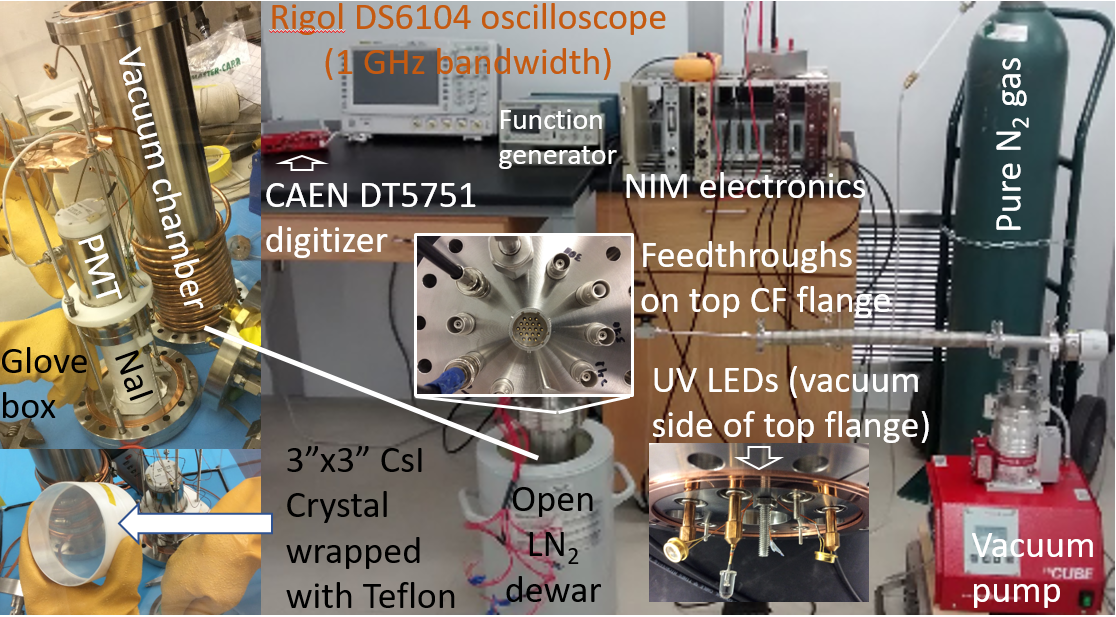}
  \caption{Conceptual sketch and pictures of existing general purpose liquid nitrogen cryostat to measure light yields of cryogenic crystals and characterize SiPMs and front-end electronics.} \label{f:fusd}
  \vspace{0.5cm} \noindent
  \begin{tabularx}{\linewidth}{rc>{\raggedright\arraybackslash}X}\hline
    Item & Quantity & Comment\\\hline
    Customized 2.75" \& 6" CF flanges & 5 & for mounting structures \& feedthroughs\\
    Long 2.75" \& 6" CF nipples & 2 & main parts of vacuum chambers\\
    UV LEDs & 5 & peaking at 300~nm, 340~nm, 375~nm, 420~nm \\
    PT100 connected to Raspberry Pi& plenty & for automatic temperature logging\\
    Open LN2 dewar & 2 & for convenient cooling and warming operations\\
    Turbo pump & 2 & Pfeiffer HiCube 80\\
    \hline
  \end{tabularx}
\end{figure}

All measurements were done in a general purpose LN$_2$ cryostat shown in Fig.~\ref{f:fusd}. The cryostat offers a large internal volume and multiple electronic feed-throughs to host various crystals and light sensors.

The QF is defined as the ratio of the signal amplitudes induced by an alpha-particle and an electron of the same energy:
\begin{equation}
  \text{QF} = LY_\alpha/LY_e,
  \label{e:QF}
\end{equation}
where $LY_\alpha$ is a light yield of an alpha-particle, and $LY_e$ is a light yield of an electron. 

The QF results for the OKEN, SICCAS and AMCRYS CsI crystals at room temperature are presented in Table~\ref{t:qf}. In comparison to the alpha-particle QF measurement at 77~K in Fig.~\ref{f:aq}, where the QF is nearly 1, our measurements indicate nuclear quenching. This implies that the same energy of an incoming particle deposits less energy through nuclear recoil than through electron recoil.
 \begin{table}[ht!]\centering
 \captionof{table}{Summary of the quenching factor measurements for the OKEN and SICCAS CsI crystals with the Am-241 source.
}\label{t:qf}
 \begin{tabular}{ccccc}\hline
    Crystal size & Origin & Am-241 collimator & Quenching \\
    & & hole size / position & factor \\
    \hline
    $\varnothing 2'' \times 10$~cm & OKEN & $\varnothing$0.6~mm / Top & $0.70 \pm 0.04$\\
    $\varnothing 3'' \times 5$~cm & SICCAS & $\varnothing$0.4~mm / Top & $0.62 \pm 0.03$\\
    & & $\varnothing 1.0$~mm / Side & $0.62 \pm 0.03$\\
    & & $\varnothing 0.4$~mm / Bottom & $0.64 \pm 0.03$\\
    $\varnothing 3'' \times 3''$ & AMCRYS & $\varnothing 0.7$~mm / Top & $0.61 \pm 0.03$\\
    \hline
 \end{tabular}
 \end{table}

\subsection{Quenching factor measurement at TUNL}
\label{s:tunl}
\hspace{0.5cm} To minimize neutron scatterings on the surrounding materials, a new cryostat has been established. As depicted in Fig.~\ref{f:cryostat}, the crystal can be cooled from one end by an LN$_2$ pipe and scintillation light can be collected from the other end with a light sensor. Materials in between the TUNL neutron beam and the crystal are only two thin layers (1~mm) of Al for infrared radiation shielding and maintaining the vacuum.

\begin{figure}[ht!]
  \includegraphics[width=\linewidth]{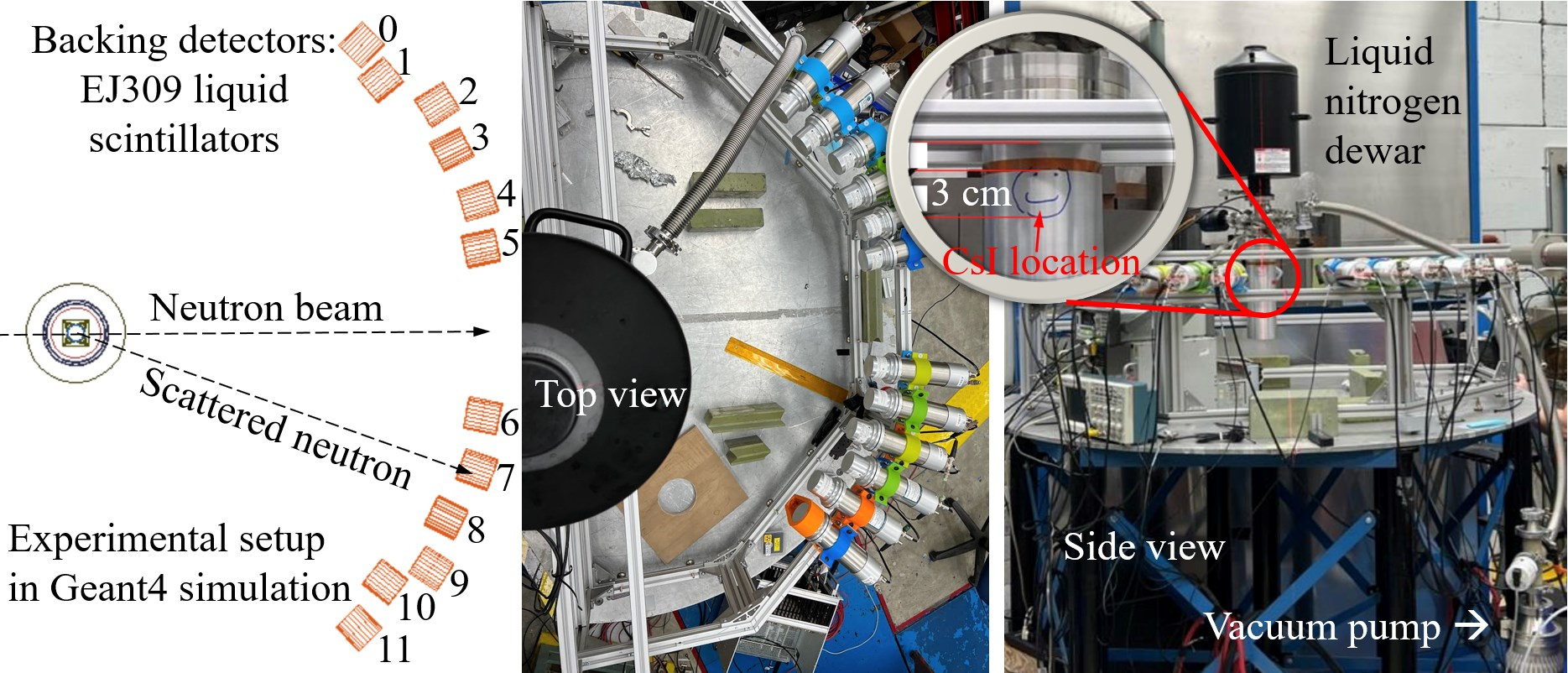}
  \caption{Geant4 simulations and measurements of recoils at TUNL.}\label{f:tunl}
\end{figure}

A comprehensive Geant4 Monte Carlo simulation of this TUNL setup for QF measurements has been performed, as depicted in Fig.~\ref{f:tunl}. A week-long measurement with a monoenergetic neutron beam at 700 keV is anticipated to provide sufficient statistics for analysis based on the simulation. Our initial measurement and analysis have been conducted, yielding a preliminary result of approximately 15\% for the QF. Details will be disclosed in a subsequent paper, which is beyond the scope of this study.

\subsection{QF measurements discussion}
\hspace{0.5cm}The significant discrepancy between the QF measured with alpha-particles and neutron beams may be attributed to differences in track lengths. Similar to electrons carrying the same energy, recoiled electrons travel farther compared to recoiled ions. Given that alpha-particles are lighter than the recoiled ions, they traverse longer distances than the ions excited by incoming neutrons, neutrinos, or DM. Consequently, this leads to a higher QF compared to measurements obtained with neutron beams. Considering that CEvNS occurs on a keV scale, measuring the QF with neutron beams would be more appropriate.

As illustrated in Figure \ref{f:aq}, both theoretical and experimental evidence suggests that QF may vary with temperature. If this variation aligns with the trend observed for light yield versus temperature, the optimal operating temperature would be around 40 K. Otherwise, the operating temperature needs to be determined by maximizing the product of the yield and the QF in the region where the afterglow rate is minimal. Therefore, the primary goal is to investigate the QF below 77~K.

\section{Designing a 10-kg prototype}
\label{s:design}
\hspace{0.5cm} The following chapter will discuss the shielding design, providing insights into the rationale behind our design choices. Additionally, I will present optical simulations aimed at optimizing the shape of the CsI crystal. It's important to note that the investigations presented here do not represent the final version, as optimization efforts are still ongoing.
\subsection{Shielding design}
\begin{figure}[!ht]\centering
    \includegraphics[width=0.8\linewidth]{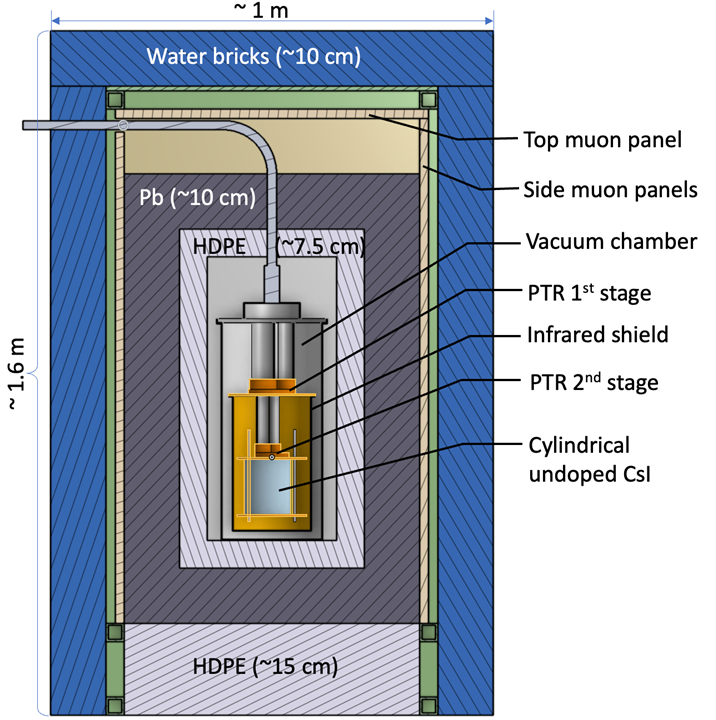}
    \caption{CryoCsI design with shielding structures.}
    \label{f:design}
  \end{figure}

\hspace{0.5cm} Shielding plays a crucial role in R\&D detection as it aims to capture desired events while minimizing contamination from background sources. Common background sources include muons, gamma rays, neutrons, and internal contamination. To mitigate these backgrounds, a full Geant4 simulation was conducted, resulting in a minimally sized structure depicted in Fig.~\ref{f:design}. The complete simulation code can be accessed at GitHub~\cite{cryocsiG4}. Our shielding design bears resemblance to that of the original COHERENT CsI\cite{CsICEvNS}. The detector is enveloped by a 1.5 cm layer of high-density polyethylene (HDPE), which serves to reduce neutron background. This HDPE layer is followed by 3 cm of low-background lead to shield against gamma rays. Surrounding the lead shield is a 1-inch-thick high-efficiency muon veto on all vertical sides and the top, anchored by a heavy-duty aluminum frame made of Bosch extrusions. This frame provides stability for the muon veto panels and facilitates the addition of aluminum frame, which are filled with 2~cm of water bricks to serve as a neutron moderator. 

\subsubsection{Muon veto}
\hspace{0.5cm} Muons, as charged particles (mu$+$ and mu$-$), interact with matter by ionization, gradually losing energy in the process. According to HyperPhysics~\cite{mu}, the average energy of muons reaching sea level is approximately 4 GeV. Upon passing through the atmosphere, a muon typically loses around 2 GeV, resulting in an original energy of about 6 GeV. Both mu$+$ and mu$-$ can originate from protons in the atmosphere, with the ratio of mu$+$ to mu$-$ estimated to be around 1.3 according to Ref.~\cite{muRatio}. For detailed momentum and angular distributions of cosmic ray muons at sea level, one can refer to Ref.~\cite{mudis}.

Muon veto detectors primarily fall into four categories: nuclear emulsion, cloud chamber, scintillator, and silicon tracker. In our setup, we opted for plastic scintillator due to its cost-effectiveness and flexibility. In our simulation, 10,000 mu$+$ particles were homogeneously shot straight and at various angles above the top panel, with muon energies of 100 MeV, 1 GeV, 4 GeV, and 20 GeV. Muons at 4 GeV were found deposited the least energy in a panel. Notably, vertically penetrating muons deposited the least energy due to their shorter ionization track length within the panel. The 2 cm thick plastic scintillation panel enables the differentiation of muon (vertical incidence, 4 GeV) and gamma energy distributions, facilitating the identification of muons at varying energies and incident angles. The efficiency of a single 1-inch-thick muon panel to detect passing through muons at sea level is nearly 100\%.

\begin{figure}[!ht]\centering
  \includegraphics[width=0.4\linewidth]{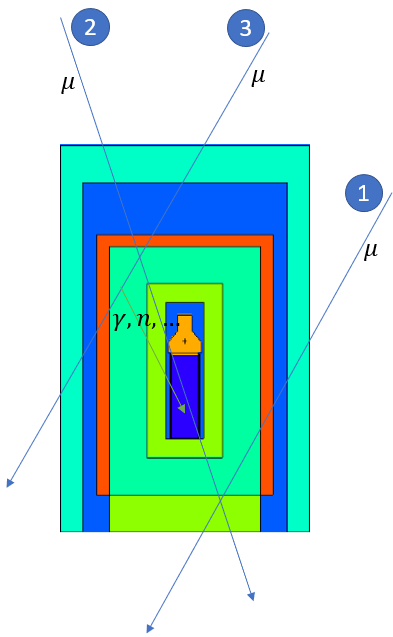}
  \caption{Three types of events in the muon panels.}
  \label{f:muD}
\end{figure}

Dead time induced by the muon veto system is a concern, as depicted in Fig.~\ref{f:muD}. Three types of events are discussed here. Event 1 is not relevant since there is no signal in the CsI. Event 2 can be vetoed, as both the panel and the CsI detect signals simultaneously. However, event 3 presents a challenge. If secondary particles reach the CsI shortly after the primary muon, a veto time window can be establish , during which any CsI event will be discarded, thereby introducing dead time. If secondary particles reach the CsI long after the primary muon, discerning a connection between the two events may become challenging. The frequency of such occurrences needs to be investigated once the entire shielding structure is established~\cite{cryocooler,pt310}.

\subsubsection{Gamma shielding}
\hspace{0.5cm} X-rays and gamma-rays primarily interact with matter through three processes: the photoelectric effect, Compton scattering, and pair production. At low energies, the photoelectric effect is predominant, while Compton scattering becomes dominant around MeV energies. Above 10 MeV, pair production prevails.

The plastic's radioactive background originates mainly from the natural decay series of uranium and thorium~\cite{gamma}, with radon being one of the daughter products. Radon is a significant concern for low-background experiments due to its widespread dispersal, and its daughter product $^{210}$Pb has a decay time of 22 years. To mitigate this, we prefer using ancient Roman lead for shielding to eliminate concerns about the 22-year decay of $^{210}$Pb. 

Lead, being a dense material, is effective for blocking gamma-rays generated in the muon panels. Lead shielding must be incorporated inside the muon veto to block gamma radiation effectively. The highest gamma-ray energy from natural decay chains is 2.6 MeV, originating from the decay of $^{208}$Tl in the $^{232}$Th decay chain. A 3 cm lead layer significantly reduces gamma events, successfully blocking 2.6 MeV gamma-rays. However, lead bricks are observed to produce a significant number of low-energy neutrons.

\subsubsection{Neutron shielding}
\hspace{0.5cm} Neutron interactions are typically categorized as neutron capture, neutron scattering, and neutron-induced fission. Neutrons originate from various sources, including beam-related neutrons, neutrino induced neutron, muon-induced neutrons, gamma-ray induced neutrons, and alpha-ray induced neutrons. To effectively block neutrons, it is advantageous to utilize light nuclei (e.g., hydrogen) to absorb the neutron's kinetic energy and slow them down through a process known as thermalization. Materials such as HDPE and water bricks are employed for this purpose. Therefore, water bricks are employed as the outermost layer to capture neutrons, while HDPE is positioned within the lead shielding to further diminish neutron events.

\subsubsection{Internal contamination}
\hspace{0.5cm} $^{40}$K is one of the most abundant sources of natural radioactivity~\cite{K40, supernova_2023} and is unavoidably present in the production of CsI crystals. To assess its impact, 10,000 events were simulated within a confined volume equivalent to the size of the CsI crystal and the distribution of energy within the crystal was analyzed. Unfortunately, low-energy events dominate below 40~keV. A future precise simulation is essential for effectively distinguishing this background from the low-energy events of interest.

\subsection{Geant4 optical simulations for CsI shape optimization}
\begin{figure}[ht!]\centering
    \includegraphics[width=0.45\linewidth]{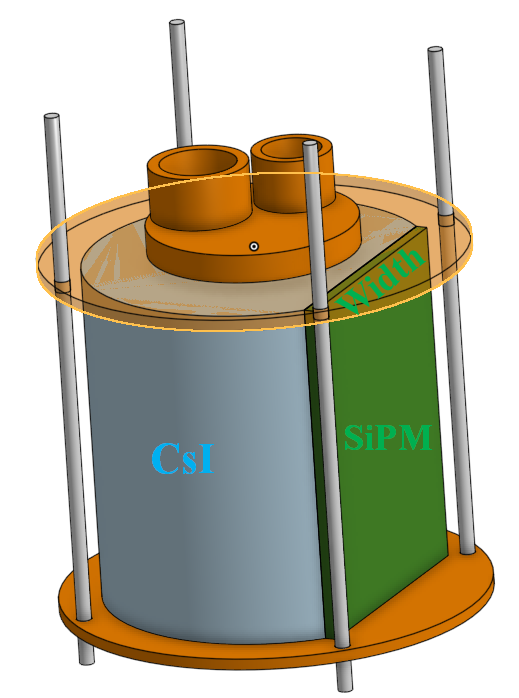}
    \caption{Potential CsI shape coupled with SiPM arrays.}
    \label{f:shape}
 \end{figure}

\hspace{0.5cm} Optimizing the shape of CsI crystal to maximize the detection of optical photons by the light sensors is also crucial for constructing CryoCsI. Geant4 optical simulations were conducted step by step to explore the optimal shape for the approximately 10 kg crystal. To build this CsI crystal, wrapped with Teflon tape and coupled with SiPM arrays, several factors need to be considered:
\begin{itemize}
  \item Material properties of CsI, Teflon tape and SiPM, such as scintillation components, absorption length, refractive index, etc.;
    \subitem these are defined after investigations, and details can be found in the GitHub repository~\cite{gears}.
  \item The model used to define the interface between CsI and Teflon wrapping;
    \subitem options include the Unified model, Glisur model, LUT, and LUTDAVIS~\cite{geant4}. 
  \item Whether to use polished, unpolished, or a combination of polished and unpolished surfaces for the crystal;
    \subitem more specifically, considerations include the roughness of the crystal surfaces.
  \item The optimal shape of the $\sim$10~kg CsI;
    \subitem the initial approach involves constructing a CsI cylinder with a diameter of approximately 6 inches and a height of approximately 6 inches. Subsequently, the side of the cylinder would be cut with rectangular cuboids to create flat surfaces for coupling SiPM arrays (see Table~\ref{t:sipm} for options). Figure~\ref{f:shape} depicts this scenario where one side of the CsI is cut. Possible combinations for one side cut include: SensL J SiPM arrays: 2 (width side) $\times$ 3 (height side), Hamamatsu S141xx SiPM arrays: 4 (width side) $\times$ 6 (height side), and Broadcom SiPM arrays: 6 (width side) $\times$ 9 (height side). If simulations indicate that doubling the light coverage significantly enhances the light yield, then cutting two sides may be considered as well.
    \subitem Alternative shapes could involve using two 3-inch diameter, 5-inch height cylinder CsI crystals and four 3-inch diameter, 2.5-inch height cylinder CsI crystals. These configurations would allow for the direct coupling of 3-inch PMTs on both the top and bottom surfaces of crystals. Such arrangements significantly reduce electronic readouts compared with SiPM arrays readouts.
  \end{itemize}

After defining the material properties, the next step is to define the optical surfaces between CsI and Teflon tapes. According to information from the Geant4 website~\cite{geant4}, the Unified model offers various interfaces and numerous parameters for customization, while the Glisur model resembles the Unified model closely. On the other hand, the LUT model is designed based on the properties of the BGO crystal, whereas the LUTDAVIS model is tailored to the characteristics of the L(Y)SO crystal. Considering that the refractive index of CsI at 340 nm (77~K) is 1.91~\cite{refractiveindexinfo}, compared to over 2.74 for BGO at the same wavelength~\cite{refractiveindexinfo}, and approximately 1.85 for L(Y)SO~\cite{LYSOrefractive}, the LUTDAVIS model appears to be more suitable for application to CsI between LUT and LUTDAVIS.

\begin{figure}[!ht]\centering
  \includegraphics[width=0.8\linewidth]{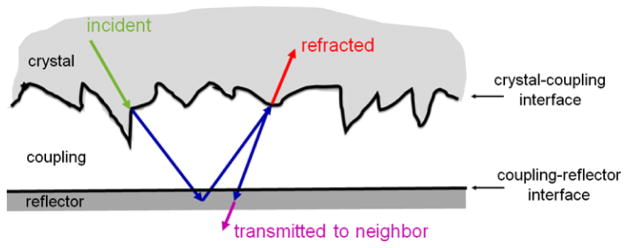}
  \caption{Crystal-reflector interface taken from Ref.~\cite{DAVIS}. The incident photon is shown in green, the refracted photon reflected by the reflector (with possible further reflections at interfaces bordering the coupling medium) is in blue. Photons ultimately transmitted through the reflector are shown in magenta, while photons re-entering the crystal are shown in red.}
  \label{f:interface}
\end{figure}

Starting with the LUTDAVIS model, one can specify a G4OpticalSurface using SetType (dielectric\_LUTDAVIS) and SetModel (DAVIS). Following this, surfaces can be selected, such as SetFinish (RoughTeflon\_LUT) and SetFinish (PolishdTeflon\_LUT), to define the rough or polished crystal surfaces coupled with Teflon. This crystal-reflector configuration is depicted in Fig.~\ref{f:interface}. 

To assess the simulations using the LUTDAVIS model alongside experimental results~\cite{polishedcsi}, two $2\times2\times2$ cm$^3$ cubes (one polished, one rough) and two $\Phi 2.5\times2$ cm$^3$ cylinders (one polished, one rough) were simulated. These samples had their side and bottom surfaces wrapped with Teflon, while the top surface was coupled with a PMT. In the experimental setups, an $^{241}$Am source was positioned at the center of the side surface of the CsI, outside of the Teflon wrapping. In the simulation, a 60~keV $\gamma$-ray was placed in the same location with a Teflon thickness of 0.2~mm. The penetration depth in the CsI was 0.24 mm. Subsequently, optical photons were generated at the point where the $\gamma$-ray stopped penetrating and were emitted isotropically inside the CsI. The experimental results and simulated results are summarized in Table~\ref{t:PoU}.
\begin{table}[ht!]\centering
  \caption{Comparison of the experimental light yields~\cite{polishedcsi}} and simulated results with different shapes and surface treatment.
  \label{t:PoU}
  \begin{tabular}{ccc}
  \hline
  Crystal & Experimental LY (PE/keVee) & Simulated result$^\dagger$ (\%)\\
  \hline
  Cubic (polished) & 35.2 & 33.1 \\
  Cylindrical (polished) & 33.9 & 34.3 \\
  Cubic (ground) & 24.8 & 73.2 \\
  Cylindrical (ground) & 22.3 & 74.4 \\
  \hline
  \end{tabular}

$^\dagger$ The simulated result was calculated by dividing the number of photons reaching the PMT by the total number of photons emitted in the CsI.\\
\end{table}

\subsubsection{Discussion of optical simulation results}
\hspace{0.5cm} The simulated results based on the LUTDAVIS model yielded significantly different outcomes compared to the experimental findings. In the experiments, more photons were detected in the polished CsI than the rough CsI. Given that the wavelength of L(Y)SO is 420 nm~\cite{lysoWL}, and the roughness of the rough surface is 480 nm~\cite{geant4}, while the wavelength of CsI is 340 nm~\cite{mikhailik15}, with the roughness of the rough surface being 800 nm~\cite{polishedcsi}. This may suggest that photons in the CsI are more likely to be trapped in the rough surface~\cite{polishedcsi}, leading to discrepancies between the simulated and experimental results using the LUTDAVIS model. To verify this assumption, future experiments need to be designed to confirm and measure this absorption effect based on the surface roughness of the CsI. 

Lacking experimental measurements on CsI surfaces, tuning parameters in the Unified model seems impractical. However, after conducting future experiments, we would be able to set the absorption coefficient on the surface accurately. Using a GroundBackPainted surface could be a viable option. Concerns arise regarding the unified model because it defines an ensemble of micro-facets determined by the distribution of normals, typically a Gaussian distribution, which may not accurately represent real surface conditions. A possible solution, as demonstrated in the LUTDAVIS model~\cite{davisSurface}, is to conduct 3D measurements of the surface using techniques like atomic force microscopy, and subsequently scan and compute the surface reflectance properties.

Proposed work, such as measuring the absorption effect on the surface of CsI corresponding to different roughness levels, and building a more accurate Geant4 surface model for CsI, would significantly impact the use and simulations of undoped CsI.

\section{Summary and outlook}
\hspace{0.5cm} Neutrinos, elusive yet abundant, represent fundamental particles in the SM, distinguished by their infrequent interactions with matter. CEvNS emerges as a vital tool for investigating these enigmatic particles. The COHERENT collaboration's observation of CEvNS has ignited substantial interest, not only for confirming a long-predicted standard neutrino interaction but also for its potential to probe various realms of new physics, including supernova dynamics, neutrino electromagnetic interactions, sterile neutrinos, non-standard neutrino interactions NSIs, and more. Chapter~\ref{s:cevns} delves into these aspects further, with particular emphasis on sensitivities of the proposed cryogenic undoped CsI detector, CryoCsI.

The global scientific community is increasingly invested in CEvNS research, evident from the proliferation of CEvNS experiments worldwide. Neutrino sources utilized in CEvNS detection can be categorized into two types: accelerator-based sources and nuclear reactors. These sources are selected for their high flux and appropriate energy range. To detect the subtle energy signatures of CEvNS events, researchers employ detectors based on phonon, scintillation and ionization signatures. The technical intricacies of these detectors, their interactions with different neutrino sources, and the significance of the sources, particularly the pion decay-at-rest source at the SNS, are detailed in chapter~\ref{s:map}.

The COHERENT experiment has made pioneering measurements in CEvNS using detectors employing CsI(Na), Ar, and Ge (soon to be released). Leveraging the intense neutrino flux and effective background rejection capabilities of the SNS beam, COHERENT achieved world-first observations of CEvNS. Detailed descriptions of the SNS and COHERENT detectors are provided in chapter~\ref{s:coherent}.

The proposed CryoCsI detector, an advancement over existing CsI(Na) detector, offers several key improvements, including the replacement of PMTs with SiPMs, cryogenic operation of SiPM arrays, and the substitution of CsI(Na) with undoped CsI at cryogenic temperatures. These enhancements detailed in chapter~\ref{s:yummy} aim to improve the energy threshold and sensitivity of CEvNS detection, ensuring better performance and broader applicability.

The development of CryoCsI involved meticulous measurements to enhance the light yield as described in chapter~\ref{s:e} and~\ref{s:esipm}. Over several years, significant progress has been made, with the light yield improving from 20 to 50 PE/keV$_{ee}$ (see Table~\ref{t:ly}). 

Despite the promising results, challenges persist with cryogenic SiPMs, including inferior energy resolution, optical cross-talk, and potential limitations on detecting rare events. Chapter~\ref{s:esipm} explores potential solutions to address these challenges.

Distinct pathways for energy transfer within materials involve electron recoils and nuclear recoils. Understanding the light yield of scintillating detectors for nuclear recoils is crucial for detecting incoming neutrino energies. Chapter~\ref{s:qf} details the progression from alpha-particle QF measurements to the more challenging neutron QF measurement ($\sim$15\%), necessitating the design of a new cryostat to minimize neutron scattering effects. Solutions were developed to address issues like the overshoot effect observed in PMTs. The future objective is to explore the QF at temperatures below 77 K to determine the optimal temperature for CryoCsI.

Chapter~\ref{s:design} delves into the design considerations behind the shielding and the optimization of the CsI crystal's shape. Optical simulations were conducted with the aim of minimizing background noise and identifying the most suitable Geant4 surface model for optimizing the crystal's geometry, which will provide valuable insights for enhancing CryoCsI's detection efficiency. The ongoing optimization efforts hold promise for further improving the performance of the CryoCsI detector. The proposed work, including measuring the absorption effect on the surface of CsI with varying roughness levels and creating a more precise Geant4 surface model for CsI, has the potential to greatly influence the utilization and simulations of undoped CsI.

In conclusion, the comprehensive investigations detailed in this thesis underscore the multifaceted nature of CEvNS research and the remarkable potential of the CryoCsI detector. Continued advancements in detector technologies and collaborative efforts suggest a promising future for CryoCsI in CEvNS detection, opening doors to groundbreaking discoveries.

\newpage
\addcontentsline{toc}{section}{Abbreviations}
\begin{center}
    \section*{}
    \textbf{\large Abbreviations}
\end{center}
\doublespacing

\textbf{APDs} avalanche photodiodes

\textbf{BRN} beam-related neutron

\textbf{BSM} beyond the Standard Model

\textbf{CC} charged-current

\textbf{CEvNS} coherent elastic neutrino-nucleus scattering

\textbf{CF} ConFlat

\textbf{CryoCsI} cryogenic undoped CsI 

\textbf{DCR} dark count rate

\textbf{DM} dark matter 

\textbf{FTS} First Target Station

\textbf{HDPE} high-density polyethylene

\textbf{IBD} inverse beta decay

\textbf{keV$_{ee}$} keV electron-equivalent

\textbf{keV$_{nr}$} keV nuclear-equivalent

\textbf{KIDs}  kinetic inductance detectors

\textbf{LDM} low-mass DM 

\textbf{LMA} large mixing angle

\textbf{NC} neutral-current

\textbf{NINs} neutrino-induced neutrons

\textbf{NSIs} non-standard neutrino interactions

\textbf{NTD} neutron transmutation doped

\textbf{NA} Neutrino Alley

\textbf{ORNL} Oak Ridge National Laboratory 

\textbf{$\pi$DAR} pion decay-at-rest  

\textbf{PDE} photon detection efficiencies

\textbf{PE} photoelectrons

\textbf{PMT} photomultiplier tube

\textbf{PVES} parity-violating electron scattering 

\textbf{QF} quenching factor

\textbf{SiPMs} silicon photomultipliers 

\textbf{SM} Standard Model

\textbf{SNS} Spallation Neutron Source

\textbf{STS} Second Target Station

\textbf{TES} transition edge sensor

\textbf{TPB} 1,1,4,4-Tetraphenyl-1,3-butadiene

\textbf{TPC} Time Projection Chamber 

\textbf{TUNL} Triangle Universities Nuclear Lab

\textbf{WIMP} Weakly Interacting Massive Particle

\newpage
\singlespacing
\addcontentsline{toc}{section}{References}
\printbibliography[]




\end{document}